\documentclass[pra,onecolumn,english,nofootinbib]{revtex4}

\usepackage[T1]{fontenc}
\usepackage[latin9]{inputenc}
\usepackage{babel}
\usepackage{graphicx,amsmath,amssymb,color}
\usepackage[normalem]{ulem}
\graphicspath{{./Images/}}
\usepackage{amsfonts}
\usepackage[toc,page]{appendix}
\usepackage{hyperref}
\usepackage{latexsym}
\usepackage{amsfonts}
\usepackage{algpseudocode}
\usepackage{amsthm}
\usepackage{mathrsfs}
\usepackage{natbib}

\usepackage{color,verbatim}
\usepackage{psfrag}
\usepackage{comment}
\usepackage{array}
\usepackage{subcaption}
\usepackage{float}
\usepackage[para,online,flushleft]{threeparttable}
\usepackage{tablefootnote}
\usepackage[export]{adjustbox}

\usepackage{pbox}

\usepackage[svgnames]{xcolor}
\usepackage{tikz}
\usetikzlibrary{quantikz}

\usepackage{multirow}

\usepackage{pdflscape}
\usepackage{afterpage}
\usepackage{capt-of}

\usepackage{lipsum}

\begin{document}

\title{ 
Using a quantum computer to solve a real-world problem -- what can be achieved today?}
\author{Robert Cumming$^\dagger$, Tim Thomas$^\dagger$}


\email{robert.cumming@appliedquantumcomputing.co.uk}
\address{$\dagger$ Applied Quantum Computing,    London, UK }

\begin{abstract}
Quantum computing is an important developing technology with the potential to revolutionise the landscape of scientific and business problems that can be practically addressed. The widespread excitement derives from the potential for a fault tolerant quantum computer to solve previously intractable problems.  Such a machine is not expected to be available until 2030 at least.  Thus we are currently in the so-called NISQ era where more heuristic quantum approaches are being applied to early versions of quantum hardware. In this paper we seek to provide a more accessible explanation of many of the more technical aspects of quantum computing in the current NISQ era exploring the 2 main hybrid classical-quantum algorithms, QAOA and VQE, as well as quantum annealing.  We apply these methods, to an example of combinatorial optimisation in the form of a facilities location problem. Methods explored include the applications of different types of mixer (X, XY and a novel 3XY mixer) within QAOA as well as the effects of many settings for important meta parameters, which are often not focused on in research papers. Similarly, we explore alternative parameter settings in the context of quantum annealing.  Our research confirms the broad consensus that quantum gate hardware will need to be much more capable than is available currently in terms of scale and fidelity to be able to address such problems at a commercially valuable level. Quantum annealing is closer to offering quantum advantage but will also need to achieve a significant step up in scale and connectivity to address optimisation problems where classical solutions are sub-optimal.    
\end{abstract}
\maketitle

\vspace{1 cm}

\tableofcontents  

\newpage

\section{Introduction}


Quantum computing has been seen as a potentially revolutionary technology ever since it was first suggested as an exciting possibility by Richard Feynman back in 1982 \cite{Feynman82a}.  At that time it was little more than an exotic concept albeit with enormous attractions which, if it could be realised, would be able to address many intractable problems that were beyond the scope of classical computation techniques.  In the nearly 40 years since then enormous strides have been made initially in terms of developing a theory of how quantum computation could work and over the last $\sim${}15 years in the practicalities of realising first qubits and more recently small scale quantum computers.  Now however as we move into the 2020s, the prospect of quantum computers as tools of commercial and scientific significance is starting to look like a reality.  Both Google and IBM have laid out so-called 10 year ``road maps'' to the production of large scale machines with $\sim${}1 million physical qubits and an, as yet, unspecified number of logical qubits.  Each is looking to produce quantum computers with $\sim${}200-1000 physical qubits by 2023 \cite{IBM20a, Google21a}.  At such a size, if hybrid classical-quantum algorithms can be further developed, there is a good possibility that quantum computers will be able to solve problems of commercial value beyond the scope of classical computers.  In parallel, quantum annealing technology has also been advancing with the latest generation in the form of the D-Wave Advantage released at the end of September 2020 \cite{DWave20a}.  Its capabilities, although less universal than those of a quantum gate computer, are considerable and again, may open a pathway to addressing commercially valuable problems by 2025.  Of course none of these future developments is certain but we are definitely entering a period of huge importance for quantum computing technology and the industry developing it.

Simultaneously there has been a major research effort to advance the algorithmic techniques that can be deployed using quantum computers and identify specific applications where near-term commercially valuable advantage  may be achieved. Quantum algorithmic techniques have been developed with potential application to the fields of optimisation, quantum simulation and quantum machine learning \cite{Preskill18a}. Further, many example proof-of-concept applications have been demonstrated including in the areas of scheduling \cite{Venturelli15a}, air traffic control \cite{Stollenwerk19a}, optimising radar waveforms \cite{coxson14a}, wireless networks \cite{Choi20a}, seismology \cite{Souza20a},  hydrology \cite{omalley18a}, modelling small molecules  \cite{Kandala_2017a}, transport optimisation \cite{crispin13a, Feld19a, Neukart17a,Yarkoni20a}, and finance \cite{Orus18a,marzec16a}. 



For the commercial user, there is a great challenge in going from a problem of commercial interest to solving this effectively using the new tools and capabilities of quantum computing. Specifically there is the need to 
\begin{itemize}
    \item Identify problems which lend themselves to potential solution making use of a quantum computer
    \item Formulate each problem of interest in a way conducive to solution using a quantum computing techniques
    \item Choose the most appropriate quantum hardware and/or algorithmic techniques to ensure the best fit with the problem definition and hence speed and ease of solution
    \item Select and implement the most suitable classical/quantum hybrid algorithm including specifying items such as the choice of classical optimizer, encoding approach and a variety of important meta-parameters so as to maximise the quality of the solution
\end{itemize}

From the perspective of a potential commercial user much of the technical literature on quantum computing is opaque and hard to understand except to those possessing an advanced relevant mathematics, computing or physics background and preferably a combination of all three!  In this paper we explore the application of some of the most promising quantum algorithms to a real world problem, albeit at a small scale, and do so in such a way as to offer greater transparency than is typical and, we hope, a deeper understanding to interested commercial actors on how quantum computing may be applied given the current state of quantum computing development. The problem we select is that of ambulance location and dispatch, an example involving at its heart optimisation: where should a finite number of ambulances be positioned within a geographical area in order to optimise the service provided?.   This is self-evidently a critical service of universal relevance but one where the computational challenges are such that classical computing is unable to provide proven optimal answers.  We work through the journey of taking the high level ambulance problem and solving it using the new tools and capabilities of quantum computing.




\subsection{Quantum computing - relevant hardware and algorithmic background} \label{QC background}

There are 2 very distinct types of quantum computing hardware platforms which can be used to generate potential solutions to combinatorial based problems such as the ambulance location problem -- these are quantum gate computers and quantum annealers.  We will make use of both hardware styles of platform to compare and contrast their respective capabilities and performance.

Quantum gate computers are the type of computer which implement quantum information science \cite{nielsen00a} which was largely developed in the 1990s  and which in its ideal form gives rise to error corrected quantum computation that can complete algorithms requiring very large numbers of operations (completed by quantum gates) and which is not constrained by the number of qubits nor their connectivity.  Such computers would unleash the full potential of quantum computing being able to deliver polynomial and/or exponential gains in performance including, for example the ability  to implement Shor's famous factorising algorithm \cite{365700}, which fundamentally undermines the security of most current internet encryption methodologies. However, despite recent rapid progress in the development of quantum gate computers by companies such as Google, IBM, IonQ, Quantinuum, and Rigetti \cite{Google21a, Quantinuum22a, IBM20a, ionq21a, rigetti21a} the quantum computing industry is still many years away from offering practical error corrected quantum computers.  This is due to a shortfall in the required gate fidelity, relatively limited qubit connectivity and the largest machines currently having ${\sim}100$ qubits.   

We are thus in the so-called NISQ (Noisy Intermediate-Scale Quantum) era as identified by John Preskill \cite{Preskill18a}.  This requires the use of noise compatible or noise tolerant computing techniques. Of these the most promising for combinatorial optimisation problems are the Quantum Approximate Optimisation Algorithm \cite{Farhi14a} and its more general evolution, the Quantum Alternating Operator Ansatz \cite{Hadfield17a}; both approaches are typically abbreviated to ``QAOA''.  In addition researchers are also working with the Variational Quantum Eigensolver (``VQE''), a similar but less complex form of quantum minimisation algorithm. Both are hybrid quantum - classical algorithms which make use of parameterised rotation gates to explore the solution space of a given problem and a classical optimiser to determine the values of the parameters that generate the minimum value.  By examining the component states in the solution wavefunction corresponding to the minimum, it is possible to obtain the optimal state --  this corresponds to the solution to the targeted problem.  This is of course a probabilistic approach, so that determining the optimal state using this method is not certain although it can be achieved with a high or at least good probability depending on the specifics of the problem. In the current NISQ era and with the current capabilities of the quantum gate hardware, the size of problems that can be addressed is inevitably constrained to the demonstration/proof of concept arena.  However it is possible to compare the effectiveness of different implementation techniques to ensure that the highest performance is extracted from these hybrid algorithm approaches.  Further there is the reasonable prospect that as the size and quality of NISQ-era quantum gate computers grows, it will be possible to use techniques such as QAOA and VQE to deliver improved solutions relative to those available from classical computing approaches to real world commercially valuable problems.  We will use both approaches to investigate solutions to a problem related to the deployment of ambulances.  Further details on the QAOA and VQE methodologies are provided in section \ref{QAOA}.

The second type of quantum computer available today is the quantum annealer, which from a practical perspective means the D-Wave Advantage system, which as mentioned previously became available for universal access remotely in September 2020.  A quantum annealer makes use of the concepts of Adiabatic Quantum Computing \cite{ farhi00a, Albash18a} to find the ground state of a Hamiltonian which encodes the problem to be solved. 
Quantum annealers are intrinsically ``noisy'' and complete their mechanism of operation too quickly to be able to come close to an adiabatic path.  Accordingly they will only rarely result in the annealer taking on the configuration of the ground or lowest cost function value state; so many thousands of run cycles and output state measurements are needed to obtain the ``best'' solution, which will not necessarily be the true optimal solution for the problem concerned. Further details of the quantum annealing approach to solving optimisation problems is provided in section \ref{QA_Method}. Quantum annealers have been available as working machines since 2011 when the 128 qubit D-Wave One was released \cite{DWave11a}.  Although other organisations have maintained an interest in quantum annealing only D-Wave has offered machines which are available to access externally. The latest generation of machine in the form of the Advantage offers 5000+ qubits and 15 connections between each qubit \cite{DWave20a}, which makes it possible to consider much larger scale problems than is currently possible with quantum gate hardware.  As a result we will explore larger ambulance deployment problems in section \ref{QA_impl}.   Notwithstanding we should acknowledge that the scale, connectivity  and noise resilience of the Advantage hardware is still currently short of what is needed to tackle ``commercial'' scale problems beyond the capability of classical approaches although progress towards that goal is being made on a regular basis.

\subsection{What is addressed in this paper}
The objectives of this paper are to:

\begin{itemize}
    \item Show in substantial detail how an important real-world problem can be addressed using quantum computing techniques
    \item Investigate the performance of QAOA and VQE approaches on a quantum gate simulator 
    \item Compare the performance of a hybrid classical-quantum  algorithm on quantum gate simulator and on currently available small-scale quantum computers
    \item Investigate the performance of a quantum annealing approach for the same problem
    \item Provide some ``new'' results on the effectiveness of different established  QC techniques
    \item Comment on when this may be achieved taking account of current known hardware development road maps
\end{itemize}

We have not looked at the potential of quantum inspired approaches; these are classical methods which have been ``inspired'' by quantum computing.  One such area is the use of a Digital annealer which uses a quantum annealing style of calculation simulated using a classical computer \cite{Fujitsu19a}.


\section{Description of the Ambulance problem}

\subsection{High level description of the problem in everyday language}

The ambulance logistics problem encompasses many difficult-to-solve components ranging from issues at the strategic level (how many ambulances are needed and what bases should be used to meet demand over the next 10 years?), tactical level (how many ambulances and crew should be deployed tomorrow or next week?) to the operational level (which ambulance service resources should be sent to a specific urgent incident?).  The challenges range across demand forecasting, roster planning, resource planning and location to actual ambulance dispatch \cite{Reuter17a}.

\begin{figure}[ht]
 \begin{centering}
 \includegraphics[width=0.65\textwidth]{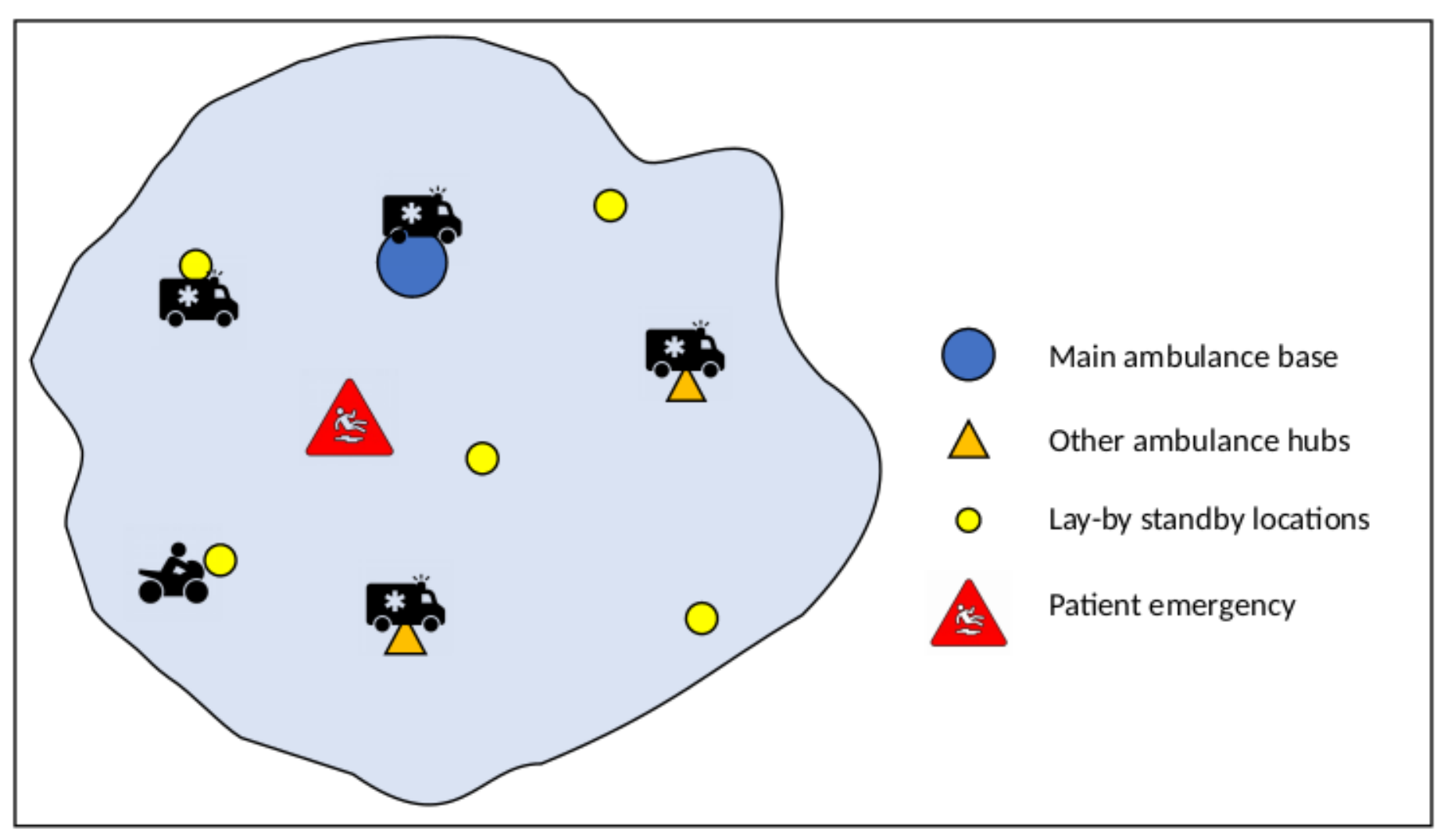}
 \end{centering}
 \caption{Cartoon illustration of the ambulance configuration problem, with hubs and other locations where ambulances and related resources can be stationed. \label{fig:ambulance_illustration}}
\end{figure}

The ambulance configuration problem at the general level is the problem of deciding how to deploy a variety of assets ranging from full sized ambulances to motorcycles to emergencies which may arise, and to assign those assets locations where they are ready to respond to emergencies so that an area is effectively covered. An example of the type of situation under consideration is illustrated in Figure \ref{fig:ambulance_illustration}.  It depicts a territory covered by an ambulance service which has a main base location, 2 other operational hubs and a number of other locations where emergency response resources can be located.  A new incident requiring attendance from the ambulance service is shown in the red triangle.  There are 4 standard ambulances available to attend based at various locations as shown and in addition a paramedic on a motorcycle is also available.  The challenge is which emergency resource should be stationed at each possible location.  When the scale of the problem is small, this may seem to be straightforward to address; however, once the problem is extended to include a large service area and an available resource of ${\sim}100$ units, it quickly becomes intractable to solve optimally using classical computing capabilities.

\subsection{Operations Research and Mathematical perspective}

As might be expected, there is a significant Operational Research literature on methods to optimally locate Emergency Response facilities/resources and a variety of increasingly sophisticated performance objectives and associated methodologies have been developed \cite{Li11a}. Most of these go well beyond the scope of this paper by including greater granularity of the operational requirements for a well-functioning emergency services network.  However at its core the Ambulance location problem is a version of the Facilities location problem (``FLP'') which itself is a form of the set cover problem \cite{Taha17a}.  The latter forms one of the 21 NP-complete problems identified by Karp in 1972 \cite{Karp72a}. In this paper we will focus solely on optimal ambulance location framed in a simplified operational landscape reflecting the current limited capabilities of quantum computing hardware.  In particular we seek to find the optimum location of a small number of ambulances within a grid structure so as to minimise the aggregate journey time to each location from the closest ambulance.  Expressed in this way the problem becomes a version of the weighted set cover problem, which can be formulated as follows.

First define a grid with $N$ nodes with the $i^{th}$ node being $s_i$. For each node, $i$, define an associated collection of sets comprised of all the possible combinations of the $N$ nodes, i.e. the power set, $P(S^i)$, of the set of nodes $S^i = \{s_i|i \in (1,2,...,N)\}$. Now use the indicator function, $I(s_j) \in \{0,1\}$, to define if node $s_j$ is included in any particular subset of $P(S^i)$, e.g. for the set which includes only the $N^{th}$ node, the indicator function takes the form $ I = (100...000)$, where the nodes are denoted in the order $(s_N, s_{N-1},...,s_2, s_1)$.  Let $S_k^i$ be the $k^{th}$ set within $P(S^i)$ where $k$ is the binary numbering equivalent of the Indicator function. The overall set of sets we are considering, $S$ is therefore $S = \{P(S^1), P(S^2), ..., P(S^N) \}$.  For each subset, $S_k^i$, there is an associated weight $w_k^i$ which is the sum of the distances from the ambulance location, node $i$, to the other nodes represented in set $S_k^i$.  The ambulance location problem becomes to minimise, $F$, the aggregate distance measure for the included sets, such that:
\begin{equation}
    F_{sol} = \text{min} \sum_{j, k}w_k^j x_k^j  \qquad   x_k^j \in \{0,1\}
\end{equation}
where $x_k^j$ indicates if the relevant subset, $S_k^j$, is included and  subject to the constraint that all nodes are covered:
\begin{equation}
    s_i \in \{S_k^j| x_k^j=1\} \quad \forall ~ i = {1,2,...,N}  
\end{equation}


In the context of the ambulance problem this means that we will seek to identify a collection of sets, one for each ambulance, which between them cover all the nodes of the grid and which minimise the aggregate travel distance to the grid nodes. Most of the possible set combinations will either not cover all of the nodes or will cover some nodes more than once, so cannot be optimal, so only a small number of set combinations have the potential to be viable solutions. In practice, we will not explicitly create sets \textit{a priori} and test them as possible solutions but rather will infer the sets \textit{post facto} based on the closest ambulance for any given node. 




\section{What can be done using classical optimiser approaches} \label{classical_opt}


In general there are 3 types of approach to solving optimisation problems \cite{Kalinin18a}:

\begin{itemize}
    \item Exact methods which find the precise answer subject to machine precision and the available computing resource/time including brute force methods
    \item Algorithms which promise an approximate answer and are subject to a performance or approximation guarantee
    \item Heuristic algorithms such as Simulated annealing or Tabu search which deliver solutions which may be close to optimal and often are but which offer no guarantee as to solution quality 
\end{itemize}

We focus on the the heuristic algorithms which are of most relevance as problem sizes increase towards typical real-world levels.  We make use of Simulated annealing and Tabu search algorithms to find classical solutions to a specific set of examples of the Ambulance location problem: these are locating 2 ambulances so as to optimally serve an $n$ x $n$ grid of locations where the distance metric is the Euclidean distance squared.  We choose this problem as it is one of the problem variants we seek to model using a quantum computer later in this paper. The calculations were performed using the D-Wave classical optimiser capability provided within the Ocean optimizer software.  The results from the classical calculations are provided in Table \ref{Classical sim results}.

\begin{table}[ht]
\begin{center}
\caption{\textbf{Performance of classical heuristic algorithms on sample Ambulance location problems}}
\begin{tabular}{| >{\centering\arraybackslash}m{3.25cm}| >{\centering\arraybackslash}m{2.25cm} | >{\centering\arraybackslash}m{2.25cm} |>{\centering\arraybackslash}m{2.25cm} | >{\centering\arraybackslash}m{2.25cm} | >{\centering\arraybackslash}m{2.25cm} |}
 \hline
 \textbf{Algorithm} & \textbf{Location grid size} & \textbf{Lowest aggregate distance ($d_{sol}$)} & \textbf{Probability of finding lowest distance} & \textbf{True minimum distance ($d_{min}$)} & \textbf{Performance measure ($d_{sol}$)/($d_{min}$)}\\
 \hline
 Tabu search & 5 x 5 & 65 & 99.99\% & 65 & 1.00 \\
\hline
 Tabu search & 6 x 6  & 134 & 6.85\% & 134 & 1.00 \\
\hline
Tabu search & 7 x 7 & 252 & 8.27\% & 252 & 1.00 \\
\hline
Tabu search & 8 x 8 & 432 & 0.01\% & 432 & 1.00 \\
\hline
Tabu search & 9 x 9 & 684 & 0.17\% & 684 & 1.00 \\
\hline
Tabu search & 10 x 10 & 1042 & 0.01\% & 1038 & 1.004 \\
\hline
Tabu search & 11 x 11 & 1553 & 0.01\% & 1529 & 1.016 \\
\hline
Tabu search & 12 x 12 & 2257 & 0.01\% & 2172 & 1.039 \\
\hline
Tabu search & 15 x 15 & 5473 & 0.01\% & 5336 & 1.026 \\
\hline
Tabu search & 20 x 20 & 17115 & 0.01\% & 16700 & 1.025 \\
\hline
Tabu search & 30 x 30 & 115006 & 0.01\% & 84338 & 1.364 \\
\hline
\multicolumn{6}{|c|}{} \\
\hline
Simulated annealing & 5 x 5 & 83 & 0.01\% & 65 & 1.28 \\
\hline
 Simulated annealing & 6 x 6  & 190 & 0.01\% & 134 & 1.42 \\
\hline
Simulated annealing & 7 x 7 & 353 & 0.02\% & 252 & 1.40 \\
\hline
Simulated annealing & 8 x 8 & 632 & 0.01\% & 432 & 1.46 \\
\hline
Simulated annealing & 9 x 9 & 1017 & 0.01\% & 684 & 1.49 \\
\hline
Simulated annealing & 10 x 10 & 1623 & 0.02\% & 1038 & 1.56 \\
\hline
Simulated annealing & 11 x 11 & 2209 & 0.01\% & 1529 & 1.44 \\
\hline
Simulated annealing & 12 x 12 & 3367 & 0.01\% & 2172 & 1.55 \\
\hline
Simulated annealing & 15 x 15 & 8391 & 0.01\% & 5336 & 1.57 \\
\hline
\end{tabular}
\label{Classical sim results}
\caption*{\small{This table shows the results from using 2 classical heuristic algorithms to solve for the location of 2 Ambulances on a square grid of a given size and to allocate each grid location to one of the 2 ambulances. The objective is to minimise the total travelled distance. Each of the algorithms were run 10,000 times with the best (smallest distance) reported together with frequency it was found. The distance between grid locations was based on the Euclidean distance squared using the coordinate location of each 2 grid points. The Tabu algorithm clearly performs much better than Simulated annealing (which could not even solve the 5x5 grid problem) and is able to find the optimal configuration up to grid sizes of 9x9.  Above that size the quality of solution gradually deteriorates as measured by the performance measure $d_{sol}$/$d_{min}$. Although this is a upper bound on the quality of solution obtainable as it does in any way seek to customize the solution approach to the problem specification, it does give an indication of the size of problem where these 2 classical algorithms start to break down as effective solvers.}}  
\end{center}
\end{table}

The results clearly show that Tabu search performs much better than Simulated annealing for all sizes of problem considered. It is able to find the optimal solution up to a 9 x 9 grid size based on 10,000 attempts although the success rate rapidly declines to only 1 in 10,000 for the 8 x 8 problem and 17 in 10,000 for the 9 x 9 grid. For problem instance sizes of 10 x 10 and larger, Tabu search is unable to find the true minimum and the quality of the solution gradually deteriorates - the best solution found is still within a few percentage points of the true minimum until grid sizes in excess of 20 x 20 are considered.  Simulated annealing fails to find the true minimum even for a 5 x 5 grid and the quality of solution as measured by the ratio of the aggregate distance for the proposed solution to the true minimum ($d_{sol}$)/($d_{min}$) is in excess of 1.28 and rapidly reaches 1.5 or more as problem size increases.

We see clearly that even with relatively modest problem sizes that these heuristic algorithms are no longer able to generate optimal results.  As a result, even allowing for improved performance from ``bespoke'' algorithms more carefully adapted to the characteristics of this problem, it should come as no surprise that in practice these algorithms are not the main tool used for such location problems.  In fact a simulation based approach is generally used whereby trial location configurations and other relevant operational information are tested using large datasets of previous activity data to generate expected real-world performance. 
In this way an empirical modelling assessment of the "best available" configuration is achieved although there is no optimality guarantee.


\section{Quantum computing methods} \label{QC_methods}

\subsection{Variational Quantum Algorithms} \label{VQA}

We will describe the methods most relevant to the NISQ era of quantum computing as regards quantum gate approaches in relation to optimisation.  There are other methods (such as Quantum phase estimation \cite{Lloyd97a}) which rely on quantum error correction and will ultimately be more powerful when appropriate quantum hardware is available but given the the focus of this paper, we will not discuss these. Instead we focus on so-called variational methods which are based on the variational principle from quantum mechanics, which we briefly explain below.  Suppose that we have an eigenvalue equation such as 
\begin{equation}
    H |\psi\rangle = E |\psi\rangle
\end{equation}
where $H$ is a Hamiltonian, $|\psi\rangle$ a wavefunction or eigenfunction, and $E$ is the eigenvalue.  In general there will be a set of eigenvalues and eigenfunctions which satisfy the equation. We can rank the eigenvalues so that we obtain a minimum value which is conventionally called the ground state energy and we can designate as $E_{min}$.  This corresponds to the wavefunction taking an associated form $|\psi_{min}\rangle$, so that in this case, we would have
\begin{equation}
    H |\psi_{min}\rangle = E_{min} |\psi_{min}\rangle
\end{equation}
and 
\begin{equation}
    \langle\psi_{min}| H |\psi_{min}\rangle = \langle\psi_{min}|E_{min} |\psi_{min}\rangle = E_{min}
\end{equation}

Now by the variational principle, a general wavefunction (not necessarily an eigenfunction) must always give a value for $E$ such that $E\ge E_{min}$. As a result, if we define the general wavefunction so that it is itself a function of one or more parameters, $\theta_0, \theta_1,..., \theta_n$, we can start with an initial wavefunction, $|\psi_{init}(\theta_0, \theta_1,..., \theta_n)\rangle$ and then seek to vary the parameters $\theta_0, \theta_1,..., \theta_n$, so as to minimise the value, $E$, confident that it can never go below $E_{min}$. If we are able to formulate a Hamiltonian in a suitable form so that it encapsulates the problem we are seeking to solve (and the solution is itself the represented by the minimum of that Hamiltonian), then the variational method provides a route to the solution or rather an approximation thereof. The process makes use of a classical optimiser to determine how best to evolve the values of the $\theta_i$ parameters to obtain a minimum estimate for $E$; hence such methods are termed hybrid classical-quantum algorithms.

There are two main algorithms that make use of the variational principle: the Quantum Approximate Optimisation Algorithm (``QAOA'') and the Variational Quantum Eigensolver (``VQE'').  These are described in the next two subsections.

\subsubsection{Quantum Approximate Optimisation Algorithm method} \label{QAOA}

In this subsection we set out the details of the Quantum Approximate Optimisation Algorithm and its more general evolution the Quantum Alternating Operator Ansatz.  The former was described by Farhi et al in 2014 \cite{Farhi14a} while the latter was introduced by Hadfield et al in 2017 \cite{Hadfield17a}. In the description which follows we have drawn on some of the presentational forms used by Zhou et al \cite{Zhou18}.

Suppose that the problem to be solved can be represented as a Cost function, $C$, which is a function of a set of $n$ variables, $s_i, i \in \{0,1,...,n-1\}$ each of which can take on the value `0' or `1', so that:  

\begin{equation}
    C = C(s_0, s_1, ..., s_{n-1}) 
\end{equation}

In the context of the Ambulance problem, the variables could signify if an ambulance is positioned at location $i$ or not.  Since the values of the variables are simply `0' or `1', they can be captured in a bit string $s =s_0s_1...s_{n-1}$. Now we can convert the problem into a quantum form by promoting each binary variable, $s_i$, to a quantum spin operator, $\sigma_i^z$, which is the Pauli Z operator whose eigenvalues are `-1' or `+1'.  The resulting problem Hamiltonian, $H_C$, has the form:

\begin{equation}
    H_{C} = H_C(\sigma_0^z, \sigma_1^z, ..., \sigma_{n-1}^z) 
\end{equation}


The basic idea is to seek to minimise (or, if appropriate, maximise) the expectation value of $H_C$ by finding an appropriate setting of the quantum spin variables.  In order to do this we need to explore all the possible quantum spin values that can be taken by the quantum wavefunction, $\psi$, characterised by the possible quantum numbers for each variable.  We begin the process by initialising each quantum variable in the $|+\rangle$ state, i.e. $\frac{1}{\sqrt{2}}(|0\rangle + |1\rangle)$ which ensures that $\psi_{init} =  |+\rangle_0 |+\rangle_1...|+\rangle_{n-1}$ is an equally weighted superposition of all the possible eigenstates. We now need a method to alter the weightings of the possible states so as to minimise the measured expectation value of $H_C$. This is achieved by applying in exponential form the problem Hamiltonian and a so-called mixer Hamiltonian, $H_B$ and this is repeated $p$ times. This has the effect of evolving the starting state or ``ansatz'':

\begin{equation}
     \psi (\boldsymbol{\beta, \gamma}) =  e^{-i \beta_p H_{B}}e^{-i \gamma_p H_{C}}... e^{-i \beta_1 H_{B}}e^{-i \gamma_1 H_{C}}|\psi_{init} \rangle
\end{equation}
where $\boldsymbol{\beta} = (\beta_1,...\beta_p)$ and  $\boldsymbol{\gamma} = (\gamma_1,...\gamma_p)$  are variable real parameters which will be optimised in order to generate a low expectation value for $H_C$. The mixer Hamiltonian in the Quantum Approximate Optimisation Algorithm of Farhi et al. takes the form
\begin{equation}
    H_B = \sum_{i=0}^{n-1}\sigma_i^x      
\end{equation}
where $\sigma_i^x$ is the Pauli X matrix acting on the $i^{th}$ qubit and has the effect of changing the weight of $|0\rangle$ and $|1\rangle$ in the $i^{th}$ qubit 
\footnote{Due to the matrix form of the Pauli operator $\sigma_i^z$ being $\begin{pmatrix}1 & 0\\0 & -1\end{pmatrix}$, the eigenstate$|0\rangle$ has an eigenvalue of 1 and the eigenstate $|1\rangle$ has an eigenvalue of -1 so that the state $|0\rangle$ maps to `1' and the state $|1\rangle$ maps to `0' in terms of the $s_i$ variables.}
The terms of the form 
$e^{-i \gamma_i H_{C}}$ have the effect of changing the relative phase of the ansatz, 
$\psi$, and is known as the phase separating Hamiltonian for this reason. For example, if the cost Hamiltonian has the simple form 
$H_C = w_{01}\sigma_0^z \sigma_1^z$, 
where $w_{01}$ is a weight then the effect of the first phase separating Hamilton operator in matrix form would be:

\[
\psi_{init} = \frac{1}{2}
\begin{pmatrix}
1 \\
1\\
1\\
1
\end{pmatrix}
\overset{\text{$e^{-i \gamma_1 H_{C}}$}}\longrightarrow
\frac{1}{2}
\begin{pmatrix}
e^{-i \gamma_1 w_{01}} \\
e^{+i \gamma_1 w_{01}}\\
e^{+i \gamma_1 w_{01}}\\
e^{-i \gamma_1 w_{01}}
\end{pmatrix}
\]

\vspace{0.25cm}

To obtain the expectation value of the Hamiltonian we use the standard expression from quantum mechanics which gives:
\begin{equation}
\begin{split}
\langle H_C \rangle = &\langle \psi_p (\boldsymbol{\beta, \gamma}) |H_C| \psi_p (\boldsymbol{\beta, \gamma}) \rangle \\
= & \langle \psi_{init}|e^{i \gamma_1 H_{c}}e^{i \beta_1 H_{B}}...e^{i \gamma_p H_{c}}e^{i \beta_p H_{B}}H_{c}e^{-i \beta_p H_{B}}e^{-i \gamma_p H_{c}}...e^{-i \beta_1 H_{B}}e^{-i \gamma_1 H_{c}}|\psi_{init}\rangle
\end{split}
\end{equation}


We now use a quantum gate computer to perform repeated measurements of the output in the computational basis, $\psi(\boldsymbol{\beta, \gamma}) $   to calculate the expectation value of $H_C$. With the assistance of a classical optimiser we try successive values of  $\{\boldsymbol{\beta, \gamma}\}$ to deduce their optimum values which we denote $\{\boldsymbol{\beta^*, \gamma^*}\}$;  these give the minimum expectation value. To obtain the best approximation to the problem we are seeking to solve, for each distinct measured state we classically calculate the cost function value and find the minimum; the measured state corresponding to the minimum is the solution from the QAOA algorithm.

\begin{equation}
     \psi (\boldsymbol{\beta^*, \gamma^*}) =  e^{-i \beta_p^* H_{B}}e^{-i \gamma_p^* H_{C}}... e^{-i \beta_1^* H_{B}}e^{-i \gamma_1^* H_{C}}|\psi_{init} \rangle
\end{equation}

From the methodology, it can be seen that the final result is an approximation, but with the selection of optimum parameters, it has the potential to be a good approximation and one that has the possibility to outperform a classical approach.  Its effectiveness relative to a classical approach is currently not known. We should also note that the final wavefunction is very unlikely to be a pure state and so multiple measurements are required in order to increase the probability of accessing the state which gives rise to the lowest cost function value.  A key objective of the QAOA process by minimising the cost function value is to increase the probability of measuring low cost function states and, of course, ideally to access the \textit{lowest} such state.

In theoretical terms Farhi et al \cite{Farhi14a} demonstrated that QAOA can be considered as a Trotterisation (splitting into many small steps) of the evolution of the wavefunction under the operation of the Hamiltonian.  As the number of steps, $p$, tends to infinity, then we would have a form of  adiabatic evolution. As explained in section \ref{QA_Method}, an adiabatic trajectory ensures that the ground state can be achieved.  Thus there is a strong theoretical underpinning for the QAOA methodology.

The Quantum Alternating Operator Ansatz is an evolution of the Quantum Approximate Optimisation Algorithm in which both the starting ansatz and the mixer Hamiltonians are varied.  The rationale for these changes is driven by two main considerations:

\begin{itemize}
    \item The original QAOA starting as it does with a superposition of all possible states is drawing from the whole possible space of states including those that do not meet any constraints built into the problem - this is inefficient
    \item Many other mixer Hamiltonians are possible, some of which may be more suited to the context of the problem and in particular may ensure compliance with problem constraints 
\end{itemize}

One particular implementation which we will make use of is the XY mixer \cite{Hadfield17a, Hen_2016} combined with the Dicke state  
\footnote{Also known as a generalised W-state}
ansatz.  The XY mixer can be represented as:

\begin{equation}
H_B = \frac{1}{2}\sum_{i=0}^{n-1}(\sigma_i^x\sigma_{i+1}^x + \sigma_i^y\sigma_{i+1}^y)
\end{equation}

where in the case $i=n$, it should be understood to mean $i+1=n+1=0$ and results in a so-called ring mixer. It has the key benefit that it maintains the Hamming weight which can be relevant for many problems, for example in the case of the Ambulance problem where there are a fixed number of ambulances to deploy.  In this context it would be appealing to only investigate possible solutions with a given fixed Hamming weight - this is provided by the Dicke state \cite{Dicke_54a} which is an equally weighted superposition of all possible states with a specified Hamming weight. 

There are also potentially some drawbacks.  For example one of the attractions of the Farhi et al version is that the initial ansatz and the mixer Hamiltonian are both easy to set up in the context of a quantum computer.  By contrast the Hadfield et al version can draw on more complex ansatzes and  mixers whose set up may offset much of the potential gain.  In the current NISQ era this may be particularly serious due to the limited number of gate operations that can be completed before noise and low gate fidelity becomes significant.  One way to reduce the length of circuit required is to randomly choose as the starting ansatz a \textit{single} state from the set of possible fixed Hamming weight states albeit that this reduces the effectiveness of the method.  Nonetheless the Hadfield QAOA offers additional potential and is likely to become of growing importance as the capability of quantum gate computers grows.

\subsubsection{Variational Quantum Eigensolver method} \label{VQE}

\begin{figure}[ht]
 \begin{centering}
 \caption{VQE overview \label{fig:VQE overview}}
\includegraphics{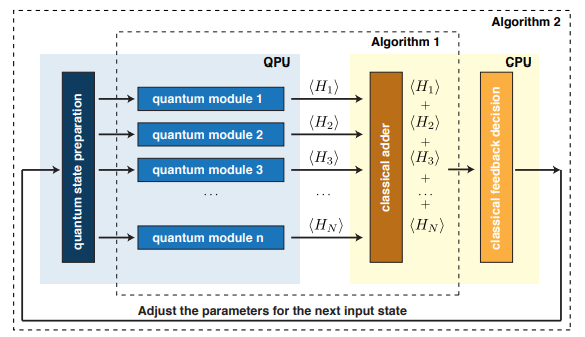}
 \end{centering}
 \caption*{\small{This VQE overview is taken from the original paper by Peruzzo et al \cite{Peruzzo13a} and depicts the different stages of the algorithm. Starting from the left, first a parameterised initial quantum state is prepared.  Each of the terms of the problem Hamiltonian in a separate quantum circuit and the output state measured and converted to give the value of each term.  The total of the terms (added classically) is fed to a classical optimiser (algorithm 2 in the diagram) which identifies new parameters for the initial quantum state.  The process is repeated until a minimum of the of the cost function is found.}}
\end{figure}

The Variational Quantum Eigensolver method was originally proposed in a chemistry context as an efficient quantum computing means of finding the lowest energy state of a molecule but it can also be applied to optimisation problems.  It is an inherent part of how the QAOA method works but uses the problem Hamiltonian in a different way, as we will see. An overview of the algorithm is shown in Figure \ref{fig:VQE overview} sourced from the original paper which proposed the VQE by Peruzzo et al in 2013 \cite{Peruzzo13a}. 

Let us begin with the Hamiltonian, $H_{C}$, which as for the QAOA method captures the problem as a cost function.  For the type of optimisation problem we will be considering the Hamiltonian will be the sum of a series of discrete terms:
\begin{equation}
    H_{C} = H_1 + H_2 + H_3 + ... + H_N
\end{equation}

Each of the  terms is encoded into a separate quantum circuit using appropriate quantum gates reflecting the associated Pauli operators (as for QAOA). A parameterised initial state, $|\psi_{init}(\theta_0, \theta_1,..., \theta_n)\rangle$, is prepared which acts as the starting ansatz for the circuit.  Usually parameterised rotation gates such as $RX(\theta_i)$ and (unparameterised) entangling gates such as CNOT or control-Z are used to prepare the ansatz. Next the value of each term of the Hamiltonian is calculated based on measuring the output of the (multiple) circuits.  The overall value of the Hamiltonian is calculated classically (simply summing the output values) and a classical optimiser then used to generate new parameters for the start state. The process continues until convergence in the expectation value of $H_C$ is achieved or some other stopping condition satisfied, e.g. a certain number of iterations.

The depth of the circuit depends on the complexity of the ansatz which can contain as little as one set of rotation and entangling gates per qubit or multiple sets per qubit.  As the size of the problem and the depth of the ansatz increases, so does the number or parameters which are subject to optimisation, which both increases the potential to obtain a good approximation to the cost function minimum but also increases the parameter search space, potentially unhelpfully.  One of the key attractions of the VQE approach is that the depth of quantum circuit is relatively low which is particularly useful given the challenge of noise for a circuit of any significant depth.  Against this there is no theoretical guarantee that the VQE method will generate a good estimate for the cost function minimum; this is contrast to QAOA which does have a theoretical backing.

\subsection{Quantum annealing (Adiabatic Quantum computing) } \label{QA_Method}

The Quantum annealing approach to solving optimisation problems was first proposed by Finnila et al in 1994 \cite{finilla94a}.  Its name is inspired by the classical technique of simulated annealing which is an implementation based on the ideas of thermal fluctuations and the tendency as a physical body is cooled for it to settle in the lowest potential energy state.  In the case of quantum annealing, the relevant process is that of quantum tunneling which allows the process to escape local minima in search of the global minimum.  The methodology used by Quantum annealing depends to a large degree on the theory of Adiabatic Quantum Computing \cite{Kato51a, Albash18a}.  In essence this theory suggests that if a quantum system is initially in the ground state of a given Hamiltonian and then the system evolves by some means to be subject to a different Hamiltonian, provided the evolution is sufficiently slow it will settle in the ground state of the second Hamiltonian.  There are conditions on the rate of evolution and these may mean that the rate of evolution may need to be so slow as to not be achievable in a practical setting.  Nonetheless it offers a means to solve an optimisation problem for if the final Hamiltonian encodes the problem to be solved, then the measured state should represent the problem solution provided the necessary conditions can be satisfied.

We now present the methodology more formally \cite{farhi00a, Hen_2016}.  Suppose that the quantum system is prepared in the ground state of a chosen Hamiltonian, $H_{init}$. The initial Hamiltonian of the system is then slowly modified, say linearly for simplicity over time $T$, into the Hamiltonian encoding the problem, $H_{problem}$:

\begin{equation} \label{anneal eqn}
H(s) = (1-s)H_{init} + sH_{problem}
\end{equation}

where $s$ is a parameter varying with time $s = t/T$, so that when $t=0$, $H(s) = H_{init}$ and when $t=T$, at the end of the evolution process  $H(s) = H_{problem}$. In a manner analogous to the Farhi et al QAOA method, $H_{init}$ is chosen so that the initial ground state is an equally weighted superposition of all possible states, i.e. $\psi_{init} = |+\rangle_0 |+\rangle_1...|+\rangle_{n-1}$.  This is achieved by setting:

\begin{equation}
H_{init} = \sum_{i=0}^{n-1} \sigma_i^x
\end{equation}
where $\sigma_i^x$ is as before the Pauli X operator.  For the quantum system to remain in its evolving ground state it is typically necessary for the time of evolution, $T$, to be ${\sim} k/\Delta^2$ where $k$ is a constant and $\Delta$ is the energy gap between the ground state and the first excited state \cite{Jansen07a}. The latter condition is in practice difficult to apply as for a general problem Hamiltonian it is not possible to calculate the relevant energy gap; it does however provide comfort that the required evolution time is not infinite. Given that we cannot be certain that the system will end up in the ground state of the problem Hamiltonian, $H_{problem}$, it is necessary, as for QAOA, to undertake multiple measurements of the final state in the computational basis and again adopt as the problem solution the one that gives the lowest cost function value when this is calculated.  We have illustrated quantum annealing using $ H_{init} = \sum_{i=0}^{n-1} \sigma_i^x$ but it is possible to use other initial Hamiltonians as suggested by Hen et al. \cite{Hen_2016} in order to constrain the search space (as for QAOA); however such alternative Hamiltonians are not currently available on the D-Wave quantum annealing platform, so we do not pursue this possibility further.

As can be seen there are great similarities between quantum annealing and the QAOA approach.  QAOA represents a discretised approach to the evolution of the Hamiltonian while quantum annealing is a smooth analogue approach albeit in both cases current implementations use only a few steps and relatively rapid annealing, so are some way from meeting the conditions for adiabatic behaviour.

\section{Encoding the problem for use in a QC algorithm}
\label{Encoding the problem}

\subsection{Challenges arising from nature of current quantum computing hardware}
Currently available quantum gate computers, as mentioned earlier in section \ref{QC background}, are far from the universal computers they can potentially become in the future. They have a significant number of operational limitations which impact the ability to run quantum algorithms. These challenges have an impact on the algorithmic techniques used (which we have already covered) and how the problem needs to be encoded in a suitable Hamiltonian.  The main limitations of quantum gate computers are:

\begin{itemize}
    \item A relatively limited number of qubits (though steadily increasing every 6-12 months)
    \item Limited coherence times which constrains the length of quantum circuit that can be run (i.e. the number of successive gate operations before quantum coherence is lost)
    \item Imperfections in the reliability of the quantum computer internal operations so that the outcome of each step in the operation of the computer is subject to error resulting in an increased probability of calculation error as the depth (run-time) of the quantum algorithm increases
    \item Very limited connectivity between qubits (in the computers with the largest number of qubits) which results in a need to use ``swap'' gates to allow 2-gate interactions for most of the computer's qubits - this introduces a potentially large overhead impacting effective circuit depth and also further reducing effective gate fidelity
\end{itemize}

Quantum annealers also have limitations:
\begin{itemize}
    \item Limited connectivity between qubits which means that the effective number of "logical" qubits in the sense of being able to represent a problem variable of interest can be significantly reduced from the number of physical qubits
    \item Manufacturing challenges mean that not all of the theoretically available qubits are in practice usable
    \item Limited control over some of the key operating parameters that has an important impact on the ability to solve problems successfully
\end{itemize}

Thus there are significant challenges for both types of quantum computer. As a result, a key part of the problem solving lies in converting the problem of interest into a quantum computing suitable form. and, given the limitations of current hardware, an efficient encoding of the problem.  This means encoding the problem so as to require the fewest possible number of qubits and additionally, in the case of a quantum gate computer, the fewest associated number of gate operations and an allocation of variables to qubits which minimises the number of swap gates needed.

\subsection{Choosing suitable variables} \label{Choosing variables}
 
A first question is how to choose the variables we are trying to solve for.  An obvious approach is to use a so-called ``one-hot'' encoding methodology. Under this approach each possible location for each ambulance becomes a variable.  Since we also want to know which ambulance serves a given location we also need an additional set of variables for this purpose. Specifically if we assume there are $m$ ambulances and $n$ locations we require for each location:

\begin{itemize}
    \item A variable (qubit) which indicates if each ambulance is located at that location - requires $m$ variables for $m$ ambulances per location and $nm$ variables in total  
    \item A variable (qubit) which indicates if each ambulance serves that location - requires $m$ variables for $m$ ambulances per location and $nm$ variables in total 
\end{itemize}
Overall this results in a need for $2nm$ variables/qubits.

In specifying possible configurations of locations with an ambulance and a location without an ambulance a natural approach is to associate a `1' with the location of an ambulance and a `0' with no ambulance. So the ``state'' `01001000' has ambulances at locations 1 and 5 using a notation convention that the location most to the left is location `0'.  We adopt this convention when we have 2 ambulances but for the single ambulance problems we choose to use the \textit{opposite} convention so that a `0' indicates the location of an ambulance as it turns out this simplifies the required calculations.  Of course both conventions are possible as it is trivial to convert between the two. 

An alternative variable encoding  which requires fewer qubits is the domain wall encoding approach due to Chancellor \cite{Chancellor19}.  This reduces the number of qubits per location by 1 both for the start locations and for the destination locations although it may increase the total number of terms in the cost function that need to be evaluated\footnote{We have not checked for this problem whether a greater number of terms is required.}.

\subsection{Selecting the distance norm}

In the ambulance problem we are concerned with finding the ambulance configuration that provides the overall minimum time to reach each location from the initial starting locations.  Rather than time specifically we will consider ``distance'' as a proxy instead.  However, once we have made that transformation, we can consider alternative distance measures, for example the Manhattan distance, the Euclidean distance and the square of the Euclidean distance.  Each have potential merits.  We choose to use the square of the Euclidean distance as this has the benefit of not simplifying the problem so much that it might not prove to be a realistic challenge to a Quantum computer solution.  We also undertook simulations on a subset of the techniques we present using each of the other two measures and found there were no significant differences to the nature of results obtained.  
For simplicity we have considered a regular graph of locations (nodes) which are equally spaced on either a single line or a 2D array. This has the advantage that there is significant symmetry so that the solutions can be solved analytically. The distance between any 2 locations is encoded as the value of the edge between the relevant variables (qubits).  We also assume that the edge is undirected so that each edge only needs to appear once in the cost function and the associated ``Adjacency matrix'', which encodes the distances and any problem constraints, is also simplified.


\subsection{Encoding the constraints} \label{Encoding the constraints}
Encoding the constraints is an important step and takes advantage of the variable encoding approach we described in section \ref{Choosing variables}.  First we set out the constraints that need to be captured
\begin{itemize}
    \item The total number of ambulances deployed needs to equal the specified number of ambulances available
    \item Every destination is serviced by at least one ambulance
    \item Only 1 ambulance is to serve any location
\end{itemize}

The means of encoding the constraints differs depending on the method of solution used. In general the means of encoding the constraints involves 2 approaches: 1) the use of penalty terms to disfavour possible configurations which do not meet the required constraint and 2) the use of implementation design to build-in constraints so that they are automatically satisfied. First let's look at an example of a penalty term that captures the requirement that there are precisely $m$ ambulances to be deployed. The total number of ambulances is given by
\begin{equation}
    \sum_{l=1}^{m} \sum_{i=0}^{n-1} s_i^l = m
\end{equation}
where $s_i^l$ is a variable indicating if ambulance $l$ is located at location $i$.  For the 2 ambulance problem, $s_i^l = 1 $ when ambulance $l$ is located at location $i$ and $s_i^l = 0 $ otherwise.  
If we now introduce the term, $P$, defined as
\begin{equation}
    P = \lambda\big( (\sum_{l=1}^{m} \sum_{i=0}^{n-1} s_i^l) - m)\big)^2
\end{equation}
it will be a minimum (zero) when $\sum_{l=1}^{m} \sum_{i=0}^{n-1} s_i^l = m$ and positive otherwise.  By suitably choosing $\lambda$, a positive constant,  we can ensure that the possible configurations (or states) which do not satisfy the constraint will have higher cost function values and so will be disfavoured when we search for the lowest cost function value.  Similar formulations can be used to encode the other 2 constraints.  This approach can be used to generate a cost function for use with either a Quantum annealer or Quantum gate computer.  However in the case of the QAOA approach used with a quantum gate machine, an additional approach is possible which in the problem set up reduces the search space and ensures the constraint is met throughout the search (optimisation) process; as a result no cost function penalty term is required for the relevant constraint.  This involves using as the initial state a Dicke state and an XY mixer rather than the X mixer originally defined by Farhi et al.  A Dicke state is an equally weighted superposition of all the states with a specified Hamming weight while an XY mixer has the property that its operations maintain the Hamming weight.  So if we initially prepare the quantum computer with the appropriate Dicke state, then it will have as its starting position only the states which satisfy the required constraint(s) and will seek to find the lowest state amongst them.  Potentially this is a huge gain as there is a massive reduction in the size of the search space, a gain which grows exponentially as the size of the problem increases \cite{Wang19a}. In the case of the ambulance problem it is possible to encode all but one of the constraints by building the initial state as a set of several conjoined Dicke states each of which encodes one of the constraints.  Despite the elegance of this approach, there is a drawback in that the Dicke state(s) take many gate operations to construct which given current limitations of quantum gate computers is a major disadvantage.  We will nonetheless investigate this approach using a simulator for a small scale problem instance.

\section{Quantum Gate implementation}

In this section we will explore in more detail how well the techniques described in section \ref{QC_methods} perform in practice on a quantum gate simulator and on a true quantum gate computer.  First we specify the problem in detail. Then we explore the degree of success in obtaining the optimal answer as we vary the various user-configurable parameters within the problem as well as the results generated by some extensions of the previously described variational techniques.  Finally we report the results obtained from a true noisy quantum computer. 

\subsection{The Ambulance problem we actually consider}

Hitherto we have described the ambulance configuration problem in size agnostic terms.  
Now we will specify  5 ``small-scale'' problem variants which can be accommodated on a current simulator and which facilitate the exploration of the different aspects of the problem solving methodology. For some of the variants we relax some of the criteria in order to explore a slightly larger scale problem. These problem variants are set out in Table \ref{Problem specifications}.

For variants A, B, and D the distance metric used for the minimisation is for the \textit{closest} ambulance to each destination location, whereas for variant C it is the sum of the distance for \textit{each} of the ambulances to all of the destination locations.

For problem variant B we use the encoding approach outlined in section \ref{Encoding the problem}, which requires 2 x 2 x 4 = 16 qubits to implement.
In variants A, C and D where only 1 ambulance is involved or the distance metric is relaxed to include all ambulances to all destinations, the number of qubits required to implement the problem is the same as the number of locations to be served, namely 5, 8 and 17 respectively.
\newline


\begin{table}[ht]
\begin{center}
\caption{\textbf{Ambulance problem variant specifications}}
\vspace{0.3cm}
\begin{tabular}{| >{\centering\arraybackslash}m{1.5cm}| >{\centering\arraybackslash}m{2.5cm} | >{\centering\arraybackslash}m{2.5cm} |>{\centering\arraybackslash}m{6cm} |>{\centering\arraybackslash}m{2.5cm} |}
 \hline
 Problem variant & Number of ambulances & Number of locations & Distance metric minimised & Number of qubits required\\
 \hline
 A & 1   & 5    & Squared Euclidean distance   &   5\\
  \hline
 B & 2   & 4    & Sum of Squared Euclidean distance to all locations from closest ambulance  & 16\\
 \hline
 C & 2   & 8    & Sum of Squared Euclidean distance to all locations for both ambulances &   8\\
 \hline
D & 1   & 17    & Squared Euclidean distance   &   17\\
\hline
E & 2   & 5    &  Sum of Squared Euclidean distance to all locations from closest ambulance    &   20\\
\hline
\end{tabular}
\label{Problem specifications}
\end{center}
\end{table}

\subsection{QAOA results and how they are affected by implementation choices}

We will now work through the effect of the various user-configurable parameters on the results generated by the QAOA as implemented on the ambulance configuration problem.
\subsubsection{Role of \texorpdfstring{$\lambda$}{lambda} and impact on performance}


The first parameter we will explore is the penalty weight, $\lambda$, which determines the relative contribution of the penalty weight term(s) in the overall cost function.  As its role is to ensure that candidate solutions which do not satisfy the constraint conditions are disfavoured, it is clear that $\lambda$ should be sufficiently large so that any candidate solutions which do not meet the conditions should have a cost function value which is higher than all the candidate solutions which do meet the conditions.

If we consider problem variant A (5 locations, 1 ambulance), the cost function, $C_A$, is composed of the distance metric, $D$ and the penalty term, $P$:
\begin{equation} \label{pen1}
   C_A = \sum_{i\leq j, i, j = 0}^{4} s_iD_{ij}s_j - 
   \lambda\sum_{i\leq j, i, j = 0}^{4} s_iP_{ij}s_j
\end{equation}

where the distance metric takes the form 
\begin{equation}
\begin{split}
D_{ij}  &=   -(j-i)^2   \qquad   i < j \\
 &= 0  \qquad \qquad \quad \text{otherwise}
\end{split}
\end{equation}

and the penalty term is defined by

\begin{equation} \label{pen2}
P_{ij} =
\begin{cases}
    1-2c & \text{ $i=j$}\\
    2 & \text{ $i<j$}\\
    0 & \text{ $i>j$}
\end{cases}
\end{equation}

where $c$ is the Hamming weight of the desired solution. Note that in equation \eqref{pen1} $\lambda > 0 $ and is real. For problem variant A, $c$ has the value 4.  In this simple case we can classically calculate the values of the cost function for all 32 ($2^5$) possible configurations for a range of the penalty weight, $\lambda$.  The results are shown in Table \ref{Cost function values q5}.  When $\lambda$ has a value less than 10, it can be seen that the configuration state $|11111 \rangle$ gives the lowest value for the cost function even though it does not satisfy the constraint that there should be one ambulance (as indicated by including a single `0' in the state).  Further $\lambda$ needs to exceed 30 for all the constraint satisfying states to have a lower value cost function than the $|11111 \rangle$ state.  Note that as $\lambda$ increases the cost function values for the states of interest decrease.
\newline

\begin{table}[ht]
\begin{center}
\caption{\textbf{Cost function values for selected configuration states and values of the penalty weight ($\boldsymbol{\lambda}$)}}
\vspace{0.3cm}
\begin{tabular}{| >{\centering\arraybackslash}m{4cm}| >{\centering\arraybackslash}m{1.75cm} | >{\centering\arraybackslash}m{1.75cm} |>{\centering\arraybackslash}m{1.75cm} |>{\centering\arraybackslash}m{1.75cm} |>{\centering\arraybackslash}m{1.75cm} |>{\centering\arraybackslash}m{1.75cm} |}
 \hline
 \multirow{2}{4cm}{\textbf{Configuration state(s)}} & \multicolumn{6}{c|}{\textbf{Value of penalty weight} $\boldsymbol{\lambda}$} \\
 \cline{2-7}
 & \textbf{0} & \textbf{10} & \textbf{20} & \textbf{30} & \textbf{40} & \textbf{100}\\
 \hline
`Non-compliant' state $|11111\rangle$* &	-50 (1)	&-200 (1=)	&-350 (4)  & -500 (4=)	& -650 (6)	&-1550 (6)\\
\hline
`Solution state' $|11011\rangle$*	&-40 (2)	&-200 (1=) &	-360 (1)	&-520 (1)	&-680 (1)	&-1640 (1) \\
\hline
$|10111\rangle \text{ and } |11101\rangle$* &	-35 (3=) &	-195 (3=)	& -355 (2=) &	-515 (2=) &	-675 (2=) &	-1635 (2=) \\
\hline
$|01111\rangle \text{ and } |11110\rangle$*	&-26 (5=)	&-180 (5=) &	-340 (5=) &	-500 (4=) &	-660 (4=)	&-1620 (4=) \\
\hline
Gap from `Non-compliant' $|11111\rangle$ state to highest value constraint satisfying state	 &-24 &	-20	&-10	&0	&+10 &	+70\\
\hline
Gap from `Non-compliant' $|11111\rangle$ state to the state with the lowest  value which satisfies the constraint &	-10&	-0	&+10	&+20	&+30	&+90\\
\hline

\end{tabular}
\label{Cost function values q5}
\caption*{\small{* Figures in brackets indicate the rank of the state relative to the lowest cost function state for the specified $\lambda$. 

The table shows how the cost function values and the relative rankings of the feasible states and the non-constraint compliant state $|11111\rangle$ evolve as a function of the penalty weight.  As the penalty weight increases the non-compliant state moves from being the lowest cost function state to becoming a state with a cost function value which is higher than all the feasible states (for $\lambda > 40$).}}
\end{center}
\end{table}

\begin{figure}[ht]
\caption{\textbf{Impact of $\lambda$ on mean probability of obtaining the  $|11111 \rangle$ state.}}
\includegraphics[width=14cm, height=7cm]{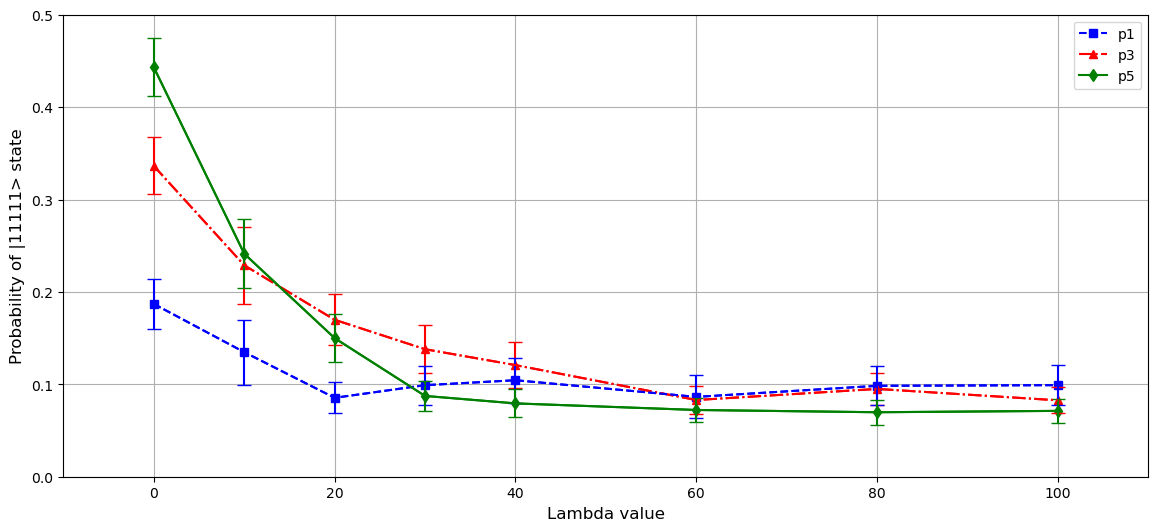}
\centering
\label{Impact of lambda on mean probability of obtaining the  11111 state}
\caption*{\small{QAOA was run using 100 randomly drawn initial angles for $\beta$ and $\gamma$ in the range $0-2\pi$ for steps($p$) 1, 3 and 5 for Problem A (5 locations and 1 ambulance).  The classical optimiser used was Nelder-Mead. Error bars at the 2 standard deviation level are shown. As the penalty coefficient, $\lambda$, increases the probability of obtaining the |11111$\rangle$ declines consistent with it no longer being the lowest cost function state for $\lambda >30$.}}
\end{figure}
\begin{figure}[h!]
\caption{\textbf{Impact of $\lambda$ on mean probability of obtaining the problem solution state}}
\includegraphics[width=14cm, height=7cm]{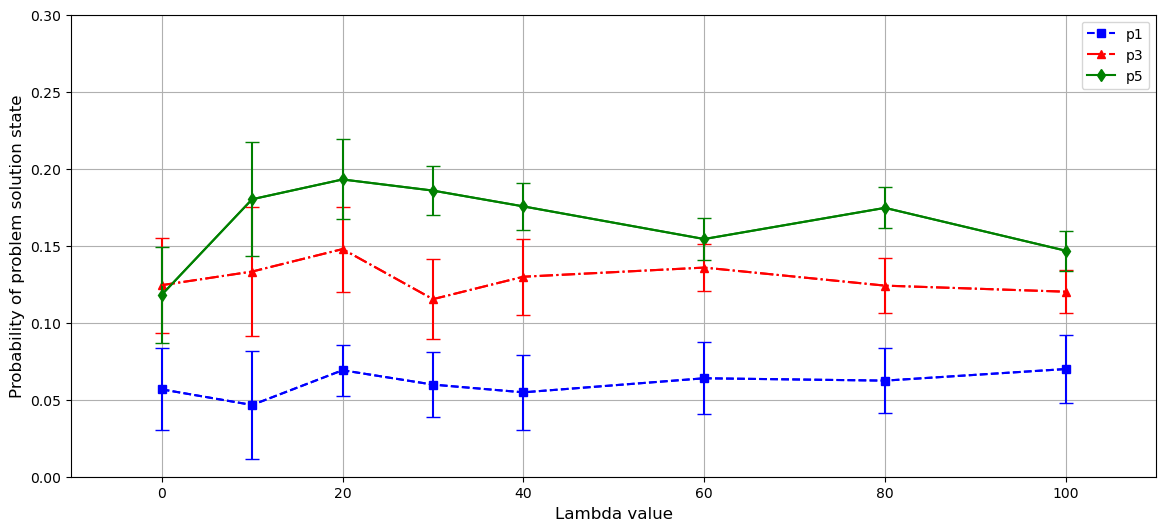}
\centering
\label{Impact of lambda on mean probability of obtaining the  ground state}
\caption*{\small{QAOA was run using 100 randomly drawn initial angles for $\beta$ and $\gamma$ in the range $0-2\pi$ for steps($p$) 1, 3 and 5 for Problem A (5 locations and 1 ambulance).  The classical optimiser used was Nelder-Mead. Error bars at the 2 standard deviation level are shown. As the penalty weight, $\lambda$, increases the lowest cost function state becomes the solution state, |11011$\rangle$, and this is reflected in the probability of this state increasing.  Comparing with Figure \ref{Impact of lambda on mean probability of obtaining the  11111 state}, the probability of obtaining the |11011$\rangle$ state exceeds that of the |11111$\rangle$ state in the case of $p=5$ for $\lambda \ge 20$ as expected. In the case of $p=3$, the probabilities are the same subject to experimental variance.  For $p=1$, the probability of obtaining the |11111$\rangle$ state remains above that pf the |11011$\rangle$ state indicating the less discriminatory performance of the algorithm on this problem instance.}}
\end{figure}

\begin{figure}[ht]
\caption{\textbf{Impact of $\lambda$ on mean probability of obtaining solution satisfying constraint}}
\includegraphics[width=14cm, height=7cm]{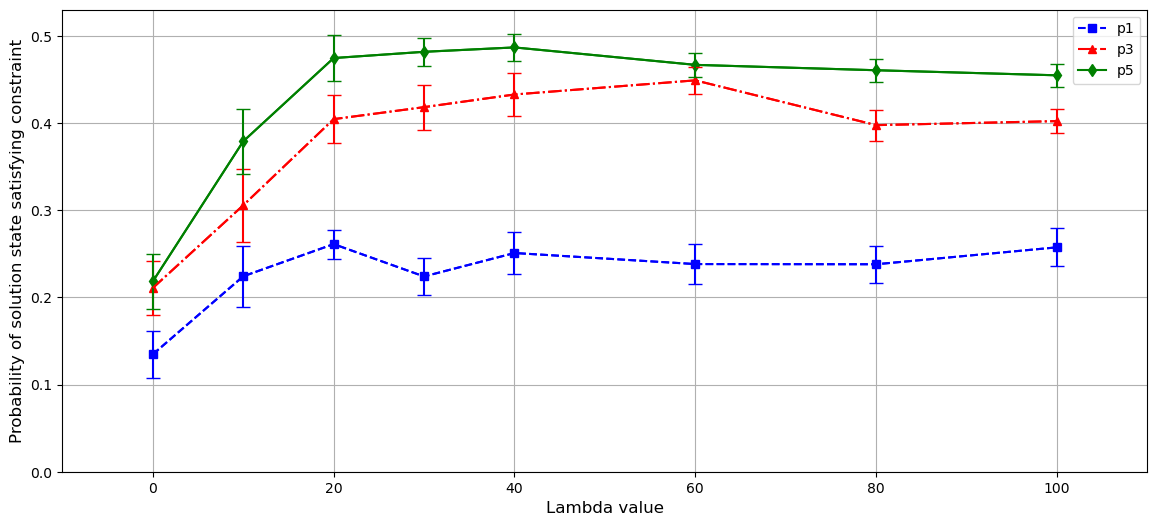}
\centering
\label{Impact of lambda on mean probability of solution satisfying constraint}
\caption*{\small{QAOA was run using 100 randomly drawn initial angles for $\beta$ and $\gamma$ in the range $0-2\pi$ for steps($p$) 1, 3 and 5 for Problem A (5 locations and 1 ambulance). Error bars at the 2 standard deviation level are shown. The classical optimiser used was Nelder-Mead. As the penalty weight, $\lambda$, increases the probability of obtaining a state satisfying the problem constraint increases for all values of $p$ until $\lambda = 40$ after which the probability is unchanged (subject to experimental variance).}}
\end{figure}

\subsubsection{Improved performance as the number of steps, \textit{p}, increases}
\label{performance as p increases}

 Figures \ref{Impact of lambda on mean probability of obtaining the  11111 state} to 
\ref{Impact of lambda on mean probability of solution satisfying constraint} also show how performance varies as the number of steps, $p$, included in the QAOA methodology is increased.  Performance, as measured by the probability of finding the state with the lowest cost function, clearly improves as the number of steps increases.  This is as expected given that there is always the option to set the $\beta$ and $\gamma$ angles to zero so that any additional steps have no impact on the algorithm performance.  Accordingly we expect:
\begin{equation}
    P(\psi_{sol})^{p+1} \geq  P(\psi_{sol})^{p} 
\end{equation}
where $P(\psi_{sol})^{p}$ represents the probability of obtaining the lowest cost function configuration state using a $p$ step QAOA routine. This is most clearly demonstrated in Figure \ref{Impact of lambda on mean probability of obtaining the  ground state}.  However in the case where the penalty weight is zero, the performance of $p=3$ and $p=5$ QAOA is very similar (especially after allowing for QAOA result variance - see error bars), which might appear to be a problem until we recall that the state $|11111\rangle$ has the lowest value for this value of the penalty weight.  As a consequence the probability of obtaining the $|11111\rangle$ will be highest for the $p=5$ optimisations and this is sufficient to cause the probability of obtaining the state with the lowest cost value which also meets the constraints to reduce to a comparable level to that of the $p=3$ optimisations. 
It might be expected that as the value of $p$ increases, the probability of obtaining the lowest cost function state might increase rapidly.   However since the number of parameters which are being optimised also increases with $p$, the improvement in performance is not as great as might initially be hoped for. 
For this reason, it is important to make use of techniques which will improve the speed with which good choices of $\beta$ and $\gamma$ are found; some such techniques are considered later in this section.

\subsubsection{Comparing performance across different techniques - figures of merit}

Before continuing to examine the performance of different methodologies in addressing the ambulance problems, we need to introduce suitable figures of merit to enable a comparison of performance when the raw data outputs are not directly comparable.  This arises in part to the use of the penalty weight, $\lambda$, which affects the value of the cost function and makes direct comparisons based solely on expected cost function values of little value.  The first figure of merit we will use is the Approximation ratio.   We define the Approximation ratio as:
\begin{equation} \label{r_appr}
    r_{approx} = \frac{\langle \psi_{feas}| H_{cost} | \psi_{feas} \rangle - C_{max}} {\langle \psi_{feas}|\psi_{feas}\rangle (C_{min}- C_{max}) }
\end{equation}


where $\psi_{feas}$ is that part of $\psi_{sol}$, the output ``solution'' of the QAOA algorithm\footnote{Note that $\psi_{sol}$ in general is composed of many states and should not be confused with the ``solution state'', which is a single state which generates the lowest cost function value; in principle there could be more than one ``solution state'' but for simplicity we assume here that is not the case}, which meets the hard constraints of the problem, $H_{cost}$ is the cost function expressed in Hamiltonian form, $C_{min}$ and $C_{max}$ are respectively the minimum and the maximum value of the cost function which satisfy the problem constraints. . This is similar to but slightly different to the definition used by Wang et al \cite{Wang19a}.   Referring to the previous section where $\lambda $ and the penalty term were introduced we note that:

\begin{equation} \label{Basic Objective equation}
    H_{cost} = H_{core} + \lambda H_{pen} 
\end{equation}

where $H_{core}$ is the underlying objective or cost function we are trying to minimise (and simultaneously find the associated state configuration) and $H_{pen}$ is the penalty term or terms.  As has been mentioned already  in section \ref{Encoding the constraints} and is explored further in the next section, it is not always necessary to include a penalty term in $H_{cost}$ to ensure that the constraints are met. In order to make the outputs comparable in such cases, we need to remove the effect of the penalty term in the numerator of equation \eqref{r_appr}. For $\psi_{feas}$ we can calculate classically the value of penalty term component and thereby remove this from the associated $H_{cost}$ effectively generating a cost function value where $\lambda = 0$. By this means we are now able to generate an $r_{approx}$ value which is comparable across the different methods we employ. 

The second figure of merit, $\text{Pr}(\psi_{sol}^{C_{min}})$, is the probability of obtaining the state with the lowest cost function value solution in $\psi_{sol}$. We are able to calculate this figure as we are making use of a simulation throughout this paper except where otherwise indicated. It is worth noting as pointed out by Amaro et al \cite{Amaro21a}, that a high probability of sampling the lowest cost function value state implies a high approximation ratio while the reverse is not necessarily true. However as $r_{approx} \rightarrow 1$, so does $\text{Pr}(\psi_{sol}^{C_{min}}) \rightarrow 1$. 

Additionally we define $\text{Pr}_{feas}$ as the probability of obtaining a feasible solution in $\psi_{sol}$, when sampled This is the proportion of the state configurations within $\psi_{sol}$ which when repeatedly sampled meet the hard constraints of the problem. This is the third figure of merit.

Finally we also make use of the number of calls that would need be made to the quantum computer.  This is an implied (or calculated) figure as the results in this paper are taken from a classical simulation.

\subsubsection{X mixer and XY mixer}

The original QAOA methodology as described by Farhi et al \cite{Farhi14a} involved the use of an X mixer and an initial equal superposition of all possible states.  Subsequently it was suggested that an XY mixer combined with an initial Dicke state \cite{Hen16, Wang19a} could have significant advantages in certain cases as it offers a means to directly encode certain types of problem constraint and also to limit the search space to only those states that satisfy the constraint.  The superiority of the XY mixer in appropriate cases has been demonstrated on standardised problems such as the Max-$\kappa$-Colorable Subgraph problem \cite{Hadfield17a, Wang19a}.  However it does rely on creating an initial Dicke state, an equally weighted superposition of all the states with a given Hamming weight,  which imposes significant overhead on the calculation process. The most efficient method of preparing a Dicke state we are aware of is due to B\"artschi and Eidenbenz  requiring $\mathcal{O}(kn)$ gates and a circuit depth of $\mathcal{O}(n)$ where $k$ is the Hamming weight and $n$ the number of qubits. \cite{Bartschi19a}. 
This is still likely to represent a significant overhead for NISQ era quantum computers even those with the highest level of gate fidelity.  For this reason in this section we will also examine making use of an initial start state which is a simple pure state which satisfies the problem constraint.


We compare the performance of 3 alternative XY mixer/initial state combinations with an X mixer approach as set out in Table \ref{Mixer and initial states}.

\begin{table}[h]
\begin{center}
\caption{\textbf{Mixer and initial state combinations}}
\vspace{0.1cm}
\begin{tabular}{| >{\centering\arraybackslash}m{2.5cm}| >{\centering\arraybackslash}m{3.5cm} | >{\centering\arraybackslash}m{4cm} |>{\centering\arraybackslash}m{4.5cm} |}
 \hline
 \textbf{Mixer} & \textbf{Penalty term/ constraint satisfaction approach} & \textbf{Initial state} & \textbf{Comment}\\
 \hline
 X mixer & Penalty term added to cost function   & Equal superposition of all states    & Original QAOA approach\\
 \hline
 XY mixer & \multirow{7}{3.5cm}{\centering{Built into initial state/mixer}}    & Ideal initial state for XY mixer approach    & Assumes Dicke state can be easily prepared\\
\cline{1-1} \cline{3-4}
XY mixer &    & Single state that satisfies the constraint    & Has the advantage that the initial state can be prepared in a single step\\
\cline{1-1} \cline{3-4}
XY mixer &    & Randomly chosen single state that satisfies the constraint    & More realistic test of a single constraint satisfying state as the initial state \\
\hline
\end{tabular}
\label{Mixer and initial states}

\end{center}
\end{table}

\begin{figure}[hp]
\caption{\textbf{Comparison of alternative mixer and initial state performance}}
\vspace{0.15cm}
\caption*{\hspace{1cm} Problem A (5 locations, one ambulance) \hspace{1cm}   Problem C (8 locations, two ambulances)}
\includegraphics[width=0.85\textwidth, center]{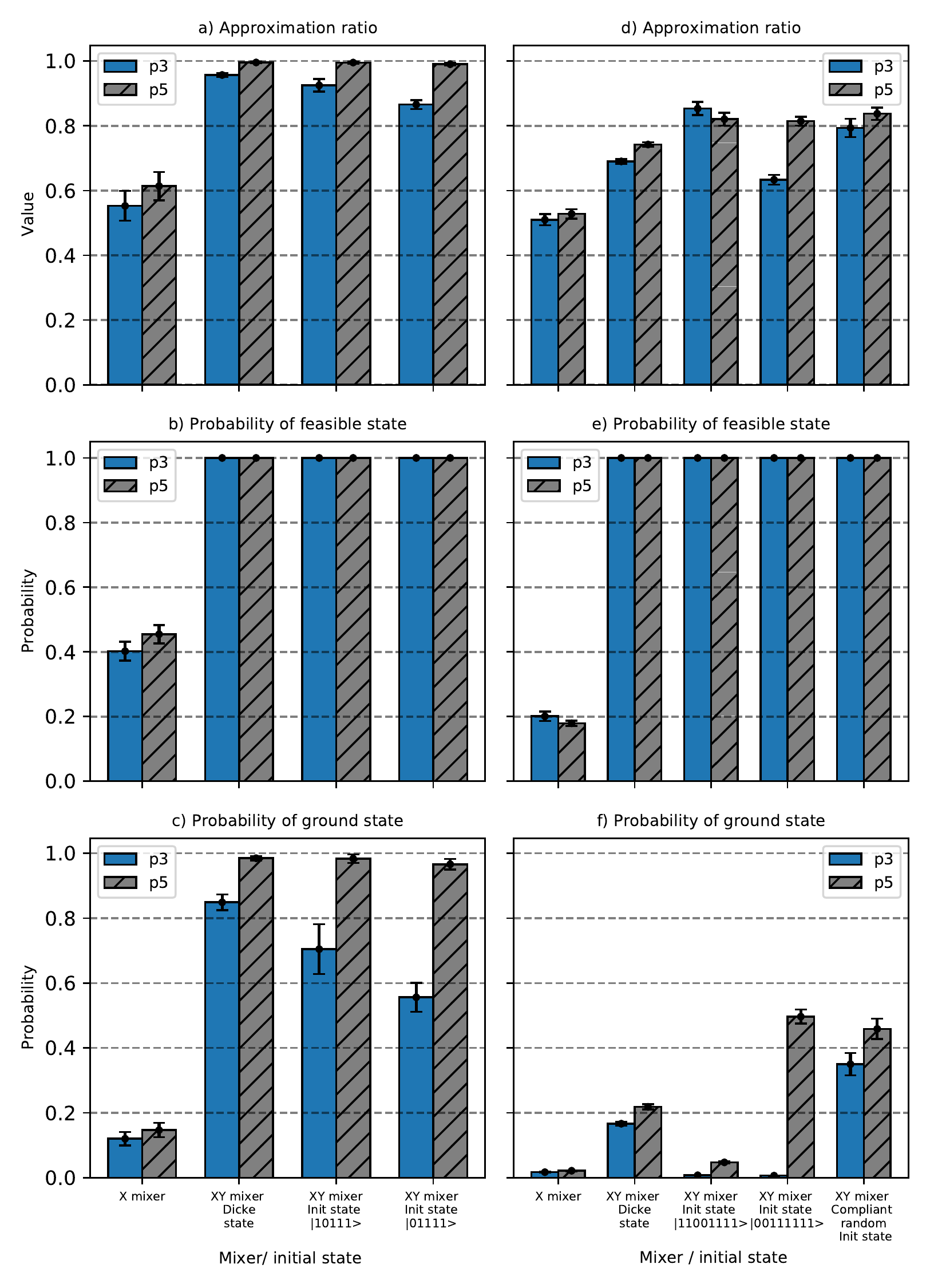}
\centering
\label{Problem A and C comparison}
\caption*{\small{The figure shows the performance of different combinations of mixer and initial states in solving Problem A (5 locations, 1 ambulance) and Problem C (8 locations, 2 ambulances). The XY mixer approach in combination with both a Dicke initial state and an easier-to-implement constraint compliant pure state performs significantly better than an X mixer approach. For each mixer/initial state 100 runs were completed with randomly selected angles for $\beta$ and $\gamma$ with the plots showing mean performance achieved.  Error bars at the 2 standard deviation level are shown. As expected, QAOA with $p=5$ outperforms $p=3$ in terms of probability of finding the lowest cost function (ground) state.}}
\end{figure}

To illustrate the different performance of the alternatives considered, we present results in Figure \ref{Problem A and C comparison} for problem variants A and C, the simple 5 site problem with a single ambulance to position and the slightly more complex 8 site problem with 2 ambulances.  In both cases, we consider $p = 3$ and $p = 5$ (number of steps). The probability of obtaining the state with the lowest value of the cost function shown in the bar charts is calculated by taking the average probability of finding the lowest value state from 100 runs each using random initial settings for $\beta$ and $\gamma$.  


As can be seen from the bar charts in Figures \ref{Problem A and C comparison}c and \ref{Problem A and C comparison}f the XY mixer using a Dicke initial state performs significantly better than the X mixer combined with a penalty term in terms of the probability of finding the ground state solution.  For the simpler case where the number of locations is 5, the probability of finding the lowest cost function state in the case of the XY mixer/Dicke state is close to 1 for both the $p=3$ and $p=5$ runs.  This compares to 0.1 \textendash{} 0.15 for the X mixer under the same conditions.  The improved performance is attributable both to the higher probability of obtaining a feasible state (100\% probability for the XY mixer, as expected) and also the higher approximation ratio for the cases using the XY mixer.
An alternative way to consider the improvement for the cases using XY mixer is to think in terms of the much smaller solution space being explored. 
For example in the 8 site, 2 ambulance instance (problem C) in the case of the XY mixer there are 28 states that are explored whilst there are 256 state possibilities in the case of the X mixer.  This important limitation of the search space becomes more extreme as the number of potential ambulance locations increases and in fact decreases exponentially as the number of locations increases \cite{Wang19a}.
For problem C, using an X mixer with a penalty term, the ground state representation in the average solution obtained when p = 3 is only 2.18x higher than it is in the initial state.  Using the XY mixer with a Dicke state as the initial state similarly boosts the ground state probability by 2.33x, a similar amount.  However the fact that the Dicke state/XY combination guarantees that the ``two ambulance'' constraint is met results in a much higher overall ground state probability at 8.33\%  vs 0.85\%.  The results for the simpler 5 site, 1 ambulance problem show a similar improvement although, due to the simpler nature of the problem, the XY mixer is able to get very close to finding the ground state whichever initial state is used, particularly when p=5.

Figure \ref{Problem A and C comparison} also includes the effect of using the XY mixer with a single pure initial state rather than a Dicke state.  Results for 2  specific constraint compliant start states are included as well as in the case of problem C a randomly chosen constraint compliant state.  It appears from these results that the much easier to initialise pure start state works almost as effectively as the equally weighted Dicke state in achieving a low cost function value. This is somewhat better than the  results found by  Wang et al \cite{Wang19a} when considering the Max-$\kappa$-Colorable Subgraph problem and by Cook et al who considered the Max-k Vertex Cover problem \cite{Cook20a}.  For problem C, we note that the results in some cases are better than for the unbiased Dicke initial state.  Although we have no clear explanation for this outcome, we attribute this  surprising result to the fact that certain initial states may be favoured in terms of generating higher probabilities of finding the lowest cost function state and the presence of inevitable statistical uncertainty arising from the random nature of the initial angles for $\beta$ and $\gamma$ (and for the initial state when this too is randomly chosen).   Notwithstanding these anomalous results we conclude that, if capable of replication at larger scales, choosing a (random) single constraint compliant state as the initial state could offer significant benefits in terms of the near-term practicality of making use of the XY mixer approach for problems of interest.

\subsubsection{Optimiser choice and performance} \label{Optimiser perf}


So far we have only been considering the impact of different parameter and methodological choices within the quantum part of the QAOA. The algorithm is however hybrid classical/quantum in nature so some consideration also needs to given to the classical part; in practice this means the choice of optimiser.  The literature shows many different classical optimisers being used including Nelder-Mead, BFGS, LBFGS, Bayesian Optimisation, COBYLA, SPSA,  RBFOpt \cite{Zhou18, Nakanishi20a, Nannicini19a}.  We do not here undertake a thorough review of the optimiser performance and suitability but rather simply comment on our experience on testing different optimisers.  

Our main observation is that the performance measured in terms of final cost function value achieved, number of iterations required and time to solution was one of varied performance between the classical optimisers with no clear ``winners''.  We found COBYLA was effective for a $p=1$ QAOA but much less so for higher step instances. Nelder-Mead often  performed well but not always. For the same problem instance (same starting angles) we found that the different optimisers could generate significantly different cost function values. These observations are regarding a noise free environment using a simulator. A number of recent reviews of optimisers for use in hybrid classical quantum algorithms have suggested that
\begin{enumerate}
  \item The tuning of optimiser hyperperameters such as the number of measurements used or the learning rate (where relevant) can be critical to performance\cite{Lavrijsen20a, Sung20a}
  \item Stochastic optimisers such as SPSA were more robust to noise and variations in the problem than non-stochastic optimisers\cite{Sung20a}
  \item In noise-free optimisation BFGS and COBYLA are fast optimisers \cite{Lavrijsen20a}
  \item Nelder-Mead performs well in a wide range of problems and can do so even when noise is present\cite{Lockwood22a}
\end{enumerate}


\subsection{Strategies for improving performance (i.e. reducing time to solution)}

In this section we look at strategies for implementing QAOA that specifically try to improve performance as measured by reducing the total problem solution time [and also ideally reducing the number of calls on the scarce quantum computing resource]. In the absence of such strategies, we would be forced to resort to simply using a random initial angle approach but for a greater number of angles ($2p$), which is both crude and inefficient in time and computing resources required \cite{Zhou18}.  In particular we will look at 2 specific areas: firstly strategies for improvement in the selection of $\beta$ and $\gamma$ angles as we increase the number of steps ($p$) included in the QAOA optimisation and secondly making effective use of XY mixers.

\subsubsection{Efficient strategies for selecting  \texorpdfstring{$\beta$}{beta} and \texorpdfstring{$\gamma$}{gamma} angles for increasing steps} \label{Efficient strategies for selecting}


Zhou et al \cite{Zhou18} proposed a relatively simple but effective approach for efficiently choosing ansatz initial angles for increasing-$p$ in relation to the Max-cut problem.  In fact they proposed two methods - one was to use a form of interpolation when moving between $p_{n}$ and $p_{n+1}$ which they called INTERP and a second related approach using a Fourier component technique though it was claimed both gave similar results in terms of quality of performance.  In this paper we look at the performance of the INTERP approach as well as two variants of a simple naive alternative; we also make use of 2 separate classical optimisers. Further we compare the performance using both the X-mixer and XY-mixer approaches. 

A choice needs to be made as to the seed angles used to initiate the increasing-$p$ methodologies.  In this case we completed 100 QAOA optimisations using 100 randomly selected angles for $p=1$ and used the angles with the lowest cost function result as the seed angles for subsequent $p>1$ QAOA optimisations.  We found that the COBYLA optimiser was an effective choice for completing this initial phase due to its short run time when $p=1$; it was less ``runtime-efficient'' for other values of $p$ and was not used in these cases.


The 3 different approaches examined are:
\begin{enumerate}
  \item The INTERP strategy in which the angles to be used as the starting angles for an optimisation at the $p_{n+1}$ level are based on the classically optimised angles found at the $p_{n}$ level. The details are set out in the Appendix, section \ref{INTERP}. 
  \item A simpler EXTRAP1 strategy whereby the starting angles for an optimisation at the $p_{n+1}$ level are, for the first $p$ angles, those from a previous optimised run at the $p_{n}$ level with the ${(n+1)}^{th}$ beta and gamma angles set to 0.0. 
  \item The EXTRAP2 strategy, which is the same as EXTRAP1 but where the increase in $p$ level is 2 rather than 1. This means that the starting angles for an optimisation at the $p_{n+2}$ level are, for the first $p$ angles, those from a previous optimised run at the $p_{n}$ level with the ${(n+1)}^{th}$ and ${(n+2)}^{th}$ $\beta$ and $\gamma$ angles all set to 0.0. 
\end{enumerate}

The idea behind EXTRAP1 mirrors that of the increasing $p$ strategy itself, namely that the starting $\beta$ and $\gamma$ angles for each value of increasing $p$ should generate the same cost function expectation values as from the previous lower value of $p$ and that the extra angles, initially set to 0.0, provide an extra degree of freedom over which to improve the classical optimisation. EXTRAP2 is just a variation of EXTRAP1, which offers the potential to reduce the number of separate optimisations needed to reach a given level of $p$.

The two classical optimisers used and for which results are presented here are Nelder-Mead and BFGS; other optimisers were also tried but were found to be less effective. For EXTRAP1 and EXTRAP2 only the Nelder-Mead optimiser worked well so results using Nelder-Mead as the sole optimiser are presented.   We compare the different strategy and optimiser combinations on problems C (8 locations, 2 ambulances)  and D (17 locations, 1 ambulance). 

\begin{figure}[H]
\caption{\textbf{ Performance of different increasing $p$ strategies - Problem C (8 locations, 2 ambulances)}}
\caption*{X mixer \hspace{7cm}   XY mixer}
\begin{subfigure}{.5\linewidth}
\centering
\caption{Approximation ratio}
\includegraphics[width=.9\linewidth]{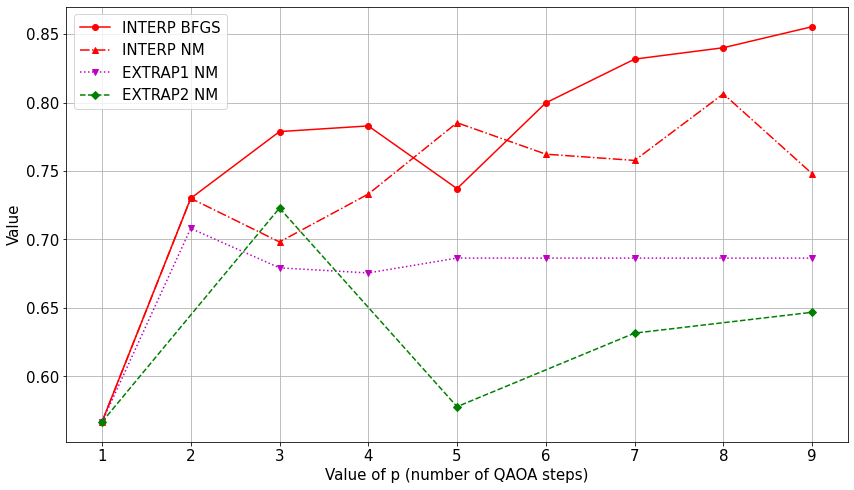}
\label{fig:sub1}
\end{subfigure}%
\begin{subfigure}{.5\linewidth}
\centering
\caption{Approximation ratio}
\includegraphics[width=.9\linewidth]{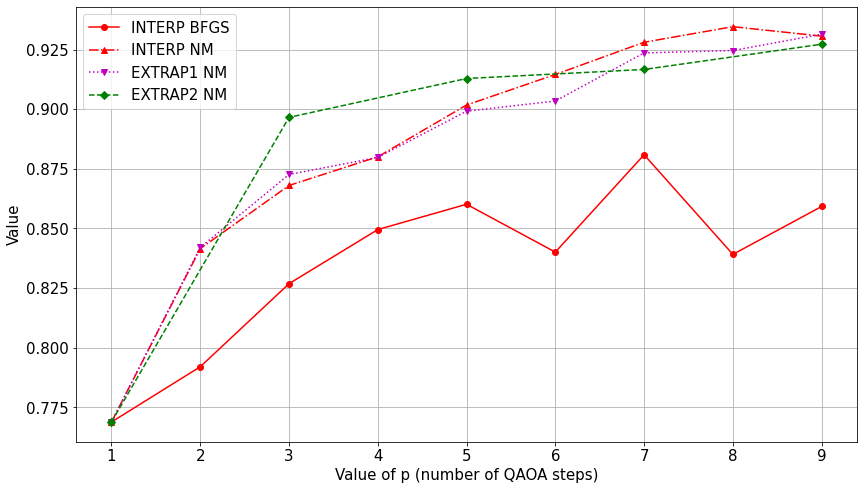}
\label{fig:sub2}
\end{subfigure}\\[1ex]
\begin{subfigure}{.5\linewidth}
\centering
\caption{Probability of feasible state}
\includegraphics[width=.9\linewidth]{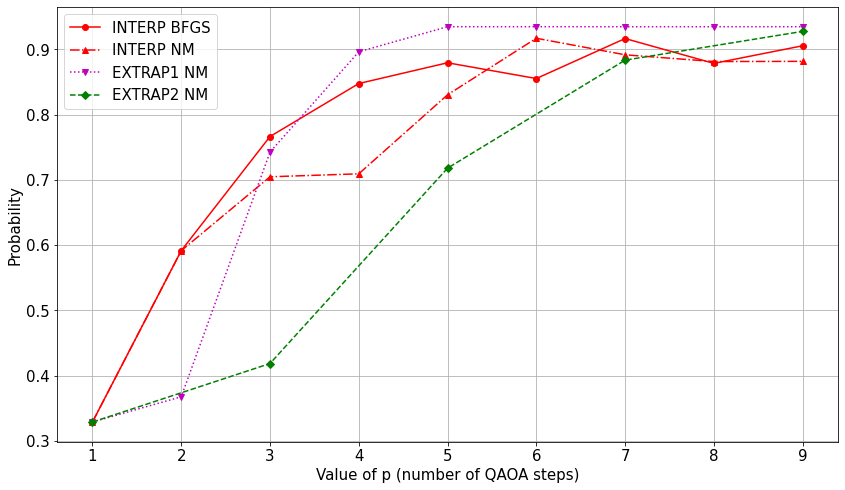}
\label{fig:sub3}
\end{subfigure}
\begin{subfigure}{.5\linewidth}
\centering
\caption{Probability of feasible state}
\includegraphics[width=.9\linewidth]{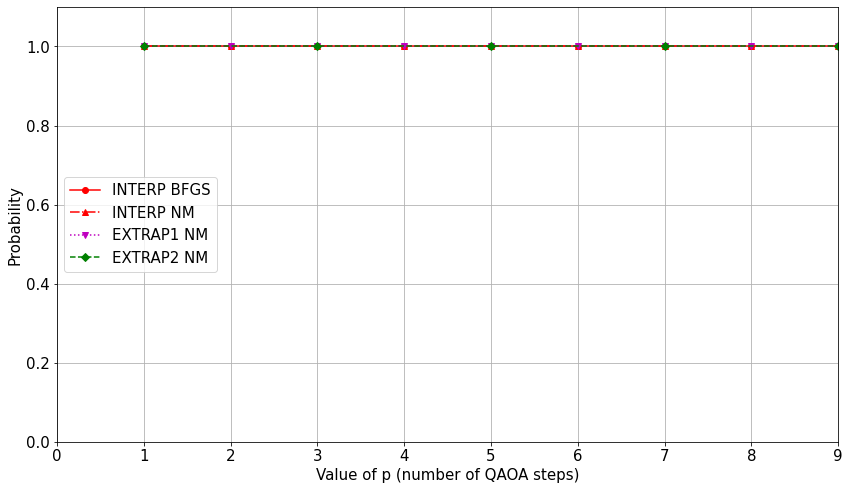}
\label{fig:sub4}
\end{subfigure}\\[1ex]
\begin{subfigure}{.5\linewidth}
\centering
\caption{Probability of lowest cost function state}
\includegraphics[width=.9\linewidth]{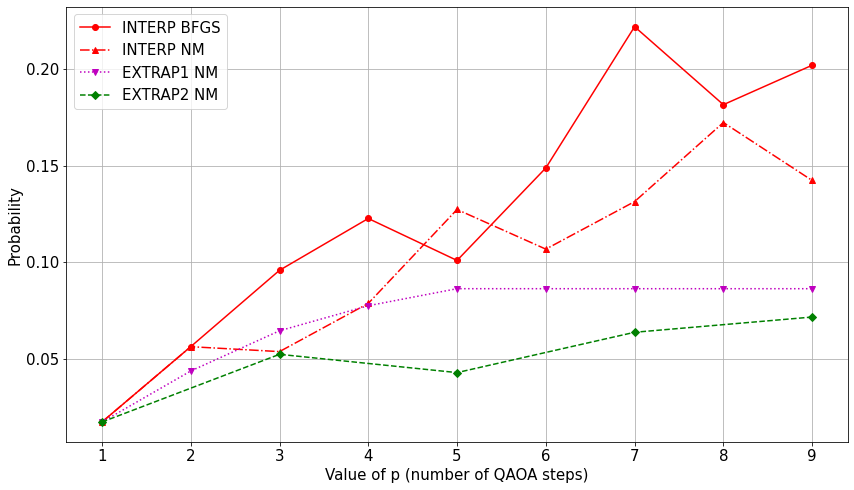}
\label{fig:sub5}
\end{subfigure}
\begin{subfigure}{.5\linewidth}
\centering
\caption{Probability of lowest cost function state}
\includegraphics[width=.9\linewidth]{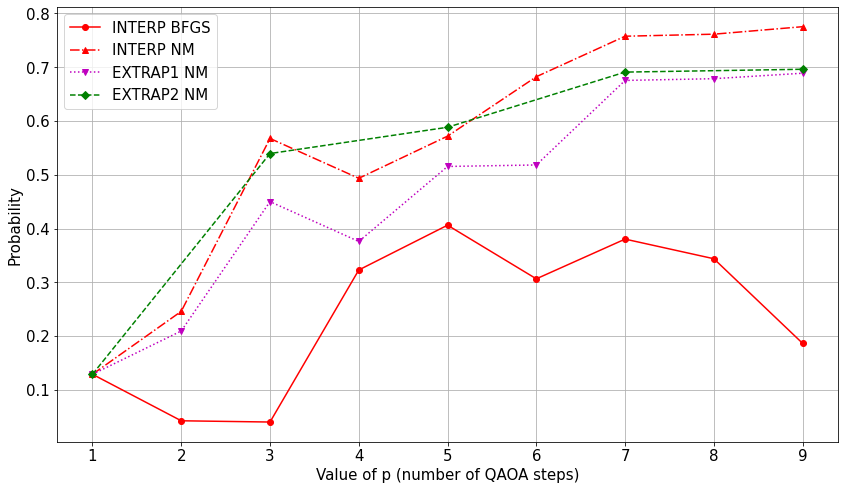}
\label{fig:sub6}
\end{subfigure}\\[1ex]
\label{Increasing p strategy comparison 1}
\caption*{\small{The figure shows the performance of the INTERP and EXTRAP1/2 strategies for a QAOA optimisation with both X mixer and XY mixer approaches for increasing values of $p$. The problem instance used is for 8 locations and 2 ambulances.  Nelder-Mead (``NM'') is used as the classical optimiser with BFGS also used with the INTERP strategy. The metrics used are the approximation ratio, probability of feasible state and, as the prime metric, probability of lowest cost function state.  Each of the strategies is seeded with initial angles sourced from the best outcome for 100 runs using random $\beta$ and $\gamma$ angles for a $p=1$ QAOA.  Considered with the results in Figure \ref{Increasing p strategy q17} we find no one strategy is universally superior although the INTERP strategy overall performs best in this limited comparison.  Overall the strategies deliver an efficient method of achieving a $\sim$5-10x gain in the probability of obtaining the lowest cost function state in these instances.}}
\end{figure}
\begin{figure}[ht]
\caption{\textbf{Performance of different increasing $p$ strategies - Problem D (17 locations, 1 ambulance)}}
\begin{subfigure}{.5\linewidth}
\centering
\caption{Approximation ratio}
\includegraphics[width=.9\linewidth]{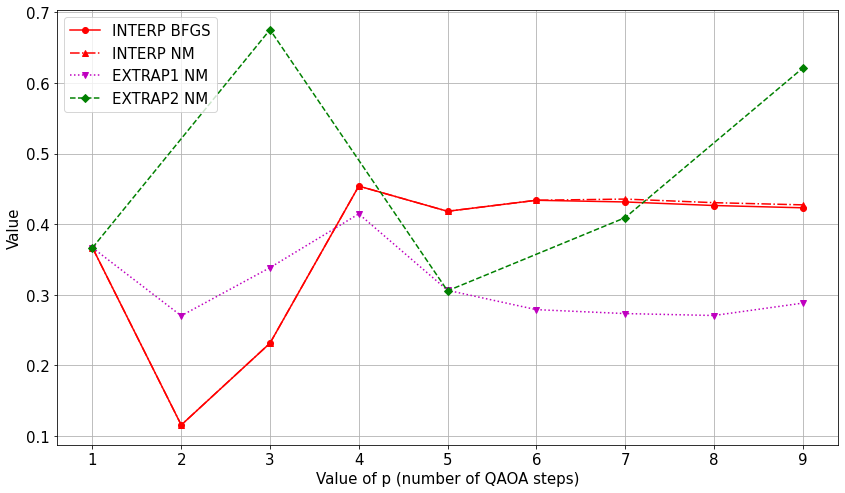}
\label{fig:sub12}
\end{subfigure}%
\begin{subfigure}{.5\linewidth}
\centering
\caption{Probability of feasible state}
\includegraphics[width=.9\linewidth]{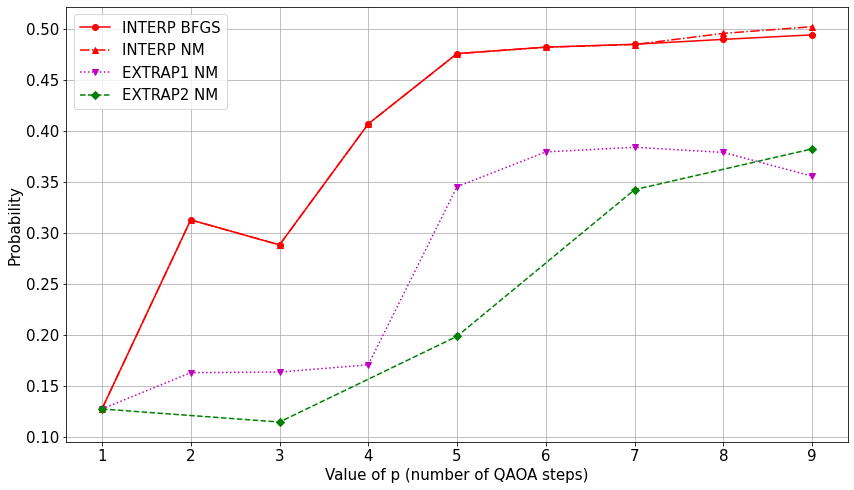}
\label{fig:sub22}
\end{subfigure}\\[1ex]
\begin{subfigure}{1.0\linewidth}
\centering
\caption{Probability of lowest cost function state}
\includegraphics[width=.5\linewidth]{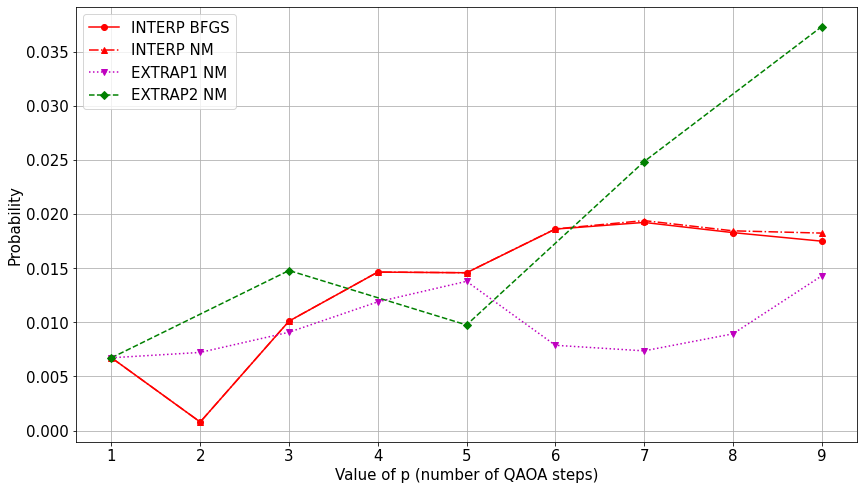}
\label{fig:sub32}
\end{subfigure}\\[1ex]
\label{Increasing p strategy q17}
\caption*{\small{This figure shows the performance of the INTERP and EXTRAP1/2 strategies for a QAOA optimisation using the X mixer approach for increasing values of $p$. The problem instance used is for 17 locations and 1 ambulance. See the text provided with Figure \ref{Increasing p strategy comparison 1} for further details and comment on this figure.}}
\end{figure}

In Figures \ref{Increasing p strategy comparison 1} and \ref{Increasing p strategy q17} the relative performances of the different  increasing $p$ strategy and classical optimiser combinations are shown on Problems C and D respectively. It can be seen that all of the strategies are broadly successful in significantly improving the probability of finding the lowest cost function value state although the degree of improvement can vary substantially between them and the improvement observed as $p$ increases is not monotonic. The overall increase observed in the amplification in the probability of finding the lowest cost function state ranges from ${\sim}5.5$x to ${\sim}12.5$x for the most successful strategy for each problem, indicating that the increasing $p$ strategies are of useful benefit adding  a factor of ${\sim}10$ to the probability of obtaining the ``solution'' state.  

A visual inspection of the results shows the INTERP strategy overall performing best  on 2 of the 3 problem instances (Problem C X and XY mixer methodologies) but still exhibits significant variation in the degree of improvement achieved and in its performance relative to the other strategies. For example in the case of Problem C using the X-mixer, it performs best with the BFGS optimiser but using the XY-mixer the Nelder-Mead optimiser delivers a much better result.  In the case of Problem D, there is very little difference between the 2 optimisers when INTERP is used.  This is in line with the observations on optimiser performance made in section \ref{Optimiser perf}.  Further these results imply that actual performance for any of these strategy/optimiser combinations is strongly instance dependent.  The simpler EXTRAP1 and EXTRAP2 strategies actually perform competitively and together produce a better overall result for problem D and a similar result for problem C using the XY mixer.  Interestingly EXTRAP2, which increases $p$ by 2 each time, performs overall as well as EXTRAP1 and is much better for Problem D where it generates the best probability overall of finding the lowest cost function state by a factor of ${\sim} 2$.  In conclusion, we see that for these problem examples, no one approach is universally superior, so a multi-strategy approach would seem advisable in a more general optimisation setting unless there is specific knowledge regarding the problem concerned that indicates a particular approach is indicated.  Further, we conclude on this limited evidence that both INTERP and EXTRAP are potentially useful increasing-$p$ techniques.

Finally, it is worth reflecting on the different sources of gain in improving the probability of finding the lowest cost function state.  There are effectively 5 contributing aspects: a) choice of mixer, b) using randomly selected angle runs to identify good seed angles for $p>1$ QAOA runs, c) improving the solution proportion in the feasible space, d) improving the approximation ratio and e) an increase in the relative probability of obtaining the  lowest cost function state in the mix of states \textit{beyond} that due to the approximation ratio improvement.  The final element arises from the observation that there is a non-linear boost to the lowest cost function states as the approximation ratio approaches 1.    For the problems considered the contributions of each source is shown in Table \ref{Performance contributions new}. The  most important sources of gain are the use of the XY mixer and good seed angle selection but all sources are material. We see that the increasing - $p$ technique is a valuable improvement but, at least for these small scale examples, it sits third in the hierarchy of significance for finding a good solution.  Notwithstanding, using an efficient strategy for increasing $p$ is much more efficient than simply conducting many instances using random angles selection for ever higher values of $p$; this observation seems to arise due to the difficulty of optimising the angle parameters over ever larger sizes of parameter space, which itself grows exponentially with $p$. 

\begin{table}[ht]
\begin{threeparttable}
\caption{\textbf{Contributions of different sources of performance improvement}}
\vspace{0.3cm}
\begin{tabular}{| >{\centering\arraybackslash}m{3cm}| >{\centering\arraybackslash}m{1.25cm} | >{\centering\arraybackslash}m{1.25cm} |>{\centering\arraybackslash}m{1.5cm} | >{\centering\arraybackslash}m{2.5cm} | >{\centering\arraybackslash}m{2.5 cm} |>{\centering\arraybackslash}m{1.5 cm} |}
 \hline
\multirow{2}{2cm}[-5mm]{\centering{\textbf{Sample problem}}} & \multirow{2}{1.25cm}[-5mm]{\centering{\textbf{Mixer choice}}} & \multirow{2}{1.25cm}[-5mm]{\centering{\textbf{Seed angle}}} & \multicolumn{3}{c|}{\textbf{Increasing-$p$ strategy contribution}} & \multirow{2}{1.5cm}[-5mm]{\centering{\textbf{Overall gain}}} \\ 
  \cline{4-6}
 &   &  & \textbf{Feasible state} & \textbf{Approximation ratio}   & \textbf{Lowest cost function state mix} & \\
 \hline
 
Problem C, X mixer & 1.0 & 4.09 & 2.75 & 1.51 & 3.04 & 51.72 \\
\hline
Problem C, XY mixer & 9.14 & 3.62 & 1.0 & 1.21 & 4.98 & 198.50\\
\hline
Problem D, X mixer & 1.0 & 882.8 & 3.0 & 1.69 &  1.09 & 4,893.24\\
\hline
\end{tabular}
\label{Performance contributions new}
\begin{tablenotes}
      \item \small{All numbers represent factor contributions to the improvement achieved, as measured by the probability of finding the lowest cost function state.  The benefit from the seed angle arises from running multiple sets of randomly drawn angles for a QAOA $p=1$ and selecting the angle set generating the lowest expected cost function value. We see that the greatest specific gains derive from the mixer choice and the gain from seed angle. However overall the increasing-$p$ strategy element (comprised of the gains from a more probable feasible state, higher approximation ratio and improved state mix) is an important further addition.}
    \end{tablenotes}
\end{threeparttable}
\end{table}

\subsubsection{3 XY mixers}

\begin{figure}[h]
\caption{\textbf{Problem B - Encoding 2 ambulances and 4 locations using XY mixers}}
\includegraphics[width=14cm, height=3.5cm]{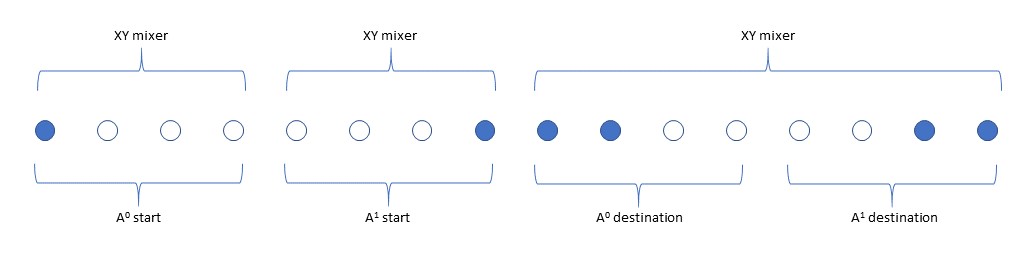}
\centering
\label{Problem B - Encoding of 2 ambulance 4 locations}
\caption*{\small{The figure shows the encoding of a 4 location, 2 ambulance problem using start and destination qubits for each of the 2 ambulances.  Three XY mixers are used: 1 each for the start positions of each ambulance, where the XY mixer ensures that a Hamming weight of 1 is maintained and 1 for the set of 8 destination qubits where a Hamming weight of 4 is required.}}
\end{figure}

An alternative approach considered in this subsection is to make use of the XY mixer's property of confining the search to a limited part of the overall solution space.  We make use of problem variant B which uses 16 qubits to specify for each of the 4 locations if each of two ambulances is initially placed at that location and if that location is served by either Ambulance  0 or Ambulance 1.  Due to the one-hot encoding approach used this results in 4 x 2 x 2 qubits being required.  The encoding approach is set out in Figure \ref{Problem B - Encoding of 2 ambulance 4 locations}.  The figure also shows 3 separate XY ring mixers being used:
\begin{enumerate}
    \item An XY mixer operates on the 4 qubits defining the starting location of Ambulance 0 so as to ensure that the ambulance is placed at precisely 1 location
    \item Similarly an XY mixer operates on the 4 qubits defining the starting location of Ambulance 1 so as to ensure that the ambulance is placed at precisely 1 location
    \item A third XY mixer is used to ensure that exactly 4 locations are served by the two ambulances taken together - in this case the XY mixer operates on 8 qubits
\end{enumerate}


In Figure \ref{Problem B - Encoding of 2 ambulance 4 locations} an example solution is shown noting that a filled circle denotes the start or destination location of ambulance 0 or 1 as appropriate.  The starting states need to be a combination of Dicke states with Hamming weights of 1, 1 and 4 respectively.  Although 16 qubits in principle offers $2^{16}$ possible states, the use of the XY mixers reduces the number of states searched to 1120 (4x4x70\footnote{Note that $^{8}C_{4} =70$}), a factor of $\sim${}60 less; this advantage scales exponentially with the number of qubits in a given XY ring mixer.  In addition we introduce a penalty term into the cost function to ensure that only 1 ambulance serves each location; together with the Hamming weight of 4 requirement, this also achieves the requirement that all 4 locations are served.  This encoding does not prevent both ambulances starting at the same location which would require a further penalty term if that was a constraint of the problem.  The need for a penalty term means that unlike the situation in the previous section where we used an XY mixer to ensure only feasible states were found, we are not guaranteed that all solutions generated will be feasible ones.  Accordingly we expect that the Probability of a feasible  state measure will be $\leq$ 1.



We are now faced with a choice as to how many independent $\beta$ and $\gamma$ angles to use amongst the 3 separate XY mixers.  In principle we can use anything from 1 - 3 independent angles per step for each of $\beta$ and $\gamma$. We investigated 4 separate approaches for the number of 
\{$\beta$, $\gamma$\} values per step: \{1, 1\}, \{2, 1\}, \{3, 1\}, \{3, 3\}. In the case where two $\beta$ angles are used, one $\beta$ applies to the 2 XY mixers covering the start location allocation and the other $\beta$ applies to the destination part.  The other combinations of $\beta$ and $\gamma$ used and their allocation amongst start and destination qubits are self explanatory.  

The $r^{th}$ mixer term ($0 < r < p$) is $e^{i\beta^rH_B^r}$ where 




\begin{equation}
\beta^r H_B^r = \frac{1}{2}\underbrace{\sum_{i=0}^{3}\beta_1^r(\sigma_i^x\sigma_{i+1}^x + \sigma_i^y\sigma_{i+1}^y)}_\text{A} +\underbrace{\sum_{j=4}^{7}\beta_2^r(\sigma_j^x\sigma_{j+1}^x + \sigma_j^y\sigma_{j+1}^y)}_\text{B} +\underbrace{ \sum_{k=8}^{15}\beta_3^r(\sigma_k^x\sigma_{k+1}^x + \sigma_k^y\sigma_{k+1}^y)}_\text{C}
\end{equation}


When the summation index takes its maximum value, i.e. $i=3$, $j=7$, $k=15$, it should be understood to mean $i+1=0$, $j+1=0$, $k+1=0$ and results in a ring mixer in each case.  The letters A, B and C correspond to the XY mixers operating on the qubits representing the start location of Ambulance 0, Ambulance 1 and the destination locations of the 2 ambulances respectively (see Figure \ref{Problem B - Encoding of 2 ambulance 4 locations}). Note that the 3 $\beta$ angles can all be different (3 $\beta$s), all the same (1 $\beta$) or 2 the same ($\beta_1^r = \beta_2^r$) and one different  according to the choices outlined above. For the phase separating terms of the form  $e^{i\gamma^rH_C^r}$ in the case where there are 3 $\gamma$ angles, the $\gamma$ s are allocated so that where there are mixed terms involving qubits from separate blocks of qubits, e.g. start qubit and a destination qubit, the $\gamma$ angle used is that applying to the start qubit.





\begin{table}[h]
\begin{center}
\caption{\textbf{Performance of different XY mixer implementations}}
\vspace{0.3cm}
\begin{tabular}{| >{\centering\arraybackslash}m{3cm}| >{\centering\arraybackslash}m{2.5cm} | >{\centering\arraybackslash}m{2.5cm} |>{\centering\arraybackslash}m{2.5cm} | >{\centering\arraybackslash}m{2.5cm} | >{\centering\arraybackslash}m{2.5cm} |} 
 \hline
XY implementation \boldmath{\{$\beta, \gamma$\}} & \textbf{Steps} & \textbf{Minimum cost function value} & \textbf{Prob. of lowest cost function states} & \textbf{Classical optimizer iterations - average}  \\ 
 \hline
 
 \{$1, 1$\} & $p = 1$ & 15.202 & 3.24\% & 33 \\ 
\hline
\{$1, 1$\} & $p = 2$ & 13.598 & 4.03\% & 95\\
\hline
\{$1, 1$\} & $p = 3$ & 8.196 & 5.79\% & 371\\
\hline
\{$2, 1$\} & $p = 1$ & 13.951 & 3.44\% & 75\\
\hline
\{$2, 1$\} & $p = 2$ & 12.185 & 3.43\% & 400\\
\hline
\{$3, 1$\} & $p = 1$ & 13.951 & 3.44\% & 151\\
\hline
\{$3, 1$\} & $p = 2$ & 10.848 & 2.48\% & 845\\
\hline
\{$3, 3$\} & $p = 1$ & 14.947 & 2.97\% & 362\\
\hline
\{$3, 3$\} & $p = 2$ & 13.531 & 3.88\% & 999\\
\hline
\end{tabular}
\label{XY mixer results for beta gamma approaches}
\caption*{\small{The table shows the minimum cost function value and the associated probability of finding the problem's lowest cost function state for varying numbers of $\beta$ and $\gamma$ angles in the 3 XY mixer implementation of QAOA on Problem B (4 locations, 2 ambulances.  In each case a minimum of 200 optimisation runs using random initial angles for each of the $\beta$ and $\gamma$ angles using Nelder-Mead as the classical optimiser were completed. No significant improvement was found using the larger numbers of $\beta$ and $\gamma$ angles while the time to convergence increased rapidly.}}
\end{center}
\end{table}

Results making use of the Nelder-Mead classical optimiser and based on multiple runs using a minimum of 200 sets of random angles drawn from $[0,2\pi)$ for the initial angles are presented in Table \ref{XY mixer results for beta gamma approaches}. From the results, there are two clear conclusions.  Firstly we observe the expected pattern of improved results as the number of steps, $p$, increases. Secondly the best results obtained for each level $p$ is similar for each of the  \{$\beta$, $\gamma$\} combinations.   
Further we note that as the number of separate $\beta$ and $\gamma $ parameters increases the number of classical optimizer iterations required, a rough proxy for the run-time, increases rapidly; this is attributable to the curse of dimensionality as described by Wang et al \cite{Wang18a}.  This is of particular relevance when we seek to combine the use of the XY mixer(s) with the increasing-p strategy described in section \ref{Efficient strategies for selecting}, when the number of parameters grows further due to the greater number of steps involved.

As a result we conclude that when using a random angle initial ansatz approach there is no significant benefit from working with larger numbers of $\beta$ and $ \gamma$ in terms of the Cost function expectation value generated and associated probability of finding the lowest cost function state(s). This does not mean that by choosing {\{$\beta =1, \gamma =1 $\}} we can match the lowest cost function value obtained for other {\{$\beta, \gamma $\}} possibilities for low $p$ values; rather that allowing for the benefits from increasing the number of steps and a good strategy for selecting start angles for increasing $p$ optimisations, choosing a {\{$\beta =1, \gamma =1 $\}} or {\{$\beta =2, \gamma =1 $\}} approach can perform better as part of an overall strategy for finding a low cost function value when assessed in terms of resource required and time taken to achieve a given level of solution.  Accordingly in the remainder of this section we focus on \{$\beta =1, \gamma =1 $\} 
and \{$\beta =2, \gamma =1 $\} angles per step implementations of the 3 XY mixer approach.





\begin{figure}[H]
\caption{\textbf{3XY mixer approach - Performance of different increasing $p$ strategies - Problem B  (4 locations, 2 ambulances)}}
\caption*{1 beta, 1 gamma \hspace{5.2cm}   2 betas, 1 gamma}
\begin{subfigure}{.5\linewidth}
\centering
\caption{Approximation ratio}
\includegraphics[width=.9\linewidth]{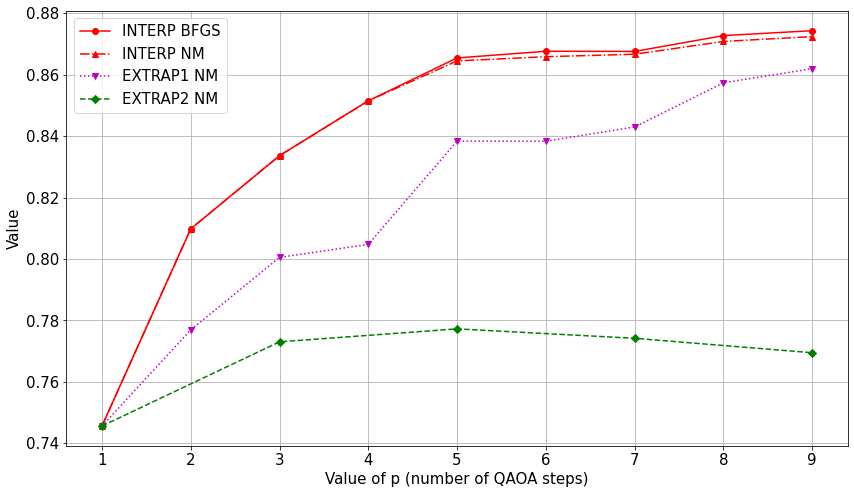}
\label{fig:sub13a}
\end{subfigure}%
\begin{subfigure}{.5\linewidth}
\centering
\caption{Approximation ratio}
\includegraphics[width=.9\linewidth]{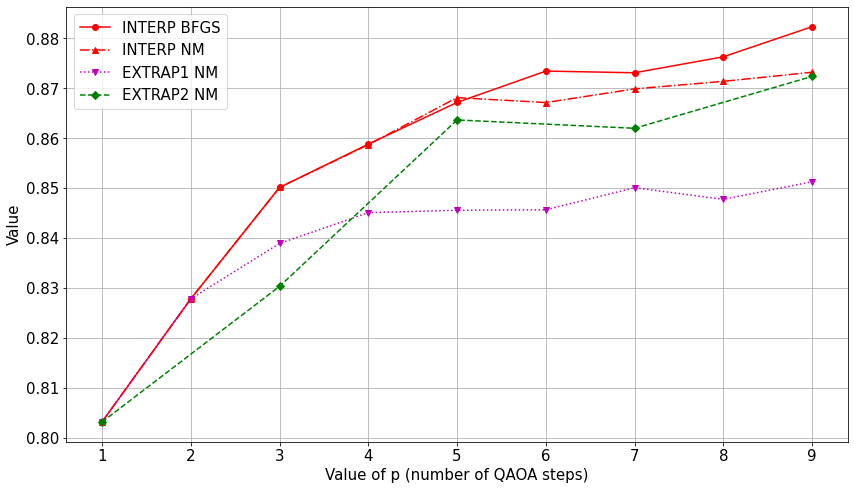}
\label{fig:sub23a}
\end{subfigure}\\[1ex]
\begin{subfigure}{.5\linewidth}
\centering
\caption{Probability of feasible state}
\includegraphics[width=.9\linewidth]{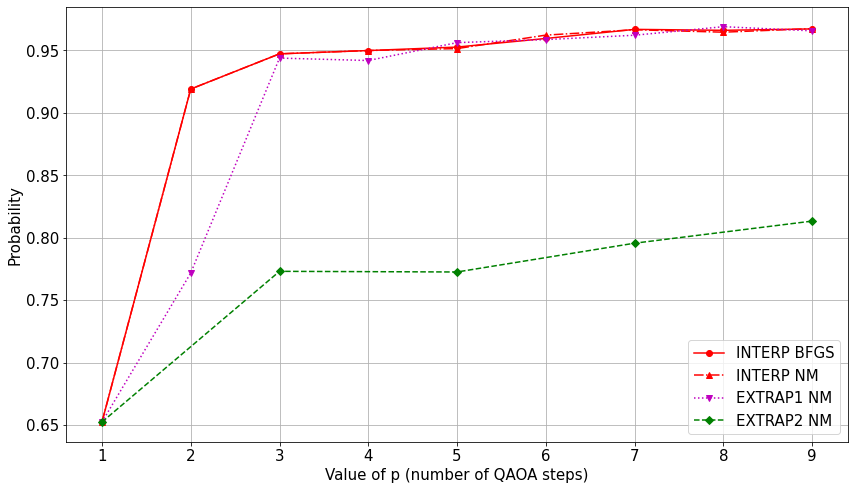}
\label{fig:sub33a}
\end{subfigure}
\begin{subfigure}{.5\linewidth}
\centering
\caption{Probability of feasible state}
\includegraphics[width=.9\linewidth]{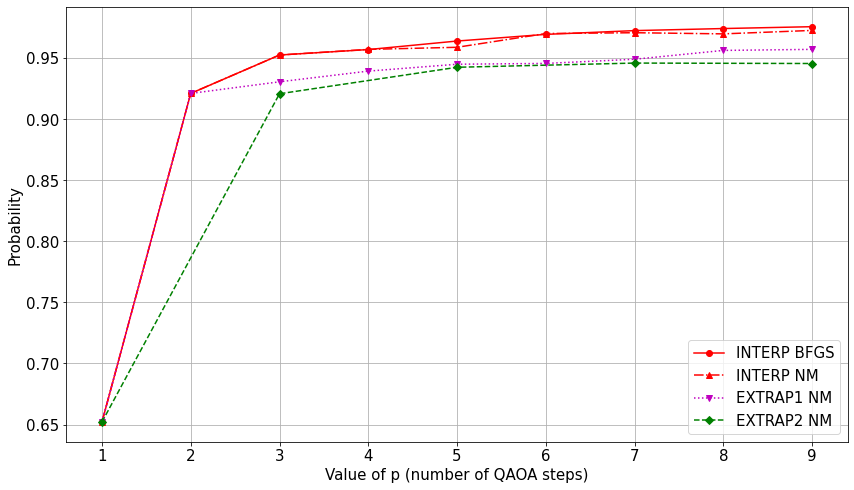}
\label{fig:sub43a}
\end{subfigure}\\[1ex]
\begin{subfigure}{.5\linewidth}
\centering
\caption{Probability of lowest cost function state}
\includegraphics[width=.9\linewidth]{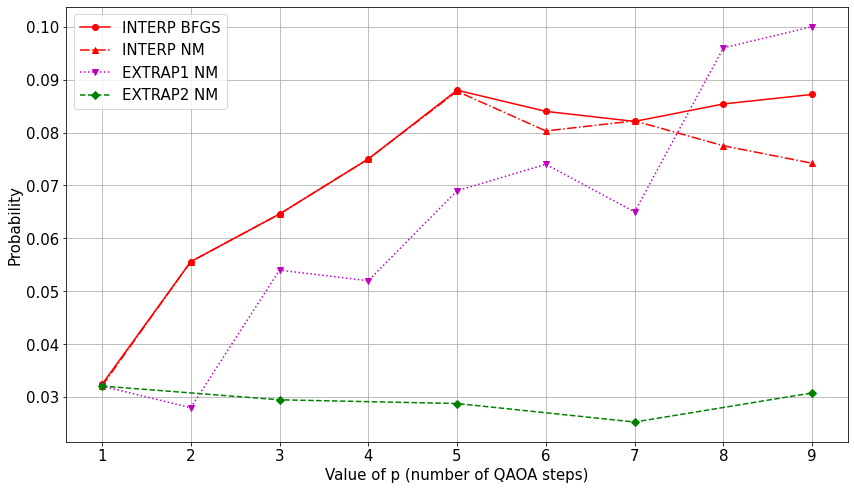}
\label{fig:sub53a}
\end{subfigure}
\begin{subfigure}{.5\linewidth}
\centering
\caption{Probability of lowest cost function state}
\includegraphics[width=.9\linewidth]{3XY_q16_g1b1_p_gnd_interp}
\label{fig:sub63a}
\end{subfigure}\\[1ex]
\label{Increasing p strategy 3XY method}
\caption*{\small{The figure shows the performance of the INTERP and EXTRAP1/2 strategies for a QAOA optimisation applied to the 3 XY mixer approach for $\{\beta, \gamma\} = \{1,1\}$  and $\{\beta, \gamma\} = \{2,1\}$. The increasing-$p$ strategies are  seeded with initial angles sourced from the best outcome (lowest cost function value) using random $\beta$ and $\gamma$ angles for a $p=1$ QAOA as reported in Table \ref{XY mixer results for beta gamma approaches}.  The problem instance used is for 4 locations and 2 ambulances.  Nelder-Mead (``NM'') is used as the classical optimiser with BFGS also used with the INTERP strategy. The metrics used are the approximation ratio, probability of feasible state and, as the prime metric, probability of lowest cost function state.  The best strategy (EXTRAP1 for $\{\beta, \gamma\} = \{1,1\}$  and INTERP for $\{\beta, \gamma\} = \{2,1\}$) in each case provides an efficient method of increasing the probability of obtaining the lowest cost function state  by  a factor of $\sim$3-3.5.}}
\end{figure}

We now apply the increasing steps methodology used in the previous section \ref{Efficient strategies for selecting} to the 3 XY mixer approach. In Figure \ref{Increasing p strategy 3XY method}  we show the results for increasing $p$ of applying the INTERP approach (with both Nelder-Mead and BFGS optimisers), EXTRAP1  and EXTRAP2 to problem variant B using as the initial angles those generating the lowest cost function from a set of 100 random angle instances.   The increasing $p$ strategies are again successful in generating lower expected Cost function values efficiently by comparison with simply using a random angle approach.
The INTERP strategy performs best of the 3 strategies in terms of improving the Approximation ratio and boosting the Probability of a solution falling within the feasible space.  EXTRAP1 performs comparably to INTERP in the  {\{$\beta =1, \gamma =1 $\}} case with EXTRAP2 exhibiting little gain from increasing $p$.  In contrast for the  {\{$\beta =2, \gamma =1 $\}} case EXTRAP2 performs better than EXTRAP1 and generates improvements close to those of the INTERP strategy.  Due to the indirect link between Approximation ratio and the probability of obtaining the lowest cost function state, we find that it is actually EXTRAP1 which generates the highest probability of finding the lowest cost function state for {\{$\beta =1, \gamma =1 $\}}. When we tried the 3 strategies on 3 other low cost function angle sets, we found that the results obtained were qualitatively better for the {\{$\beta =2, \gamma =1 $\}} approach versus the {\{$\beta =1, \gamma =1 $\}} approach; this is consistent with the higher number of parameters over which optimisation is occurring.  We did not however observe any clear difference in performance between the 3 increasing-p strategies.
If we break down the sources of gain for the angle instance shown in Figure  \ref{Increasing p strategy 3XY method} into the component factors as shown in Table \ref{Performance contributions 3XY}, we see that the choice of a 3 XY mixer contributes most of the improvement (factor of 58.5) with the choice of a good seed angle (factor of $\sim${}3) and the aggregate contribution from the increasing $p$ components (factor of 3.1-3.5) of a similar size.  We should note that in this case the lowest cost function state is 12 fold degenerate, which implies that the probability of finding a lowest cost function  state in the starting ansatz composed of a set of equally weighted feasible states is 12/1120 = 0.0107. 


In conclusion, these examples demonstrate that increasing $p$ strategies offer a useful improvement in boosting the lowest cost function state probability and are more effective than simply using a random angle approach with larger values of $p$.  However, at least for smaller problems, the gains from using an XY mixer are significantly greater and given the exponential relationship between the total Hilbert space and the feasible space, are likely to maintain that advantage as problem size grows.


\subsection{QAOA - challenges and developments} \label{QAOA - challenges}

The potential power of QAOA as a quantum computing algorithm has been evident from when it was first proposed as it enjoys both a strong theoretical underpinning and also is inherently noise resilient.  Notwithstanding, as its application has been studied further a number of important challenges have emerged. Perhaps the most significant is the discovery that so-called `barren plateaus' frequently arise in the main forms of variational quantum ansatz (QAOA, VQE) which are gradient free zones which make it difficult or even impossible for the algorithm to find the global minimum of the problem cost function \cite{McClean2018a}. We comment about the challenges and methods of addressing this important obstacle to implementation in section \ref{VQE challenges} in relation to VQE, although we should note that as regards QAOA progress has been modest to date and the issue of barren plateaus remains to be overcome.

A second important challenge for QAOA is the degree of connectivity of current quantum computing hardware.  Many of the leading QPU hardware producers, particularly those using a superconducting approach, only offer nearest neighbour connectivity between qubits often on a 2D grid with maximum neighbour connections of 4 and with most qubits having connectivity of only 2 or 3 \footnote{We note that certain technologies do offer all to all connectivity, at least in principle, including ion trap and photonic approaches}.  For any significant size ``real-world'' problem, the problem Hamiltonian will have in many cases nodes with connectivity that is proportional to $N$, where $N$ is the total number of problem variables.  As a result each layer of QAOA will potentially require the use of many `swap' gates to enable the necessary gate  interactions between qubits and this will necessarily introduce greatly increased noise effects, which for a sufficiently large problem would ensure the measured results would be poor at best. One approach which has recently been proposed to lessen this challenge is the use of a Parity encoding methodology which allows both the underlying cost function and any associated constraints to be mapped to a grid based QPU architecture requiring only nearest neighbour interactions \cite{Ender2021a, Ender2021b, Lechner2015a, Ender2022a}.

Studies of the properties of the parameter values obtained as the solutions to various standard problems such as Max Cut and Maximum k-vertex Cover have found that there are often repeatable patterns to the optimal parameter values.  In fact many researchers have found that the optimal set of parameters can be used unchanged for different instances of the same problem type and give high quality solutions in most or all cases.  This has given rise to a strong hypothesis that specific problem types may be solved making use of a previously discovered fixed set of parameters thereby greatly reducing the run time required for QAOA as no iterative classical loop is required.  Further, making use of the INTERP strategy we explored in section \ref{Efficient strategies for selecting}, successful test results have been found extending a set of fixed parameters to greater QAOA circuit depths without the need for further parameter optimisation \cite{Brandao2018a,  Cook2020a, Akshay2021a}. 

A further path of QAOA development has been how the parameters and the mixer gates themselves are determined.  Zhu et al \cite{Zhu2020a}  proposed an approach whereby instead of a fixed mixer Hamiltonian for all layers of a QAOA the mixer Hamiltonian is chosen from a set of one and two qubit gates at each step based on maximising the gradient of the expectation value of the cost function with the mixer so chosen.  They give it the name Adaptive Derivative Assembled Problem Tailored - Quantum Approximate Optimization Algorithm
(ADAPT-QAOA). This can be considered as a sort of evolution of the Quantum Alternative Optimization Algorithm.  They found that on the Max Cut problem, there was a significant reduction in the number of layers required to achieve convergence by comparison with standard QAOA.    Herrman et al investigated using multiple different angles (parameter values) within each layer of mixer and phase separating Hamiltonian \cite{Herrman2021a}.  The results obtained on the Max Cut problem showed a 33\% increase in the approximation ratio for the same number of layers relative to standard QAOA.  In the context of noisy QPUs, the reduction in layers has clear advantages although the classical optimisation task is of course made more difficult and therefore time-consuming.  

Research is being increasingly speedily undertaken on a range of techniques to improve QAOA based or derived performance in preparation for the arrival of more performant QPU hardware.  It is likely that this rate of advance will continue.

\begin{table}[ht]
\begin{threeparttable}
\caption{\textbf{Contributions of different sources of performance improvement}}
 \label{Performance contributions 3XY}
\vspace{0.3cm}
\begin{tabular}{| >{\centering\arraybackslash}m{3cm}| >{\centering\arraybackslash}m{1.25cm} | >{\centering\arraybackslash}m{1.25cm} |>{\centering\arraybackslash}m{1.5cm} | >{\centering\arraybackslash}m{2.5cm} | >{\centering\arraybackslash}m{2.5 cm} |>{\centering\arraybackslash}m{1.5 cm} |}
 \hline
\multirow{2}{2cm}[-5mm]{\centering{\{$\boldsymbol{\beta, \gamma}$\} \textbf{approach}}} & \multirow{2}{1.25cm}[-5mm]{\centering{\textbf{Mixer choice}}} & \multirow{2}{1.25cm}[-5mm]{\centering{\textbf{Seed angle}}} & \multicolumn{3}{c|}{\textbf{Increasing-$p$ strategy contribution}} & \multirow{2}{1.5cm}[-5mm]{\centering{\textbf{Overall gain}}} \\ 
  \cline{4-6}
 &   &  & \textbf{Feasible state} & \textbf{Approximation ratio}   & \textbf{Lowest cost function state mix} & \\
 \hline
 
\hline
 
{\{$\beta =1, \gamma =1 $\}}& 58.5 & 3.02 & 1.48 & 1.16 & 1.81 & 546.4\\
\hline
{\{$\beta =2, \gamma =1 $\}}& 58.5 & 3.21 & 1.50 & 1.10 & 2.15 & 665.3\\
\hline
\end{tabular}
\begin{tablenotes}
    \item \small{All numbers represent factor contributions to the improvement achieved, as measured by the probability of finding the lowest cost function state.  The benefit from the seed angle arises from running multiple sets of randomly drawn angles for a QAOA $p=1$ and selecting the angle set generating the lowest expected cost function value. The largest gain derives from the mixer choice with the gain from seed angle and the increasing-$p$ strategy element (comprised of the gains from a more probable feasible state, higher approximation ratio and improved state mix) both contributing similarly at a factor of $\sim$3-3.5.}
\end{tablenotes}
\end{threeparttable}
\end{table}

\newpage
\subsection{Variational Quantum Eigensolver results}

In this section we will present a small number of results from using the Variational Quantum Eigensolver algorithm. Although the VQE method is usually associated with applications in chemistry it can also be used for any optimisation problem.  However, until recently, the approach has not had the number of different implementations as we have seen with the QAOA method. We therefore just present the results obtained on a small number of the ambulance sub-problems we investigated in relation to QAOA; these are problem A (5 locations, 1 ambulance) and problem B (4 locations, 2 ambulances).

\subsubsection{VQE circuits} 

A VQE circuit is normally composed of two types of gates: single qubit rotation gates and 2-qubit entangling gates. An example is shown in Figure \ref{VQE_circ_res1}.   This shows an initial layer followed by a single entangling layer.  As can be seen the entangling layer is composed of a set of CNOT gates sandwiched between a pair of rotation gates.  In the example shown there are 5 qubits, 13 paramaterised $R_y$ rotation gates and 4 entangling gates which represents an initial layer and 1 entangling layer.  We considered 3 VQE structures:  1) no initial layer and 1 entangling layer, 2) an initial layer a 1 entangling layer  and 3) no initial layer and 2 entangling layers.
\begin{figure}[ht]
    \caption{VQE circuit diagram showing an initial rotation layer and 2 entangling layers}
    \begin{tikzcd}
       \lstick{\ket{q_{0}}} & \qw & \gate{R_y(\theta_0)} & \qw & \qw \slice{} & \ctrl{1} & \gate{R_y(\theta_5)}  & \qw  & \slice{} \qw & \ctrl{1} & \gate{R_y(\theta_{13})} &  \qw & \slice{} \qw & \qw &  \qw   \qw \\
    \lstick{\ket{q_{1}}}& \qw & \gate{R_y(\theta_1)}& \qw& \qw & \targ{} & \gate{R_y(\theta_6)} & \ctrl{1} & \gate{R_y(\theta_9)} &  \targ{} & \gate{R_y(\theta_{14})} & \ctrl{1} & \gate{R_y(\theta_{17})} & \qw & \qw  \\
    \lstick{\ket{q_{2}}} & \qw & \gate{R_y(\theta_2)}& \qw& \qw & \ctrl{1} & \gate{R_y(\theta_7)} & \targ{} & \gate{R_y(\theta_{10})}& \ctrl{1} & \gate{R_y(\theta_{15})} & \targ{} & \gate{R_y(\theta_{18})} &  \qw & \qw \\
    \lstick{\ket{q_{3}}}& \qw & \gate{R_y(\theta_3)}& \qw& \qw & \targ{} & \gate{R_y(\theta_8)} & \ctrl{1} & \gate{R_y(\theta_{11})} &  \targ{} & \gate{R_y(\theta_{16})} & \ctrl{1} & \gate{R_y(\theta_{19})} & \qw & \qw  \\
    \lstick{\ket{q_{4}}} & \qw & \gate{R_y(\theta_4)}& \qw & \qw & \qw & \qw & \targ{} & \gate{R_y(\theta_{12})} & \qw & \qw & \targ{} & \gate{R_y(\theta_{20})} &  \qw & \qw
    \end{tikzcd}
    \caption*{\textbf{Rotation gate layer \hspace{0.75cm} 1st entangling gate layer \hspace{1.25cm} 2nd entangling gate layer \hspace{1.5cm} {}}}
    \caption*{\small{The figure shows the VQE circuit(s) used for solving Problem A (5 locations, 1 ambulance). Three variations of the circuit were used: 1) a rotation gate layer and 2 entangling layers (21 paramaterised gates) being the full circuit shown, 2) a rotation gate layer and 1 entangling layer (13 paramaterised gates), 3) 1 entangling layer only (8 paramaterised gates).}}  
    \label{VQE_circ_res1}
\end{figure}
\begin{figure}[h]
\caption{A $Z_0Z_2$ edge circuit reduced to its casual cone form}
    \begin{center}
    \begin{tikzcd}
    \lstick{\ket{q_{0}}} & \gate{R_y(\theta_0)}\slice{} & \ctrl{1} & \gate{R_y(\theta_{10})} & \qw   & \qw \slice{}& \qw & \gate{Z}& \qw \\
    \lstick{\ket{q_{1}}} & \gate{R_y(\theta_1)} & \targ{} & \gate{R_y(\theta_{11})} & \ctrl{1} & \qw & \qw & \qw & \qw \\
    \lstick{\ket{q_{2}}} & \gate{R_y(\theta_2)} & \ctrl{1} & \gate{R_y(\theta_{12})} & \targ{} & \gate{R_y(\theta_{23})} & \gate{Z} & \qw & \qw \\
    \lstick{\ket{q_{3}}} & \gate{R_y(\theta_3)} & \targ{} & \qw & \qw & \qw & \qw & \qw & \qw \\
    \lstick{\ket{q_{anc}}} & \gate{H} & \qw & \qw &  \qw   & \qw & \ctrl{-2}  &\ctrl{-4}& \gate{H} & \qw
    \end{tikzcd}
    \caption*{\small{In this circuit only the qubits and gates required to measure the effect of the $Z_0Z_2$ edge are shown.  This is known as a causal cone and is smaller in size than the full circuit shown in the previous figure. As in Figure \ref{Increasing p strategy comparison 1}, the ancilla qubit, $q_{anc}$, is used to measure the value of the edge making use of the Hadamard test \cite{Bravo-Prieto19a}. }}
    \label{VQEcausalcone}
    \end{center}
\end{figure}
\subsubsection{Causal cones (a Hardware Efficient approach) and output measurement strategy} 

When a VQE circuit is run, it is possible to make use of a hardware efficient approach which reduces the number of qubits required.  In Figure \ref{VQE_circ_res1}, edge $Z_0Z_2$ is not shown.  In Figure \ref{VQEcausalcone} the same edge is shown but with only the qubits which affect the output of that edge.  We can see that only 4 of the 5 qubits are required plus the ancilla qubit, $|q_5\rangle$ used to measure the value of the edge; this is because there is no causal connection of any other qubit to the output values of  $|q_0\rangle$ and $|q_2\rangle$ as ultimately measured on the ancilla qubit.  While this is a small benefit in this case (5 reduces to 4), if the problem Hamiltonian is large and requires many tens of qubits or more, then the problem can still be evaluated using a much smaller circuit size - the precise number of qubits will depend on the circuit depth which itself is a function of the number of entangling layers.

There are 2 possible approaches to edge measurement when the problem is sufficiently small that the Hamiltonian can be incorporated onto a single circuit.  In this case it is possible to simply measure the  output value of all qubits, say $M$ times, and thereby deduce a range of mixed states.  These mixed states are substituted into the problem Hamiltonian and an expected value obtained subject to a specific level of error which is determined by the number of measurements.  Alternatively, we can follow the causal cone approach and determine the expected value for each edge and then sum the values obtained for each edge.  The disadvantage of the second approach is twofold: firstly it requires a greater number of shots as each edge value is determined individually with its own measurement error and secondly that it introduces an additional source of error which is avoided in the first approach, namely that the value measured for each qubit can vary between edges involving the same qubit in a way which is not possible when all qubit values are measured at the same time. 
For a NISQ-era QPU there is an advantage in the higher fidelity (lower gate error rate) possible from a narrower circuit. This improved fidelity must be weighed against first, the increased sampling error from sampling each cone separately and secondly, against the increased resources required to sample many different cones (one for each edge in the problem) rather than sample the complete circuit.  In Problem A, there are 15 separate edges/nodes in the Ising problem, so using cones of circuits can lead to over 15 times the number of shots to evaluate the expectation of a circuit.

\begin{figure}[ht]
    \caption{VQE circuit diagram showing an initial rotation layer and 2 entangling layers}
    \begin{tikzcd}
       \lstick{\ket{q_{0}}} & \qw & \gate{R_y(\theta_0)} & \qw & \qw \slice{} & \ctrl{1} & \gate{R_y(\theta_5)}  & \qw  & \slice{} \qw & \ctrl{1} & \gate{R_y(\theta_{13})} &  \qw & \slice{} \qw & \qw &  \qw   \qw \\
    \lstick{\ket{q_{1}}}& \qw & \gate{R_y(\theta_1)}& \qw& \qw & \targ{} & \gate{R_y(\theta_6)} & \ctrl{1} & \gate{R_y(\theta_9)} &  \targ{} & \gate{R_y(\theta_{14})} & \ctrl{1} & \gate{R_y(\theta_{17})} & \qw & \qw  \\
    \lstick{\ket{q_{2}}} & \qw & \gate{R_y(\theta_2)}& \qw& \qw & \ctrl{1} & \gate{R_y(\theta_7)} & \targ{} & \gate{R_y(\theta_{10})}& \ctrl{1} & \gate{R_y(\theta_{15})} & \targ{} & \gate{R_y(\theta_{18})} &  \qw & \qw \\
    \lstick{\ket{q_{3}}}& \qw & \gate{R_y(\theta_3)}& \qw& \qw & \targ{} & \gate{R_y(\theta_8)} & \ctrl{1} & \gate{R_y(\theta_{11})} &  \targ{} & \gate{R_y(\theta_{16})} & \ctrl{1} & \gate{R_y(\theta_{19})} & \qw & \qw  \\
    \lstick{\ket{q_{4}}} & \qw & \gate{R_y(\theta_4)}& \qw & \qw & \qw & \qw & \targ{} & \gate{R_y(\theta_{12})} & \qw & \qw & \targ{} & \gate{R_y(\theta_{20})} &  \qw & \qw
    \end{tikzcd}
    \caption*{\textbf{Rotation gate layer \hspace{0.75cm} 1st entangling gate layer \hspace{1.25cm} 2nd entangling gate layer \hspace{1.5cm} {}}}
    \caption*{\small{The figure shows the VQE circuit(s) used for solving Problem A (5 locations, 1 ambulance). Three variations of the circuit were used: 1) a rotation gate layer and 2 entangling layers (21 paramaterised gates) being the full circuit shown, 2) a rotation gate layer and 1 entangling layer (13 paramaterised gates), 3) 1 entangling layer only (8 paramaterised gates).}}  
    \label{VQE_circ_res}
\end{figure}

\begin{table}[H]
\begin{center}
\caption{\textbf{VQE results for Problem A (5 locations, one ambulance)}}
\vspace{0.3cm}
\begin{tabular}{| >{\centering\arraybackslash}m{2.5cm} | >{\centering\arraybackslash}m{2cm}|  >{\centering\arraybackslash}m{2cm} |  >{\centering\arraybackslash}m{2cm} |
>{\centering\arraybackslash}m{2.5cm} |
>{\centering\arraybackslash}m{2.5cm} |}
 \hline
  \textbf{Number of parameters (layers) }  &  \textbf{Optimiser} & \textbf{Calculation method} &  \textbf{Approx. ratio} & \textbf{ Prob. of feasible state }& \textbf{Prob. of lowest cost function state }  \\
 \hline
   8 (0+1)  & L-BFGS-B  & Wavefunction  &  $0.975 \pm 0.015$ & $0.964 \pm 0.030$ &  $0.883 \pm 0.060$\\
  \hline
   13 (1+1)  &  L-BFGS-B  & Wavefunction  &  $0.959 \pm 0.025$ &  $0.924 \pm 0.040$ &  $0.839 \pm 0.067$ \\
  \hline
   21 (1+2)  & L-BFGS-B  & Wavefunction  & $0.933 \pm 0.036$ &  $0.884 \pm 0.052$ &  $0.792 \pm 0.072$ \\
  \hline
  \hline
8 (0+1)  &  SPSA  & Wavefunction &  
$0.885 \pm  0.024$ & $0.674 \pm  0.065$ &  $0.421 \pm  0.088$\\
\hline
  13 (1+1)  &  SPSA  & Wavefunction  &  
  $0.934 \pm  0.018$ &       $0.647 \pm  0.063$ & $0.523 \pm  0.073$\\
  \hline
  21 (1+2)  & SPSA  & Wavefunction  & 
   $0.722 \pm  0.045$ & $0.462 \pm  0.048$ &  $0.194 \pm  0.044$\\
  \hline
  \hline
  8 (0+1)  & L-BFGS-B  & All qubit sampling  &  $0.879 \pm 0.027$&  $0.705 \pm 0.069$ &  $0.457 \pm 0.088$ \\
  \hline
   13 (1+1)  & L-BFGS-B  & All qubit sampling  &  $0.900\pm 0.021$ &  $0.658 \pm 0.071$ &  $0.480 \pm 0.085$ \\
  \hline
  \hline
   8 (0+1)  & L-BFGS-B  & Sampling - causal cone  &  $0.844 \pm 0.037$ & $0.675 \pm 0.071$ & $0.408\pm 0.091$ \\
  \hline
   13 (1+1)  & L-BFGS-B  & Sampling - causal cone  &  $0.771 \pm 0.041$ &  $0.568 \pm 0.057$ &  $0.238 \pm 0.059$ \\
  \hline
  \hline
  8 (0+1)  & SPSA  & All qubit sampling  &  $0.905 \pm  0.022$ & $0.61 \pm  0.07$ &  $0.435 \pm  0.083$ \\
  \hline
  13 (1+1)  & SPSA  & All qubit sampling  & $0.921  \pm  0.019$ &       $0.622  \pm  0.066$ & $0.485  \pm  0.074$ \\  
  \hline

\end{tabular}
\label{VQE_all}

\caption*{\small{The table shows a set of results from a simulator for various VQE ansatz circuits using either the L-BFGS-B or SPSA classical optimiser and running on the 5 location, 1 ambulance problem.  3 different parameterised ansatz circuits are used as described in Figure \ref{VQE_circ_res}. In each case the results presented are the means from 100 runs using randomised initial sets of parameters.  The errors quoted are 2 standard deviations of the mean.  The meta parameters used for the classical optimisers were for SPSA, $a =0.01 , c= 0.01$, and for L-BFGS-B, $\epsilon = 0.1$.  Experimental runs were completed using a full wavefunction approach, all qubit sampling (9,000 shots) and causal cone sampling (9,000 shots).}}
\end{center}

\end{table}

\subsubsection{VQE simulator results for Problem A} 

\begin{figure}[H]
\begin{center}
\caption{Figures of merit for an 8 parameter Ansatz}
    \includegraphics[width=.9\linewidth]{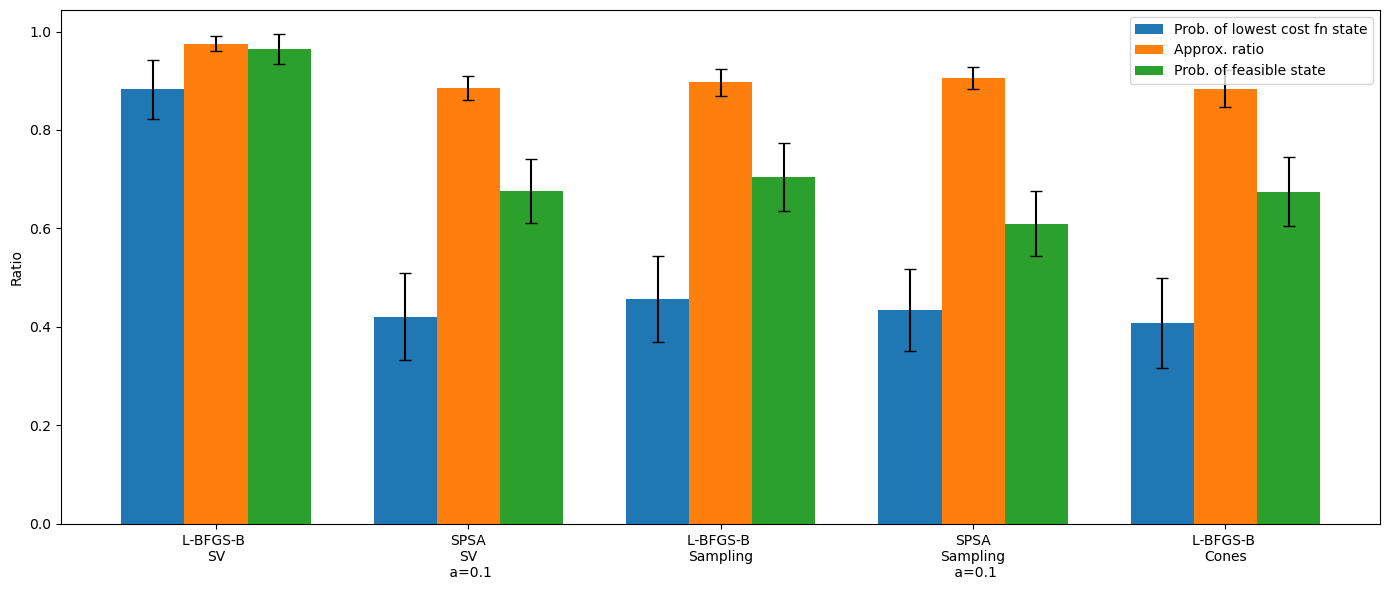}
    \caption*{\small{Samples from a quantum circuit (9000 shots) were used to evaluate the expected value (`EV') for a given set of circuit parameters.  For each random starting angle (which contains the 8 parameters for the VQE Ansatz) a final angle is selected either an SPSA optimization with 100 iterations or L-BFGS-B. The ratios shown are the average of 100  starting angles.      The SPSA meta parameters 'a' and 'c' were set to 0.1 , the L-BFGS-B $\epsilon$ was set to 0.1}}
    \label{8param_ratios}
\end{center}
\end{figure}

We made use of the VQE approach to solve Problem A (5 locations, 1 ambulance) using a simulator to examine the results from a full state-vector calculation as well as from using the 2 different shot measurement approaches described in the previous section. The VQE circuit configurations are shown in Figure \ref{VQE_circ_res}.  This shows the most complex circuit used which has an initial rotation layer followed by 2 sets of entangling layers.  Two other circuits consisting of a single rotation layer only and a single entangling layer with an initial rotation layer were also used.  The summary results are presented in Table \ref{VQE_all} and Figure \ref{8param_ratios}. From these a number of conclusions and observations can be made.

\vspace{1cm}

\begin{figure}[H]
\begin{center}
\caption{Figures of merit for a 13 parameter Ansatz}
    \includegraphics[width=.9\linewidth]{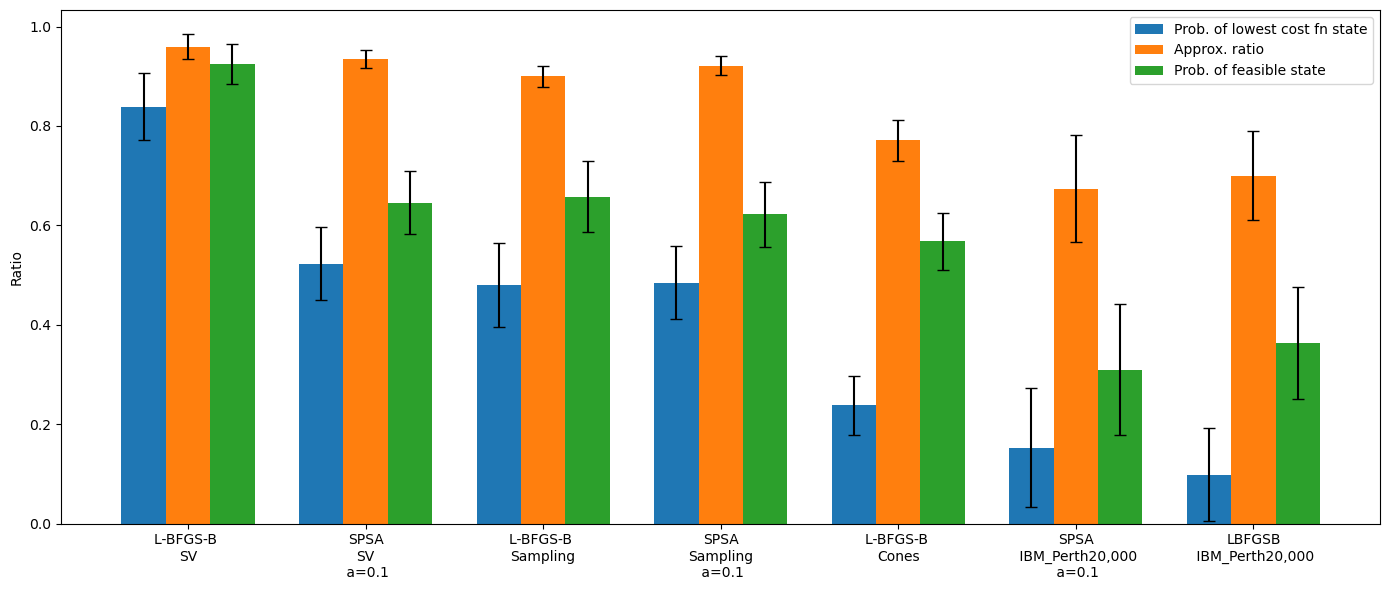}
    \caption*{\small{Samples from a quantum circuit (9000 shots) were used to evaluate the expected value ("EV") for a given set of circuit parameters.  For each random starting angle (which contains the 13 parameters for the VQE Ansatz) a final angle is selected by either an SPSA optimization with 100 iterations or L-BFGS-B. The ratios shown are the average of 100  starting angles.      The SPSA meta parameters 'a' and 'c' were set to 0.1 , the L-BFGS-B $\epsilon$ was set to 0.1}}
    \label{13param_ratios}
\end{center}
\end{figure}

\begin{itemize}
    \item \textbf{Good performance from the trialled VQE circuits}\\
    All the different VQE parameterised circuits trialled were highly successful in generating the lowest cost function state with high  probability. 
    
    \item\textbf{Too many parameters can damage the quality of optimisation result}.\\
     When using a L-BFGS-B gradient based classical optimiser and a wave function simulator, we find there is no statistically significant difference in the results obtained for VQE circuits with a single layer \textit{without} an initial rotation layer (8 parameters) and a single layer \textit{with} an initial rotation layer (13 parameters); however, a  2 layer circuit \textit{with}  an initial rotation layer (21 parameters) performed slightly less well. All obtain approximation ratios above 0.93 and a probability of measuring the lowest cost function state of 0.8-0.9.  When using the SPSA optimiser, the choice of a 21  parameter ansatz, produced significantly lower values for the approximation ratio and the probability of obtaining the lowest cost function state.
   
  \item \textbf{Using a wave function simulator L-BFGS-B optimiser produces better quality results than SPSA }\\
    One figure of merit for the VQE process is the mean probability of the lowest cost function state  being sampled by a parameter/optimised ansatz from 100 randomly chosen sets of start parameters. Figure \ref{8param_ratios} shows, with label `L-BFGS-B SV', that for the state vector (`SV') representation of the ansatz using the L-BFGS-B optimiser has an average probability of the lowest cost function state of 0.88. For the lowest expected value of the cost function (`EV') of those 100 initial parameter sets, the probability of the lowest cost function state was 0.9999. The error bars of Figure \ref{8param_ratios} (set at $\pm$ 2 standard deviations of the mean) allow comparison of statistical significance between the mean ratios of each method. 
    `SPSA SV' has an average  probability of the lowest cost function state of 0.42 (over 100 sets of parameter angles). The error bars show that this is more than 2 standard deviations lower when compared to `L-BFGS-B SV'. The same statistically significantly lower ratios are apparent for the approximation ratio and probability of a feasible state. A similarly lower quality solution from `SPSA SV' is apparent in Figure \ref{13param_ratios} which uses a rotation layer and an entangling layer in the ansatz.
  
    \item \textbf{When using the sampling methodology L-BFGS-B achieves a similar quality performance to SPSA }\\
     The results using a sampling approach, which are more directly comparable with those obtained from a real QPU (though no allowance has been made for noise), are less favourable than the full wavefunction calculation; this reflects the error introduced from the sampling approach which necessarily only enables an estimate of the full wavefunction.  When sampling, there is no statistical difference between optimisers `L-BFGS-B Sampling' and `SPSA Sampling'. This is true for VQE circuits with 8 parameters (see Figure \ref{8param_ratios}) and with 13 parameters (see Figure \ref{13param_ratios}). Hence we find the lower figures of merit for `L-BFGS-B Sampling' compared to `L-BFGS-B SV'.  

    \item \textbf{More function evaluations needed for L-BFGS-B than SPSA }\\
     Whilst the quality of solution delivered by sampling with L-BFGS-B and SPSA are similar, the number of function evaluations by `L-BFGS-B Sampling' compared to `SPSA Sampling' ( with 100 iterations) were 75\%  higher. This relative efficiency of SPSA has been observed with other combinatorial problems \cite{Gacon_2021}. On this basis we choose to use SPSA as the classical optimizer to use with the QPU (see section \ref{QG hardware}). 

    \item \textbf{Limited conclusions can be drawn in comparing the all qubit measurement approach versus the causal cone approach} \\Using the L-BFGS-B classical optimiser we see that the 'all qubit measuring' approach produces superior results to the causal cone sampling approach for a single layer VQE circuit \textit{with} an initial rotation layer (13 parameters) but the difference is not statistically significant for a single layer circuit  \textit{without} an initial rotation layer (8 parameters).  In fact the only significant difference is from the less good performance of the causal cone sampling for the 13 parameter circuit.  
     Given the single problem instance considered, it is not possible to draw any general conclusion.
     
      \item \textbf{Runtime considerations support using the least number of parameters} \\In terms of algorithm runtimes, these increase greatly as the number of parameters being optimised over increases, so this supports the view that limiting circuit design to include the smallest number of required parameters is desirable. 
      
       \item \textbf{In a simulator environment, as few as 100 shots can be sufficient to achieve convergence} Finally we found that as few as 100 shots were sufficient to generate good quality results using the all qubit sampling approach and the SPSA optimiser.  This was of course in a noise free environment and it is very likely that a higher number of shots would be needed to sample a real noisy quantum processor unit.\\


\end{itemize}

\subsubsection{VQE simulator results for Problem B and E}


\begin{table}[ht]
\begin{center}
\caption{\textbf{VQE results for Problem B (4 locations, 2 ambulances) and E (5 locations, 2 ambulances)}}
\vspace{0.3cm}
\begin{tabular}{| >{\centering\arraybackslash}m{1.5cm}| >{\centering\arraybackslash}m{2.5cm} | >{\centering\arraybackslash}m{1.5cm}|  >{\centering\arraybackslash}m{2cm} | >{\centering\arraybackslash}m{2cm} | >{\centering\arraybackslash}m{2cm} |
>{\centering\arraybackslash}m{2.5cm} |}
 \hline
\textbf{Problem} &  \textbf{Number of parameters (layers) } & \textbf{Num. of sets of parameters} &  \textbf{Optimiser} & \textbf{Approx. ratio} & \textbf{ Prob. of feasible state }& \textbf{Prob. of lowest cost function state }  \\
 \hline
 \multicolumn{7}{|l|}{Wavefunction methodology}\\
  \hline
  B & 30 (0+1) & 50 &  L-BFGS-B  &  $0.712 \pm 0.053$ & $1.0 \pm 0.0$  &  $0.12 \pm 0.092$ \\
  \hline
  B & 46 (1+1) & 50 & L-BFGS-B    & $0.806 \pm 0.050$ &  $1.0 \pm 0.0$ &  $0.2 \pm 0.113$  \\
  \hline
  B & 30 (0+1) & 100 &  NM   & $0.695  \pm 0.033$ & $0.724 \pm 0.037$ &  $0.019 \pm 0.014$ \\
  \hline
  B & 46 (1+1) & 100 & NM    & $0.698 \pm 0.031$ &  $0.409 \pm 0.043$ &  $0.019 \pm 0.018$ \\
  \hline
  E & 38 (0+1) & 50 & L-BFGS-B    & $0.697 \pm 0.062$  &  $1.0 \pm 0.0$  &  $0.04 \pm 0.055$  \\
  \hline
  \multicolumn{7}{|l|}{Sampling methodology}\\
  \hline
  B & 46 (1+1) & 50 & SPSA    & $0.743 \pm 0.038$ &  $0.984 \pm 0.004$ &  $0.042 \pm 0.049$ \\
  \hline

\end{tabular}
\caption*{\small{The table shows results for Problem B (4 locations, 2 ambulances, 16 qubits) and Problem E (5 locations, 2 ambulances, 20 qubits) using the VQE methodology run on wavefunction simulator except for the last result which uses a sampling methodology (20,000 shots) and an SPSA optimiser (500 iterations).  For each run a set of parameter angles is randomly selected and then optimised to give the minimum value for the Expected Value of the problem cost function. There is a lower probability of finding the lowest cost function state than for problem A reflecting the higher complexity of this problem instance. }}
\label{VQE_2}
\end{center}

\end{table}

In Table \ref{VQE_2}, we provide summary results applying VQE to the larger problems B (4 locations, 2 ambulances, 16 qubits) and E (5 locations, 2 ambulances, 20 qubits).  The classical optimisers L-BFGS-B and Nelder-Mead were used for the wavefunction simulations. The probability of finding the lowest cost function state is much lower than for the smaller problem A, but still gives a good overall result.  A close examination of the results reveals that the L-BFGS-B optimiser gives rise to integer values for the cost function for each set of randomised initial angles used. We attribute this to the fact that each cost function value has several states which give that value (i.e. the states are `degenerate') and that the gradient based optimiser is finding local minima composed of 2 or more of the relevant degenerate states.  As the lowest cost function state is also multifold degenerate, we find that sets of parameter angles giving the lowest cost function value are associated with a probability of unity of obtaining the global minimum while all the other sets of parameter angles  have a zero probability of obtaining the global minimum (the `solution'). The Nelder Mead  optimiser does not use a gradient method and instead obtains results for each angle which are a mixed state of many (or all) possible states; accordingly it has a much lower mean probability of finding a feasible state and the lowest cost function state but does not have the same divide between results that have a zero probability of obtaining the lowest cost function state and results with a probability of 1.0.

The final line of Table \ref{VQE_2} shows the result of an SPSA sampling circuit for problem B (4 locations, 2 ambulances, 16 qubits). The average probability of finding the lowest cost function state was 0.042 across all the 50 parameters sets, whilst the lowest EV of the 50 runs had a lowest cost function state probability of 0.93. In fact only 4 of the 50 parameter sets had a probability of finding the lowest cost function state  greater than zero. The `best' solution contains two degenerate states with probabilities of 0.62 and 0.31 respectively. The circuit was sampled 20,000 times but after optimization only 51 distinct states were sampled. This illustrates how the VQE heuristic substantially narrows down the range of solutions proposed and is able to identify multifold degenerate states. It is also noteworthy that this high probability of finding the lowest cost function state for the best angle set is high compared to any of the QAOA circuits for Problem B (4 locations, 2 ambulances, 16 qubits) (see, for example, Figure \ref{Increasing p strategy 3XY method}).


Table \ref{Q16_sim_results} shows the results for 1) the mean of all the optimised runs and for 2) the run  giving the \textit{lowest} expected value of the cost function for 3 VQE ansatzes.  It can be seen that in each of the 3 cases, the `best' run with the highest Approximation ratio gives a very much higher probability of obtaining the lowest cost function state than the mean of all the runs despite the difference in the EV values being relatively small.  For example in case c) using a sampling approach and 30 parameters (a single entangling layer), the mean ratio of the approximation ratio is 0.735 while the best set of optimised parameters gives an EV ratio of 0.999; however the best set of parameters gives rise to a probability of finding the lowest cost function state of 0.999 while the mean probability across all sets of optimised parameters is 0.045, a considerable difference.  This demonstrates the sensitivity of the VQE method for this problem instance to obtaining the lowest possible EV and thereby accessing the lowest cost function or `solution' state with high probability. We also observe that each of the three examples which use varying numbers of parameters and both sampling and wavefunction simulation approaches all give similar results.

\begin{table}[h]
\begin{center}
\caption{\textbf{Selected figures of merit for Problem B (4 locations, 2 ambulances) using VQE and SPSA optimiser}}
\caption*{a) Wave function simulator - 46 parameters (1+1 layers)}
\vspace{0.3cm}
    \begin{tabular}{| >{\centering\arraybackslash}m{2cm}| >{\centering\arraybackslash}m{2cm} | >{\centering\arraybackslash}m{2.5cm} | >{\centering\arraybackslash}m{2.5cm} |  >{\centering\arraybackslash}m{2.5cm}| }
\hline
  \textbf{Type} & \textbf{EV/(Lowest cost function value)} &\textbf{Approx. ratio} & \textbf{ Prob. of feasible state }& \textbf{Prob. of lowest cost function state }  \\ 
 \hline
 Mean of 100 &  $0.988 \pm  0.001$ &        $0.743 \pm  0.029$ &       $0.998 \pm  0.001$ &         $0.054 \pm  0.039$\\ 
 \hline
  Best of 100 & 1.0            &  1.0                 & 1.0& 1.0 \\ 
\hline
\end{tabular}\\
\label{Q16_sim_results}
\end{center}


\begin{center}
\caption*{b) Sampling simulator - 46 parameters (1+1 layers)}
\vspace{0.3cm}
    \begin{tabular}{| >{\centering\arraybackslash}m{2cm}|  >{\centering\arraybackslash}m{2cm} | >{\centering\arraybackslash}m{2.5cm} | >{\centering\arraybackslash}m{2.5cm} |  >{\centering\arraybackslash}m{2.5cm}|}
\hline
  \textbf{Type} & \textbf{EV/(Lowest cost function value)} & \textbf{Approx. ratio} & \textbf{ Prob. of feasible state }& \textbf{Prob. of lowest cost function state }  \\ 
 \hline
 Mean of 50 &    $ 0.985 \pm  0.002 $&    $0.743 \pm  0.038$ &       $0.984 \pm  0.004$ &         $0.042 \pm  0.049$\\ 
 \hline
  Best of 50 &  0.998            & 0.990                 & 0.994& 0.932 \\ 
\hline
\end{tabular}\\
\end{center}


\begin{center}
\caption*{c) Sampling simulator - 30 parameters (0+1 layers)}
\vspace{0.3cm}
    \begin{tabular}{| >{\centering\arraybackslash}m{2cm}|  >{\centering\arraybackslash}m{2cm} | >{\centering\arraybackslash}m{2.5cm} | >{\centering\arraybackslash}m{2.5cm} |  >{\centering\arraybackslash}m{2.5cm}|}
\hline
  \textbf{Type} & \textbf{EV/(Lowest cost function value)} & \textbf{Approx. ratio} & \textbf{ Prob. of feasible state }& \textbf{Prob. of lowest cost function state }  \\ 
 \hline
 Mean of 100 & $0.987 \pm  0.002$ &        $0.735 \pm  0.035$ &       $0.997 \pm  0.001$ &         $0.045 \pm  0.037$\\ 
 \hline
  Best of 100 & 0.999            &  0.999                 & 1.0& 0.994 \\ 
\hline
\end{tabular}\\
\caption*{\small{
The table shows performance metrics for 3 alternative VQE set-ups for Problem B and 16 qubits. In case a) a wavefunction simulator is used while for b) and c) a sampling approach of 20,000 shots is used.  46 parameters are specified in the ansatz circuit for cases a) and b) and 30 parameters for case c). In all cases the optimizer is  SPSA with 500 iterations ($a=0.01$, $c=0.01$). The Qiskit simulator was used to evaluate the expected value ("EV") for a given set of circuit parameters. For each set of random starting angles (which supply the parameters for the VQE ansatz), a final set of parameters was determined by minimizing EV. The Approximation ratio is the mean from 100 sets of optimised parameters. Similar results are found in each case. It is notable that the `best' result in each case corresponding to the set of optimised parameters giving the lowest EV gives a very much higher probability of obtaining the lowest cost function state ($\approx 1.0$) than for the mean EV.}}
\end{center}

\end{table}

\subsubsection{VQE - challenges and some more recent developments} \label{VQE challenges}

As mentioned in section \ref{QAOA - challenges} VQE in common with QAOA has been identified as suffering from the problem of barren plateaus. These are gradient free zones which make it difficult or even impossible for the algorithm to find the global minimum of the problem cost function \cite{McClean2018a}. The challenge is that the parameter search landscape becomes effectively flat, particularly as the number of qubits required to define the problem and the circuit depth increases, so that gradient based methods of optimisation will fail.  Further it is argued that non-gradient optimisation methods are also unable to overcome `barren plateaus' for large circuits \cite{Arrasmith2021a}.  One way to think of the general problem is that as the problem size increases, the search space increases exponentially with the  result that the barren plateaus dominate the space between locally confined minima making it difficult or impossible to have a systematic method to identify a high quality set of parameters and hence a good solution \cite{Cerezo_2021b}; this is particularly so when the initial parameters used are generated randomly.  Even the presence of noise in the QPU device has been shown to give rise to the potential for barren plateaus \cite{Wang_2021a}. Unsurprisingly the research community has been working hard to find methods to overcome this problem and a number of potential solutions have been suggested.  In most cases the mitigating methods involve departing from the standard, unrestricted VQE approach, which implies large numbers of parameter layers as problem size increases, and instead imposing some limitation that means that barren plateaus are unlikely to form in the search landscape. We describe a few examples below.  Potentially the most general approach involves ensuring that the relevant problem cost function is expressed as a series of local terms rather than a global cost function.  In effect this means that the calculation of individual terms within the cost function is always limited to a small number of qubits rather than terms being the result of circuit measurements which are affected by all qubits simultaneously.  Cerezo et al. \cite{Cerezo2021a} demonstrated that using a local cost function and with a VQE ansatz that was less than $O(\log (n))$ where $n$ is the number of qubits, they could obtain the global minimum for a system of up to 100 qubits; however using a global cost function, the algorithm encountered barren plateaus and so failed to converge.  Two other suggested approaches have involved limiting the search space by either finding a good set of initial parameters \cite{Grant2019a,Kulshrestha2022a, Rad2022a} that starts the optimisation in an area of parameter space which is not itself a  barren plateau region or limiting the number of parameters that are independently varied in the ansatz at one time, for example by iteratively introducing new layers to build up the full ansatz circuit \cite{Skolik2021a, Liu2022a}.

The approaches to mitigating barren plateaus are general in application and not limited to optimisation problems.  There have also been attempts to specifically improve VQE performance in relation to optimisation.  Two of these, Conditional Value at Risk-VQE (CVaR-VQE) and Filtering-VQE (F-VQE) amend the optimisation cost function.  The third that we mention is Layer-VQE (L-VQE) which is effectively an application of the layered approach to parameter optimisation referred to in the barren plateau mitigation methods above. 




Conditional Value at Risk-VQE \cite{Barkoutsos20a} was proposed by Barkoutsos et al and is inspired by techniques used in the financial markets to monitor and control risk.  Using the CVaR-VQE method, the measurement samples give rise to a series of associated cost function values, which are then ranked so that the cost function values are ordered by size, i.e. if the set of $N$ cost function values are \{$O_1, O_2,..., O_N$\}, then these are ordered so that $O_i \leq O_{i+1}$, where $O_i$ is the $i^{\text{th}}$ cost function value, when ordered. The key aspect of the CVaR-VQE is that rather than using the whole sample set to generate the expectation value of the cost function, just the lowest $\alpha$ proportion ($0 < \alpha < 1$) is used, where $\alpha$ is a chosen constant. By construction this has the effect of lowering the CVaR cost function value (the mean of the CVaR set of sample values) which is submitted to the classical optimiser. As a result the revised algorithm focuses on increasing the probability of sampling states with a lower cost function value and this is what was found to be the case in practice, with the probability of the lowest cost function state increasing significantly relative to basic VQE. On a variety of test problems including Max cut and Number partitioning performed on a simulator, CVaR-VQE was able to improve the probability of finding the lowest cost function state considerably.  In a test on a quantum processor, it also appeared to demonstrate good robustness to noise as well as the same improved convergence behaviour.

Filtering-VQE (F-VQE) suggested by Amaro et al. \cite{Amaro22a} is a further adaption of the basic VQE approach, which again seeks to amend the cost-function expectation value by reducing the contribution of higher cost function values to the output cost function value used by each iteration of the classical optimiser. In contrast to CVaR-VQE which simply eliminates higher value sample values, F-VQE applies a filtering function which has the effect of increasing the weight of lower value cost function samples and reducing the weight of higher value samples. The filtering function is defined such that the function $f^2(O,\tau)$ is strictly decreasing over the range of the values that the cost function can take $O \in [O_{min}, O_{max}]$. $\tau $ is a parameter which is dynamically updated as the optimisation algorithm progresses.  Amaro et al tried a number of possible filtering functions and found that the most effective of the 5 they considered is the inverse Hamiltonian, $H^{-1}$.  Of those compatible with a causal cone approach, they found the exponential function, $e^{-\tau H}$ to be best.  The improvements seen were in  a greatly reduced number of iterations to reach convergence as well  as larger
and more consistent approximation ratios on the Max cut problems that were solved.

Layer-VQE \cite{Liu21a} introduced by Lui at al is a way to improve the effectiveness of the process for optimising the parameters within VQE.  It does this by sequentially increasing the number of parameters included in the VQE circuit and optimising these as it goes.  This reduces the curse of dimensionality which makes optmising over large numbers of parameters simultaneously increasingly difficult.  L-VQE starts with a layer of single-qubit rotation gates and a randomly drawn set of angle parameters.  It seeks to minimise the relevant cost function but deliberately stops the process prior to finding a minimum.  It then adds an entangling layer composed of CNOT gates and rotation gates and initialises the optimisation process with the angles found from the initial stage; the parameter angles for the entanglement layer are all initially set to zero so as to have no effect on the starting cost function result obtained.  Again the optimisation is halted prior to finding a minimum and the process repeated with an additional (second) entanglement layer of gates.  This process of partial optimisation and then adding additional entangling layers continues until the desired number of entangled layers is present and then the optimisation is allowed to run until the termination condition is reached; this gives the final result.  Liu et al found that on their test problem (the $k$-communities modularity maximization problem \cite{Newman2006a}),  the optimisation overhead was reduced while finding the lowest cost function state more frequently; they also suggested that the method was more robust to quantum processor noise.

These improvements in VQE algorithm performance are encouraging for the potential to solve valuable problems in the NISQ era as larger, more capable quantum computers are developed. As for the QAOA algorithm, we anticipate further improvements and derivatives of the basic method to be proposed as larger numbers of researchers become involved.

\subsection{Running the ambulance problem on today's hardware} \label{QG hardware}

So far all the experimental results presented have been undertaken using simulators on classical hardware.  In this subsection we report results run on 2 quantum processing units (`QPU's) from IBM, a leading developer of superconducting based quantum computers. 

\subsubsection{What hardware, how implemented}

Due to the still small scale of most QPUs and the relatively high noise levels, we choose to focus on Problem A (5 locations, 1 ambulance)  and Problem B (5 locations, 2 ambulances) using the VQE algorithm.  We tried solving the 2 problems using the IBM Perth 7 qubit QPU (quantum volume =  32)\footnote{Quantum Volume is a measure first proposed by IBM to characterise the performance of a QPU, the higher the value, the more powerful the QPU. Further details can be found in the paper by Cross et al. \cite{Cross19a}} and the IBM Guadalupe 16 qubit QPU (quantum volume =  32).
In each case we used 13 parameters (1 rotation layer plus 1 entangling layer).  
Note that due to the relative scarcity of access to time on QPUs, the number of separate runs reported on was more limited than was possible with the runs from simulators, which comfortably run on desktop computers.

\subsubsection{Quality of QPU solutions}

\begin{figure}[ht]
\caption{Problem A - Estimated EV by iteration of CPU simulator compared to QPU IBM Perth }
    \includegraphics[width=.9\linewidth]{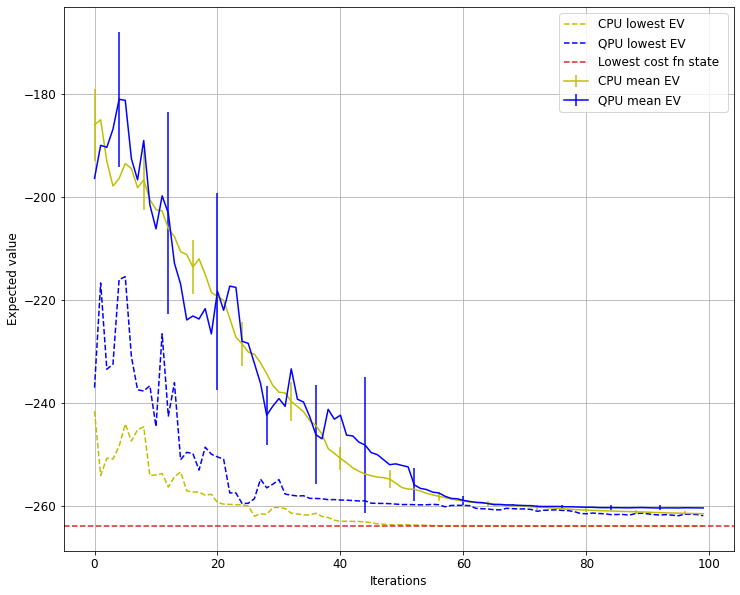}
\caption*{\small{Samples from a quantum circuit (1000 shots on CPU, 20,000 on IBM Perth QPU) were used to evaluate the expected value ("EV") for a given set of circuit parameters.  For each random starting angle set (which comprises the 13 parameters for the VQE Ansatz) a final set of parameter angles is selected by an SPSA optimization after 100 iterations. The ratio EV/(lowest cost function energy) is the average of 100 (10 on IBM Perth)  sets of starting angles. The SPSA meta parameters 'a' and 'c' were set to 0.1.  The mean CPU EV and mean QPU EV follow similar trajectories allowing for the fewer number of QPU runs giving rise to a more volatile descent path.}}
\caption*{\small{
             \begin{tabular}{c|c|c|c|c|c|c|}
                 Circuit & Shots & Parameters   & Optimizer &Iterations& "a" & "c"\\
                 \hline
                CPU     & 1,000  &   13  &      SPSA & 100 & 0.1  & 0.1\\
                IBM Perth     & 20,000  &   13  &      SPSA & 100 & 0.1  & 0.1\\
         
            \end{tabular}
        }
    }
    \label{qpu_cpu_iterations}
\end{figure}

The results from running Problem A on the IBM Perth QPU are summarised in Table \ref{VQE_3} and Figure \ref{13param_ratios}. 
Despite the presence of noise and gate infidelity, the algorithm generated overall good results.  We see that the 10 runs using the SPSA classical optimiser (20,000 shots) gave, as expected, a lower approximation ratio and other metrics than those obtained for the equivalent sampling approach using a simulator (see Table \ref{VQE_all}). Very similar results were obtained using the L-BFGS-B optimiser.  However when we look at the best 2 results of the 10 (those with the highest approximation ratio), we find that they had associated probabilities of finding one of the `solution' states of 2.4\% and 45.6\% respectively. When we note that this is after only 100 iterations while the simulator completed 500 iterations, the performance actually looks quite good.  We examined the sampling simulator and QPU performances in more detail by comparing the average cost function value or EV after each iteration of the algorithm - the results are shown in Figure \ref{qpu_cpu_iterations}. Note that the comparison is between 10 sets of parameter runs on the QPU and 100 runs on the simulator (`CPU') which affects the size of the error bars and also favours having a lower EV trajectory `best' CPU run than for the `best' QPU run.  Nonetheless we observe that :
\begin{itemize}
    \item The mean CPU EV and mean QPU EV follow similar trajectories  allowing for the fewer number of QPU runs giving rise to a more volatile descent path.
    \item For the bulk of iterations from 0 to 60 the `QPU mean EV' is within the error bars of `CPU mean EV'.
    \item Broadly, the higher number of shots used (20,000 vs 1,000 for the CPU simulator) on the IBM Perth QPU was almost enough to compensate for the lower than 100\% gate and measurement fidelity of the IBM Perth QPU.
    
\end{itemize} 

On a QPU with a higher quantum volume than the 32 of IBM Perth we would expect the required difference in shots to reduce. It is also encouraging that when run with only 1,000 shots that the SPSA optimiser also was successful in converging towards a good solution albeit at a slower rate reflecting the greater noise inherent in a smaller shot sampling of the EV for each iteration.

\vspace{0.5cm}





\begin{table}[h]
\begin{center}
\caption{\textbf{VQE results for Problem A (5 locations, 1 ambulance) using a QPU}}
\vspace{0.3cm}
\begin{tabular}{|  >{\centering\arraybackslash}m{2cm} | >{\centering\arraybackslash}m{1.5cm}| >{\centering\arraybackslash}m{2cm} | >{\centering\arraybackslash}m{2cm} |  >{\centering\arraybackslash}m{2.25cm} |
>{\centering\arraybackslash}m{2cm} |
>{\centering\arraybackslash}m{2.25cm} |}
 \hline
  \textbf{Number of parameters (layers) } & \textbf{Num of shots} &  \textbf{Optimiser} & \textbf{QPU}  & \textbf{Approx. ratio} & \textbf{ Prob. of feasible state }& \textbf{Prob. of lowest cost function state }  \\
 \hline
13 (1+1) & 20,000 &  SPSA  & IBM Perth &  0.674 $\pm$ 0.174 & 0.31 $\pm$ 0.21  &  0.153 $\pm$ 0.194 \\ 
  \hline
13 (1+1) & 1,000 &  SPSA  & IBM Perth &  0.527 $\pm$ 0.20 & 0.49 $\pm$ 0.12  &  0.10 $\pm$ 0.11 \\
  \hline
13 (1+1) & 20,000 & L-BFGS-B  & IBM Perth  & 0.690 $\pm$ 0.10 &  0.348 $\pm$ 0.15 &  0.084 $\pm$ 0.108 
       \\
  \hline
\end{tabular}
\caption*{\small{The table shows key performance metrics for the VQE algorithm run on a quantum processor (``QPU'').  In each case there were 10 separate runs initiated with random parameter angles and optimised over 100 iterations. Meta parameters, $a$ and $c$, were set to 0.1 for the SPSA optimiser with 20,000 shots and 0.01 for 1,000 shots. (See Appendix \ref{SPSA_met} for details on meta parameters) 
For the L-BFGS-B optimiser, $\epsilon = 0.1$. ($\epsilon$ is the step size used in the gradient calculation.)
The results in general show that VQE works well on real quantum hardware provided additional shots are used to compensate for the additional noise and less good gate fidelity relative to a classical simulator.}} 
  
\label{VQE_3}
\end{center}

\end{table}

\begin{table}[h]
\begin{center}
\caption{\textbf{VQE results for Problem B (4 locations, 2 ambulances) using a QPU}}
\vspace{0.3cm}
\begin{tabular}{|  >{\centering\arraybackslash}m{2cm} | >{\centering\arraybackslash}m{1.5cm}| >{\centering\arraybackslash}m{2cm} | >{\centering\arraybackslash}m{2cm} |  >{\centering\arraybackslash}m{2.25cm} |
>{\centering\arraybackslash}m{2cm} |
>{\centering\arraybackslash}m{2.25cm} |}
 \hline
  \textbf{Number of parameters (layers) } & \textbf{Num of shots} &  \textbf{Optimiser} & \textbf{QPU}  & \textbf{Approx. ratio} & \textbf{ Prob. of feasible state }& \textbf{Prob. of lowest cost function state }  \\
 \hline

  \hline
30 (0+1) & 20,000 & SPSA  & IBM Guadalupe  & 0.781 $\pm$ 0.12 &  0.061 $\pm$ 0.04 &  0.011 $\pm$ 0.02  \\
  \hline
\end{tabular}
\caption*{\small{The table shows key performance metrics for the VQE algorithm run on a quantum processor (``QPU'') with 16 qubits.  In each case there were 10 separate runs initiated with random parameter angles and optimised over 200 iterations. Meta parameters, $a$ and $c$, were set to 0.01 for the SPSA optimiser }}

\label{VQE_B}
\end{center}

\end{table}



\vspace{0.5cm}

We also tried running problem B (4 locations, 2 ambulances) using the 16 variable encoding approach on the 16 qubit IBM Guadalupe QPU.  First, we tested the success of solving problem B  with a simulated sampling circuit, an SPSA optimizer (500 iterations, $a=c=0.01$). Table \ref{Q16_sim_results} shows that using 46 parameters rather than 30 on a simulator did not improve the figures of merit, nor the average ratio EV/(lowest cost function energy). The 46 parameter circuit has greater depth than with 30 parameters, and so very probably, lower fidelity on a QPU. On this basis we used the 30  parameter circuit on IBM Guadalupe. 

The topology of IBM Guadalupe is not ideal for the Ansatz circuit we used. This Ansatz used gates between neighbouring qubits. Hence, the ideal  QPU  topology would be a sequential row of 16 qubits. On such a QPU any given qubit, say $q_4$, could connect to its neighbours, $q_3$ and $q_5$ without the need for additional swap gates and the consequent reduction in circuit fidelity. This need for additional swap gates, compared to problem A (5 locations, 1 ambulance), and the greater number of gates in the Ansatz for problem B,  both contribute to a lower circuit fidelity.  The extra swaps needed increased the count of CNOT gates from 15 to 51, and the depth of the circuit from 11 to 47 even allowing for the benefits of qubit reallocation \footnote{Qubit reallocation or mapping involves assigning variables to qubits in a way designed to minimise the need for swap gates and also to allow for different levels of gate fidelity across the quantum processor.  The resultant effect is to maximise the overall circuit fidelity}. This is high compared to Guadalupe's quantum volume of 32 which is derived from fidelity threshold for a circuit of depth 5 and width 5. If a higher quantum volume QPU had been used, or one with a topology that did not need additional CNOT gates, then the convergence of EV with iterations of the QPU would be closer to the sampled simulator.

\begin{figure}[ht]
\caption{Problem B - Estimated EV by iteration of CPU simulator compared to  QPU IBM Guadalupe}
    \includegraphics[width=.9\linewidth]{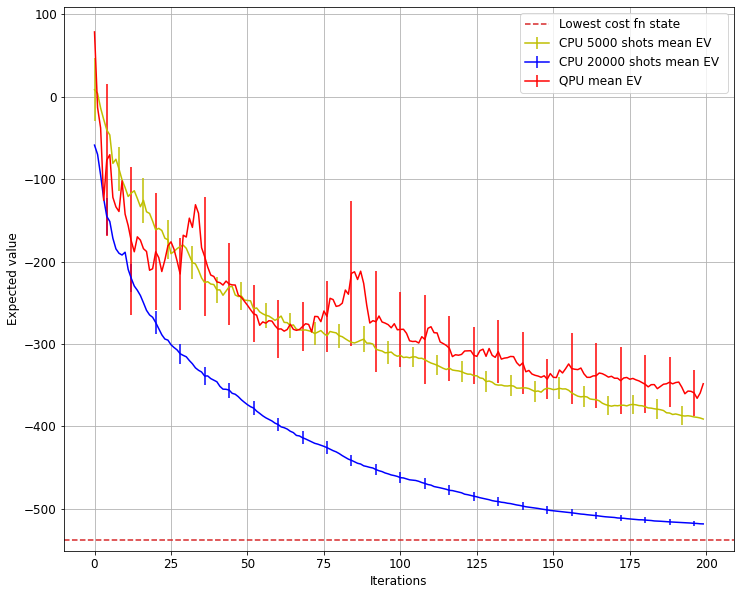}
\caption*{\small{Samples from a quantum circuit (20,000 and 5,000 shots) were used to evaluate the expected value ("EV") for a given set of circuit parameters.  For each random set of starting angles (which comprises the 30 parameters for the VQE Ansatz) a final set of parameter angles is selected by an SPSA optimization after 100 iterations. The ratio EV/(lowest cost function energy) is the average of 100 (10 on IBM Guadalupe)  sets of starting angles. The SPSA meta parameters 'a' and 'c' were set to 0.01.}}
\caption*{\small{
             \begin{tabular}{c|c|c|c|c|c|c|}
                 Circuit & Shots & Parameters   & Optimizer &Iterations& "a" & "c"\\
                 \hline
                CPU 20000\_shots     & 20,000  &   30  &      SPSA & 200 & 0.01  & 0.01\\
                CPU 5000 shots     & 5,000  &   30  &      SPSA & 200 & 0.01  & 0.01\\
                IBM Guadalupe     & 20,000  &   30  &      SPSA & 200 & 0.01  & 0.01\\
         
            \end{tabular}
        }
    }
\label{guadalupe}
\end{figure}

The lower fidelity is apparent in the results presented in Table \ref{VQE_B}. For example, the probability of a feasible state (defined as a state complying with the constraints of the problem) averaged over the 10 starting angles was only 0.061 (0.31 for problem A on IBM Perth). In Figure \ref{guadalupe} the sampling simulator ('CPU 20000 shots mean EV') shows the potential convergence of EV from a fault tolerant QPU. By contrast, after 20 iterations the QPU's error bars (with 20,000 shots) are fully clear of the `CPU 20000 shots mean EV' result. As a broad illustration of the cost of fidelity on the one hand and number of shots on the other Figure \ref{guadalupe} compares the CPU performance using only 5000 shots and the QPU using 20,000 shots but with its less than 100\% fidelity. These two converged over 200 iterations nearly always within the error bars of each other. Table \ref{Q16_sim_results}c) shows the results of a sampling simulator  (with 20,000 shots) after 500 iterations for 100 randomised starting sets of parameter angles. The parameters that achieved the lowest EV of those 100 sets of starting parameters, was an Ansatz that included one of the several lowest cost function value states (ground states) 99.4\% of the time. This shows the high potential success rate that could be achieved on a QPU with much lower levels of noise (better gate fidelity, etc) than are currently available. 
\subsubsection{QPU resource consumption - time and cost}

Two important issues for the future of quantum computing are the time taken to run quantum algorithms and the associated financial cost of solving problems using quantum computing hardware.  For the VQE algorithms running on IBM QPUs and making use of the IBM Runtime cloud based classical loop, the times taken to optimise a single trial set of angle parameters are set out in Table \ref{system_time}.  IBM charges on a per second basis and the implied cost of running each algorithm is also shown. It can be seen that the time taken is considerable even for these small problems and that the associated cost is also material.  

Execution of a quantum circuit is a small fraction of the system time used. Resetting a circuit and reading out the results can be a larger fraction of total system time. Running the same circuit for 20,000 rather than 1,000 shots just over doubled the system time consumed by our Qiskit Runtime program (see Table \ref{system_time}).

The total system time to run our modest 16 qubit problem is substantial. We used a  simulator with 20,000 shots and 500 iterations of the SPSA optimizer, for 100 randomly selected 46 parameter angles. The lowest EV in this search found the  ground state in 93\% of the 20,000 states sampled. Overall, this would consume 3 hours of system time per set of parameter angles and more than 10 days system time in total. Due to system resource limitations we ran 100 iterations for 10 different angles.


\begin{table}
\begin{center}
\caption{QPU System time and Cost}
\begin{tabular}{| >{\centering\arraybackslash}m{1.5cm}| >{\centering\arraybackslash}m{1.5cm} |
>{\centering\arraybackslash}m{1.5cm} |
>{\centering\arraybackslash}m{1.75 cm}|  >{\centering\arraybackslash}m{2cm} | 
>{\centering\arraybackslash}m{2cm} |
>{\centering\arraybackslash}m{1.5cm} |
>{\centering\arraybackslash}m{1.75cm} |}
\hline
\textbf{QPU} & \textbf{Number of qubits} & \textbf{Shots} & \textbf{Iterations} &  \textbf{Time per shot/msec}   & \textbf{Time per start angle/sec} & \textbf{Number of start angles} & \textbf{Cost U\$$^3$}\\
\hline
   IBM Perth &5 & 1,000$^1$ & 100   & 2.91  & 582     &10 & 9,312\\
   \hline
   IBM Perth &5 & 20,000$^1$ & 100  & 0.31 & 1,240    &10 & 19,840\\
\hline
    IBM Guadalupe & 16 & 20,000$^2$ & 100  &  0.55 & 2,200   & 10& 35,200\\

    \hline
    IBM Guadalupe &16 &20,000    & 500    &  0.55 & 11,000 & 100 & 1,760,000$^4$\\
\hline
\end{tabular}\\
\vspace{0.2cm}
 \caption*{ 
    \small{The table shows the estimated time taken and implied financial cost to complete 10 or 100 optimisation runs for 4 different VQE configurations using 2 IBM QPUs. In each case the SPSA optimiser is used.  For the 5 qubit estimation, Problem A is selected with 13 parameters and for the 16 qubit estimation, Problem B is chosen with 30 parameters. For each shot there are 2 function evaluations required. The total time required is calculated as (number of shots) x (time per shot) x (number of function evaluations) x (number of iterations) x (number of start angles). IBM offers access to selected QPUs and the Qiskit Runtime software for U\$1.60/second as at August 2022. 
    }
 }
 \label{system_time}
\end{center}
\end{table}

It is evident that there will need to be very significant improvements in algorithm performance if they are to become economic to use in the future on all but the most valuable problems.  Methods such as Filtering VQE and their future more efficient successors are likely to be part of making VQE a practical application to commercial problems.

\subsubsection{Comparing QAOA and VQE algorithm performance}

We have seen that both the QAOA and VQE algorithms are able to successfully solve the small scale Problems A and B.  Which is more effective?  This is a difficult question to answer based on the experimental runs we have undertaken.  The inherent differences in the way each works means it is not sufficient to simply compare approximation ratios or the probability of finding the lowest cost function state.  QAOA has a theoretical backing and offers the flexibility to reduce the search space through suitable mixer choice.  VQE benefits from simplicity and its compatibility with smaller scale QPUs using the causal cone sampling approach.  The jury is still out on this debate and we merely note that both are useful algorithms and, for now, a pragmatic approach is required to their application, i.e. trying both in many cases may be the most sensible strategy.

\section{Quantum Annealing implementation} \label{QA_impl}

In this section we turn our attention to implementing the Ambulance problem on a quantum annealer. As mentioned in the introduction and also in section \ref{QA_Method}, quantum annealing is more mature in terms of the scale of computer available and also has a simpler form of operation than quantum gate computers. There is one main approach to how to solve a combinatorial optimisation problem using a quantum annealer which is to encode the target problem in the problem Hamiltonian, $H_{problem}$ as explained in section \ref{QA_Method} and then to undertake a quantum anneal as encapsulated in equation (\ref{anneal eqn}).  Due to the relative maturity of the hardware all our runs are undertaken on a real quantum annealing computer in clear contrast to our examination of the quantum gate approach where, except for section \ref{QG hardware}, all runs take place using a simulator without noise.  Hence comparisons between the 2 technologies in terms of performance need to made with care as these differences are of great importance, in particular the presence of noise. D-Wave is the only provider of generally available quantum annealing computers, currently offering 2 types of machine; these are the D-Wave 2000Q and the Advantage with the latter being the more recent and more powerful evolution of the technology.  We report results using both of these machines.

Given the fact that quantum annealers of a usable size and reliability already exist, we will therefore in this section focus on some of the practical considerations of using a quantum annealer as well as looking at how well the quantum annealing approach performs on different size problems.  As before the problem we consider is the optimal location of ambulances seeking to efficiently service a regular grid of target destinations..

\subsection{An outline of the Quantum annealing problem solving process}

There are several stages involved in solving a problem using the D-Wave quantum annealing technology \cite{DWave21b}.  In brief these are:

\begin{itemize}
    \item Encoding the problem to be solved in a form suitable to be run using a quantum annealer 
    \item Embedding the problem into the physical architecture of the quantum processor unit (``QPU'')
    \item Optimising a range of operational parameters that affect the effectiveness of the QPU in successfully finding a good solution with a high probability
    \item Postprocessing the raw output data to obtain the best solution
\end{itemize}

Depending on the complexity of the problem the encoding stage is likely to be the most critical as through a smart formulation of the problem it will enable the current capability of a quantum annealer to be used to its best effect.  In the case of the ambulance problem the encoding is relatively straightforward as we are using a purely quantum approach for demonstration purposes.  However this will not generally be the case and is likely to always be a critical step for large, commercially valuable problems, notably those using hybrid classical-quantum approaches, where one aspect is how best to divide the problem between the classical and quantum computation.

Embedding the problem on the QPU is a key step and arises because the physical architecture of the QPU takes a particular form with only limited connectivity between the physical qubits.  Solving the "embedding problem" is itself a difficult task although there are tools available to assist.

As the quantum processor is an analogue device with its own particular features and characteristics related to its design and "how it works", there are a number of important operational parameters that need to be optimised to ensure the QPU operates optimally to solve the problem concerned.  There are quite a large number of such parameters but we will focus on just a few key ones.

Once the problem has been run using the QPU and the output obtained, it is possible to improve the solution quality by undertaking so-called postprocessing which seeks to remove apparent errors that may appear in the raw output data.  In effect it is a form of refining the data quality so that the process of converting the raw data to a final answer is as good as possible.

\subsection{The problem we will consider}

\begin{figure}[h]
\caption{\textbf{Problem B - Encoding 2 ambulances and 4 locations}}
\includegraphics[width=14cm, height=3.5cm]{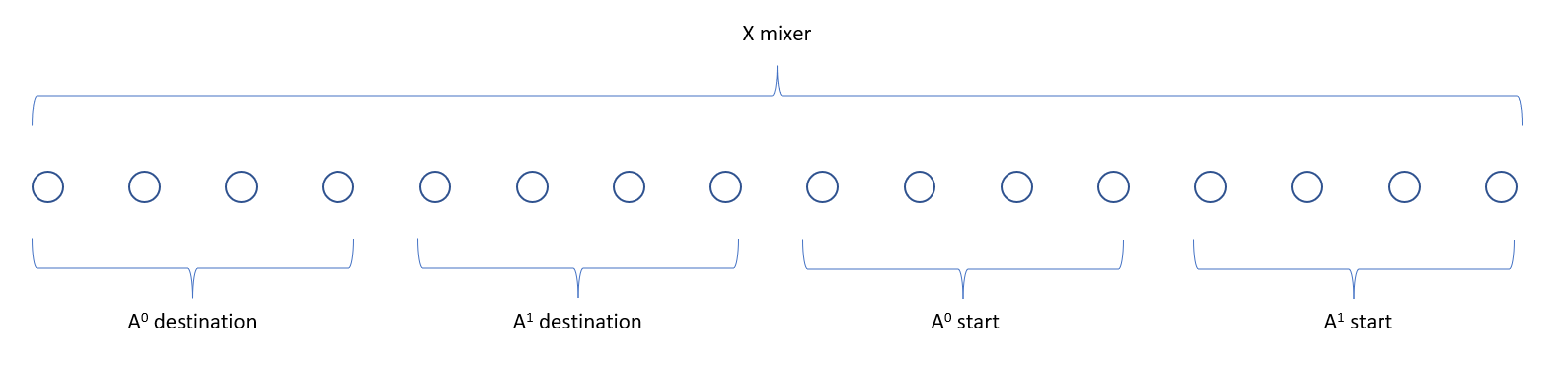}
\centering
\label{Problem B - Encoding no XY}
\caption*{\small{The figure shows the encoding of a 4 location, 2 ambulance problem using start and destination qubits for each of the 2 ambulances.  Each location is associated with 4 qubits: start and destination qubits for each of 2 ambulances.}}
\end{figure}

We will consider 6 ambulance location problem variants as set out in Table \ref{Problem specifications QA}.  Variants B and F are the same as those considered when we were considering quantum gate approaches while variants G, H, I and J are larger versions of variant B.  In all cases bar variant F we will use the encoding approach set out in Figure \ref{Problem B - Encoding no XY} which means having 4 qubits for each grid location. Note the contrast with Figure \ref{Problem B - Encoding of 2 ambulance 4 locations} where 3 XY mixers were used while here only a single X mixer is used. The larger problem sizes reflect the greater capability of the D-Wave quantum annealer although, as will see, they are still smaller than the theoretical maximum possible size the latest D-Wave annealer can accommodate.

\begin{table}[H]
\begin{center}
\caption{\textbf{Ambulance problem variant specifications}}
\vspace{0.1cm}
\begin{tabular}{| >{\centering\arraybackslash}m{1.5cm}| >{\centering\arraybackslash}m{2.5cm} | >{\centering\arraybackslash}m{2.5cm} |>{\centering\arraybackslash}m{6cm} |>{\centering\arraybackslash}m{2.5cm} |}
 \hline
 Problem variant & Number of ambulances & Number of locations & Distance metric minimised & Number of qubits required\\
 \hline
B & 2   & 4    & Sum of Squared Euclidean distance to all locations from closest ambulance  & 16\\
\hline
F & 1   & 17    & Sum of Squared Euclidean distance   &   17\\
\hline
G & 2   & 3x2 grid    & \multirow{4}{6cm}{\centering{Sum of Squared Euclidean distance to all locations from closest ambulance}}  &   24\\
 \cline{1-3} \cline{5-5}
H & 2   & 3x3 grid    &   &   36\\
 \cline{1-3} \cline{5-5}
I & 2   & 4x4 grid    &   &   64\\
 \cline{1-3} \cline{5-5}
J & 2   & 5x5 grid    &   &   100\\
\hline
\end{tabular}
\label{Problem specifications QA}
\end{center}
\end{table}

\subsubsection{Full problem with constraints}

As we are using a more complex and complete encoding approach than was used previously for problem A (5 locations, 1 ambulance),  the cost function takes a more complex form when we consider problem variant B.  The cost function, $O$, is still composed of the distance metric, $D$, and the penalty term, $P$, but now it takes the form:

\begin{equation} \label{pen1a}
   O = \sum_{i= 0}^{3} \sum_{j= 8}^{11} s_iD_{ij}s_j + \sum_{i= 4}^{7} \sum_{j= 12}^{15} s_iD_{ij}s_j - 
   \lambda\sum_{i\leq j, i, j = 0}^{15} s_iP_{ij}s_j
\end{equation}

where the distance metric has the form 
\begin{equation}
\begin{split}
D_{ij}  &=   -(J-I)^2   \qquad   \text{where } J = j (\text{mod}~ 4), ~I = i (\text{mod}~ 4)  \\
\end{split}
\end{equation}

and the penalty term is defined by

\begin{equation}
P_{ij} =
\begin{cases}
    1 & \text{ $i=j$}\\
    -2 & \text{ $i<j$, $i, j \in \{8,...,11\}$, $i, j \in \{12,...,15\}$} \\
    -2 & \text{ $i<j$, $i \in \{0,..., 3\}$, $j = i+4$}\\
    0 & \text{otherwise}
\end{cases}
\end{equation}

$\lambda$ as before is a positive constant whose value can be specified and $s_i, s_j $ are the integer variables that can take the values 0 or 1.  The complexity of the penalty term arises as it is capturing 2 constraints: a) the requirement that each ambulance should be located at a single location and b) that each grid location is served by just one ambulance. The cost function has a similar form in the cases of problem variants G - J but reverts to the simpler form (equations (\ref{pen1}) - (\ref{pen2})) we had previously in the case of variant F as only one ambulance is involved.

\subsubsection{Embedding the problem}

Once the problem of interest has been encoded into a suitable Hamiltonian, we then need to map the relevant integer variables into nodes (qubits) in the physical hardware of the quantum annealer, i.e. the QPU.  As the physical architecture has only a relatively small number of connections between the qubits it is in general unlikely that the relationships between the variables will naturally match that of the hardware\footnote{D-Wave offers quantum annealers with 2 different qubit connectivities: The D-Wave 2000Q which has a connectivity of 6 and the Advantage which has a connectivity of 15.}.  As a result a process of embedding is required to map the problem Hamiltonian to the hardware.  In mathematical terms this is equivalent to mapping a graph $H$, representing the cost Hamiltonian,  onto a sub-graph of $G$, the QPU node and edge structure represented as a graph. Owing to this mismatch it is frequently necessary to embed certain variables as chains of qubits which all take the same value during the calculation, effectively becoming a single ``logical'' qubit; this enables a variable (qubit) to connect to more variables than is allowed by the native connectivity of the hardware. 
This embedding or mapping problem is itself a challenge - it has been shown to be NP complete as a problem in its own right \cite{Cook70a}. Notwithstanding, D-Wave provides a number of embedding tools which seek to achieve a successful embedding of the problem Hamiltonian while minimising the length of any qubit chains required.  The latter is desirable as it improves the performance of the quantum annealer if chains are shorter and of similar length \cite{DWave21b}. 

In this work we made use of the tool EmbeddingComposite provided by D-Wave as part of their Ocean Tools set of software as well as using a manual embedding approach. The D-Wave software offers various options for embedding and will recalculate the embedding map for each run unless directed not to. Due to the algorithm followed, different graphical embeddings for the same problem will be calculated on each occasion, some of which will in practice be more successful in generating good quality solutions.  A key concern in finding a successful embedding is the number of chain breaks\footnote{A chain break occurs when all the physical qubits representing a single logical qubit do not all take the same value} that arise when the annealing process runs and the candidate solutions are ``calculated''.  There is also a trade-off between ensuring that the strength of coupling between chains of qubits is high enough to ensure no or few chain breaks and it being too strong resulting in inhibiting the annealer from effectively finding good quality solutions. 

The different architectures of the D-Wave 2000Q and Advantage systems means that different embeddings are required for the two systems. Notwithstanding the fact that the Advantage machine is the latest evolution of the D-Wave quantum annealer, we experimented with finding embeddings on both systems to demonstrate the degree of improvement represented by the Advantage computer.
In general we found that we were able to find good embeddings by taking the best of a series of embeddings proposed by the inbuilt embedding tool\footnote{The FixedEmbeddingComposite allows a user determined embedding, whether found using the embedding tools or otherwise, to be used.}.  In one case  we also identified a good manual embedding based on seeking to match the problem structure to the native architecture of the Advantage computer, which we found delivered superior results to the best inbuilt embedding proposed - see Table \ref{Fixed parameters for grid runs} for details. For the smallest problem, A, we used a randomly proposed embedding as the probability of finding the problem solution was high in all cases.  For the larger sized grid problems, we only used inbuilt tools, so it is likely that superior embeddings are possible and would be likely, by definition, to have delivered better results than we report.

The maximum size of problem that can be embedded is a function of the number of qubits and connections in each of the machines as well as the specific topology. Key machine data and the maximum sized clique (a fully connected graph) that can be accommodated are set out in Table \ref{Dwave specs}. The maximum theoretical fully connected problem that can be embedded is determined by the largest clique that can be mapped onto the QPU chip. Theoretically this is $K_{150}$ based on the specifications of the Advantage machine but in practice it is significantly less as not all the edges  and nodes are operational due to manufacturing issues.  D-Wave provide $K_{71}$ as a more representative figure, which is accompanied by a relatively long 10 qubit length chain.  When we sought to embed various problem sizes we found the results set out in Table \ref{Embedding result1}. The largest grid size problem we were able to embed on the 2000Q machine was a 5 x 5 grid using 100 qubits while on the larger Advantage machine an 8 x 8 grid using 256 qubits could be successfully embedded.  Although it may be possible to embed a larger problem we found that the embedding tools for larger problem instances timed out without finding a good embedding.


\begin{table}[ht]
\begin{center}
\caption{\textbf{D-Wave quantum annealer selected data }}
\vspace{0.1cm}
\begin{tabular}{|>{\raggedright\arraybackslash}m{5cm}| >{\centering\arraybackslash}m{2cm}| >{\centering\arraybackslash}m{3.5cm} |
>{\centering\arraybackslash}m{3.5cm} |} 
 \hline
\textbf{Property} & \textbf{2000Q} & \textbf{Advantage 1.1} & \textbf{Advantage 4.1}\\[2ex] 
 \hline
Nodes (theoretical max) & 2048   & 5640  & 5640\\
\hline
Edges (theoretical max) &  6016  & 40484 & 40484 \\
\hline
Nodes (max available) & 2041   & 5567  & 5627   \\
\hline
Edges (max available) & 5974   & 39452  & 40279  \\
\hline
Max clique (theoretical max) & $K_{65}$   & \multicolumn{2}{c|}{\pbox{4cm}{\hspace{1.5cm}$K_{150}$  (14~ \text{qubit chain length})}}\\ 
\hline
\pbox{10cm}{Max clique size (95\%*) \\ (10  qubit chain length)}   & $K_{28}$   & \multicolumn{2}{c|}{$K_{71}$} \\
\hline
\pbox{10cm}{Max clique size (95\%*) \\ (4  qubit chain length)} & $K_{12}$   & \multicolumn{2}{c|}{$K_{30}$} \\ 
\hline
\end{tabular}


\label{Dwave specs}
\end{center}
\text{\small{* Assumes that a maximum of 95\% of the edges are available though for Advantage 4.1 this figure is exceeded.}}
\caption*{\small{The table shows selected data for 3 D-Wave quantum annealers.  The 2000Q is the previous generation of machine while the Advantage is the current generation and was used for the experiments reported in this paper.  The actual number of nodes and edges available is less than the theoretical maximums due to manufacturing issues.  A clique is an all-to-all connected graph.} }

\end{table}

\begin{table}[H]
\begin{center}
\caption{\textbf{D-Wave embedding performance }}
\vspace{0.1cm}

\begin{tabular}{| >{\centering\arraybackslash}p{2cm}| >{\centering\arraybackslash}m{2cm}| >{\centering\arraybackslash}m{1.5cm} | >{\centering\arraybackslash}m{1.5cm} |>{\centering\arraybackslash}m{1.5cm} | >{\centering\arraybackslash}m{1.5cm} | >{\centering\arraybackslash}m{1.5cm} |  >{\centering\arraybackslash}m{1.5cm} |}
\hline
 \multirow[b]{2}{*}{\hfil\vtop{\hbox{\strut \textbf{Problem}}\hbox{\strut \textbf{grid}} }} & \multirow[b]{2}{*}{\hfil\vtop{\hbox{\strut \textbf{Number of}}\hbox{\strut \textbf{variables}} }} & \multicolumn{3}{c|}{\textbf{D-Wave 2000Q (Chimera)}} & \multicolumn{3}{c|}{\textbf{D-Wave Advantage (Pegasus)}} \\
 \cline{3-8}

 & & \textbf{Nodes} &  \textbf{Average chain length} & \textbf{Max chain length} & \textbf{Nodes} &  \textbf{Average chain length} & \textbf{Max chain length}\\
  \hline
3 x 2  & 24	&  75 & 3.125 & 5 & 36 & 1.5 & 2\\
\hline
3 x 3  & 36	& 162 & 4.5 & 7 & 79 & 2.19 & 4\\
\hline
4 x 4  & 64 & 533 & 8.33 & 12 & 204 & 3.19 & 5 \\
\hline
5 x 5  & 100 & 1213 & 12.13 & 21  & 575 & 5.75 & 10  \\
\hline
6 x 6  & 144 & - & - & - & 1069 & 7.42 & 13 \\
\hline
7 x 7  & 196 & - & - & - & 1857 & 9.47 & 17 \\
\hline
7 x 8  & 224 & - & - & - & 2578 & 11.51 & 23 \\
\hline
8 x 8  & 256 & - & - & - & 3484 & 13.61 & 27 \\
\hline
9 x 9  & 324 & - & - & - & - & - & - \\
\hline
\end{tabular}
\label{Embedding result1}

\end{center}
\caption*{\small{The table shows how problems of different sizes can be successfully embedded on the D-Wave 2000Q and Advantage quantum annealers. As expected, the greater size and connectivity of the Advantage machine enables larger problem sizes to be embedded. The largest problem grid that could be sucessfully embedded was an 8 x 8 grid.  Also of importance is the average chain length per variable as longer chains are more prone to ``chain breaks'' where the chain does not take the same value across all nodes of the physical qubit chain.}}   
\end{table}

\begin{figure}[ht]
\caption{\textbf{D-Wave quantum annealing topologies}}
\begin{subfigure}{.5\linewidth}
\centering
\caption{Chimera cell}
\includegraphics[width=.9\linewidth]{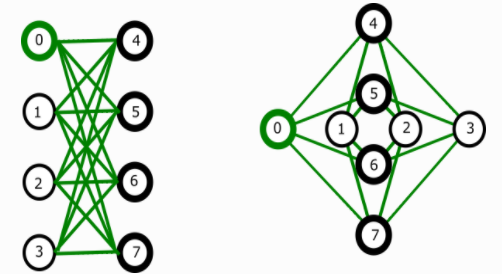}
\label{fig:sub14}
\end{subfigure}%
\begin{subfigure}{.5\linewidth}
\centering
\caption{Overall 2000Q topology}
\includegraphics[width=.75\linewidth]{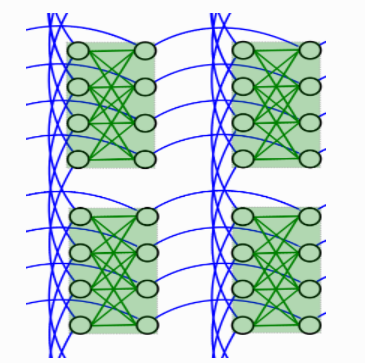}
\label{fig:sub24}
\end{subfigure}\\[2ex]
\begin{subfigure}{0.45\linewidth}
\centering
\caption{Pegasus cell structure}
\includegraphics[width=.8\linewidth]{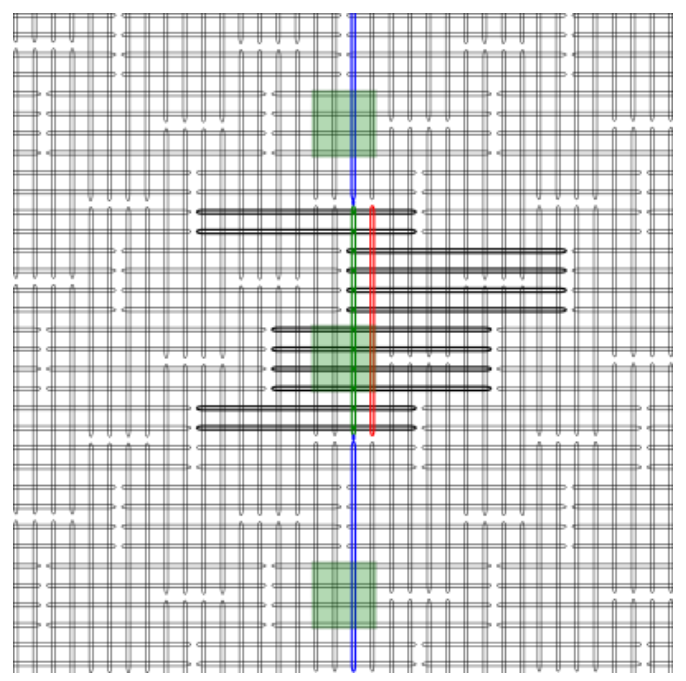}
\label{fig:sub34}
\end{subfigure}
\begin{subfigure}{.5\linewidth}
\centering
\caption{Advantage topology}
\includegraphics[width=.75\linewidth]{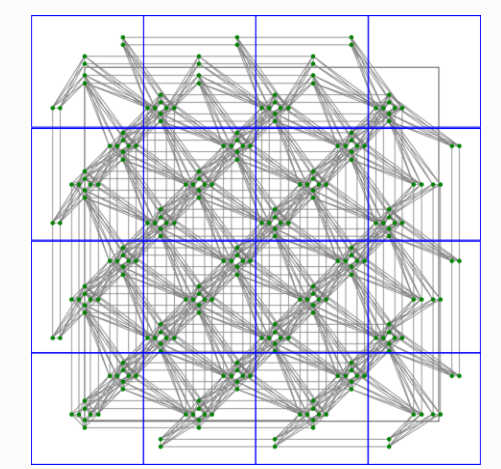}
\label{fig:sub44}
\end{subfigure}\\[1ex]
\label{D-Wave annealer topologies}
\caption*{\small{a)  and b) The D-Wave 2000Q quantum annealers are based on the Chimera cell building block which connects each qubit to 4 other qubits within the cell and 2 other qubits in other cells. This gives rise to the overall 2000Q architecture whereby each qubit is connected to 6 other qubits. c) The Pegasus topology used by the Advantage quantum annealers is more complex and involves each qubit (vertical green) making 12 internal cell connections (horizontal grey) together with 2 external connections to neighbouring qubits (blue) and 1 odd connection to a further vertical qubit (red).  This gives an overall connectivity of 15. d) The larger Advantage topology is shown - here with 16 cells of the unit structure.}}
\end{figure}


\subsubsection{Optimising QPU operating parameters}

As an analogue technology there are a number of ``operating'' parameters that need to be set in order to obtain the best results.  The main ones are the so-called chain strength, which we have already mentioned, the penalty weight and the annealing time.  

The penalty weight, $\lambda$,  needs to be sufficiently large so that solutions that do not meet the problem constraints are disfavoured by having a higher cost function value (or ``energy'' if thinking in terms of the Hamiltonian) than solutions which do.  However if $\lambda$ is too large it reduces the probability of finding the lowest cost function state.  This arises because as an analogue, physical machine there is a finite range of coupling strength within the D-Wave quantum annealers which can be applied to implement the range of coefficients of the Hamiltonian cost function - see equation (\ref{pen1a}). By undertaking a process of scaling the coefficients to fit into the available coupling strength implemented on the annealers, it can be seen that too high a value causes the available coupling strength range to be sub-optimally utilised.  It is also now possible to give some more detail on why the choice of chain strength for qubit chains that encode a single variable also impacts the success rate of running the quantum annealing process. The chain strength can be considered as contributing an additional term added to the Hamiltonian cost function of the form $-Mq_iq_j$, $M>0$ for every connected pair of qubits that encodes a given variable as a qubit chain \cite{DWave21b}.  If the chain strength is set too high then it reduces the range of coupling strength within which the other terms of the problem Hamiltonian can be accommodated. This means that the translation of the coefficients of the terms in the problem Hamiltonian into couplings on the quantum annealing device although directly related and monotonic in nature\footnote{I.e. the qubit-qubit coupling strength increases as the term coefficient increases.} is not linear.  Results obtained for a range of penalty weights, $\lambda$, are discussed in section \ref{Anneal results}.

The annealing time is, as expected, the time taken, $T$, to complete the annealing process as specified in equation (\ref{anneal eqn}) and which is reproduced here.
\begin{equation}
H(s) = (1-s)H_{init} + sH_{problem}
\end{equation}
where $s = t/T$ and $t$ is time.
The slower the annealing process the better the approximation to an adiabatic path is achieved and consequently the more likely that an output representing the problem solution will be.  This suggests that a slow anneal is desirable.  However, there are 2 offsetting factors.  The first is that the system coherence time is itself finite so any anneal needs to take place within this time. The second is that as a probabilistic process, it is necessary to repeat the annealing process many thousands of times to ensure a good probability of finding a low or the lowest cost function state; the longer the anneal time the longer it will take to complete all the repetitions and indeed, given that machine time is a scarce and expensive resource. the greater the cost.  Again therefore there is a trade off between anneal time and the total number of reads achieved and thus the overall probability of achieving a ``good'' solution in a given time.  For the ambulance problem B comparative results are reported in section \ref{Anneal results}. The annealing process described here is for so-called ``forward'' annealing. In fact there are a number of important variations which can be used to improve performance: the main ones are making use of pauses in the annealing process and ``reverse'' annealing.  Reverse annealing is discussed in section \ref{Anneal results multsize}.

There are a number of other parameters that affect the performance of a quantum annealer which include spin reversal and the detailed shape of the annealing schedule.  Spin reversal takes into account the fact that due to manufacturing imperfections there are residual biases on individual qubits, which mans they will have an increased tendency to generate an output of one or other of `0' or `1'  ; by running anneals where individual qubit spins are reversed (effectively taking $0 \rightarrow 1$ and $1 \rightarrow 0$) and then averaging it is possible to greatly reduce any impact from these residual biases.  D-Wave annealers do this as on a default basis.  

\subsubsection{Post-processing}

There is only one important feature which we will mention under this heading and that is the issue of resolving qubit ``chain breaks''.  When the output following an annealing run is reported, if the run has taken place correctly, then all the qubit values associated with a qubit chain should be the same.  However owing to the physical nature of the machine, the probabilistic character of the quantum process and the user's choice for the chain strength this is not always the case.  As an example a qubit chain of 5 qubits when read might report 4 `1's and 1 `0'.  D-Wave makes use of a ``majority voting'' protocol to resolve such internal contradictions and in this case would determine the value of the associated variable as `1'.  As this type of ``error'' is a common feature of annealing runs, it is set as a default operating protocol.

\subsection{Quantum annealing results}

\subsubsection{Effect of different choices for key operating parameters} \label{Anneal results}

In this section, we present results showing the success rate achieved in solving the ambulance location problem for a 3 x 2 grid with 2 ambulances (problem G) for different values of certain key operating parameters (penalty weight, chain strength and annealing time) discussed in the previous section.  The 3 x 2 grid was chosen due to the success rate of finding the lowest cost function value level being $\gg0$ and also $\ll 1$ so that the impact of changing parameters could be easily observed.  All the experimental data is taken from runs on the D-Wave Advantage 4.1 which is an upgraded version making use of the Pegasus architecture released in October 2021 and with a lower level of noise and a greater percentage of nodes and edges functioning relative to previous versions.  Table \ref{Dwave specs} shows key metrics of the Advantage 4.1 and its earlier Advantage 1.1 version.  Many combinations of the parameters were initially trialed.  The data presented are a subset where the parameters that are not being varied are set at or close to optimum values based on these broader trials. 

\begin{figure}[ht]
\caption{\textbf{Quantum annealing performance metrics vs $\lambda$ ratio}}
\label{fig:lambda_ratio_dependence}
\begin{subfigure}{.5\linewidth}
\centering
\caption{Probability of lowest cost function state}
\includegraphics[width=.9\linewidth]{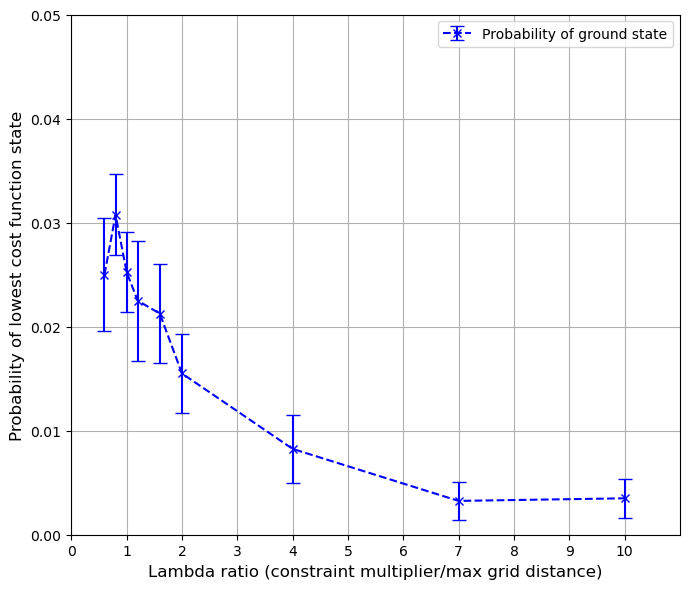}
\label{fig:lambda_ratio_dependence 1}
\end{subfigure}%
\begin{subfigure}{.5\linewidth}
\centering
\caption{Approximation ratio and probability of feasible state}
\includegraphics[width=.9\linewidth]{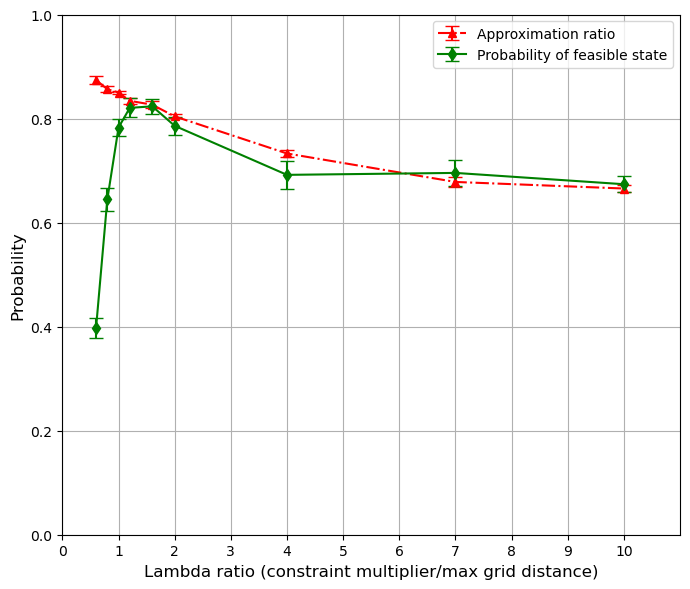}
\label{fig:lambda_ratio_dependence 2}
\end{subfigure}\\[1ex]
\caption*{\small{The results reported are for Problem variant G (a 3x2 grid instance with 2 ambulances) with 10 runs each of 400 reads using the D-Wave Advantage 4.1. The pre-factor value was 1.0 and the anneal time was 200$\mu$s. The $\lambda$ ratio varies from 0.6 to 10.0. Error bars at the 2 standard deviation level are shown. A clear peak in the probability of obtaining the lowest cost function state is seen when the $\lambda$ ratio is in the range 0.6 - 1.0.} }
\end{figure}

First we look at the impact of different values of the penalty weight, $\lambda$, which we express as a multiple of the maximum distance between any 2 grid nodes. For the 3 x 2 grid this equates to a maximum grid distance of 5, so that a $\lambda$ ratio value of 1.0 translates to a $\lambda$ value of 5. The results for $\lambda$ ratio values of 0.6 to 10.0 are shown in Figure \ref{fig:lambda_ratio_dependence}. It shows that the probability of obtaining the lowest cost function state, i.e. the optimal solution, is very sensitive to the value of the $\lambda$ ratio with larger values ($\lambda $ ratio $> 1.6$) resulting in much lower probabilities of success. In this case the highest success probability was 0.031, a factor of $\sim$10 times the lowest success rate of 0.00325.  The probability of a feasible state being found initially increases rapidly as the $\lambda$ ratio rises forming a maximum and then flattens.  In contrast, the approximation ratio gradually decreases as the $\lambda$ ratio increases.

\begin{figure}[H]
\caption{\textbf{Quantum annealing performance metrics vs prefactor value}}
\label{fig:prefactor_dependence}
\begin{subfigure}{.5\linewidth}
\centering
\caption{Probability of lowest cost function state}
\includegraphics[width=.9\linewidth]{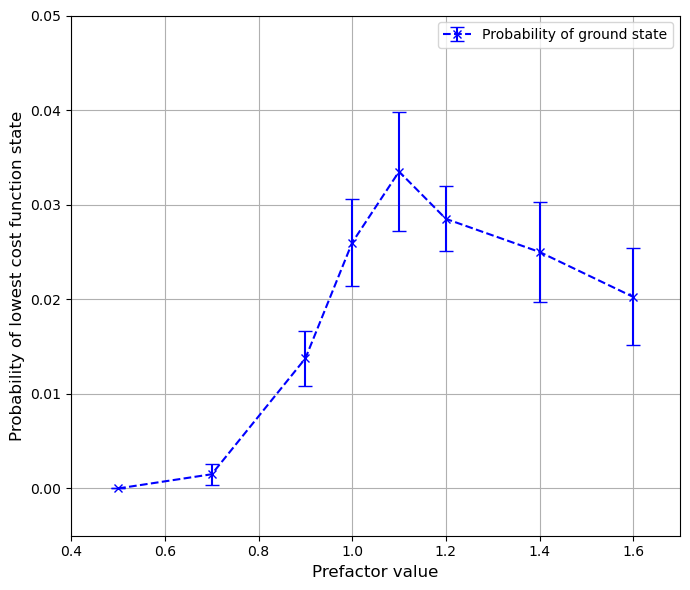}
\label{fig:prefactor_dependence 1}
\end{subfigure}%
\begin{subfigure}{.5\linewidth}
\centering
\caption{Approximation ratio and probability of feasible state}
\includegraphics[width=.9\linewidth]{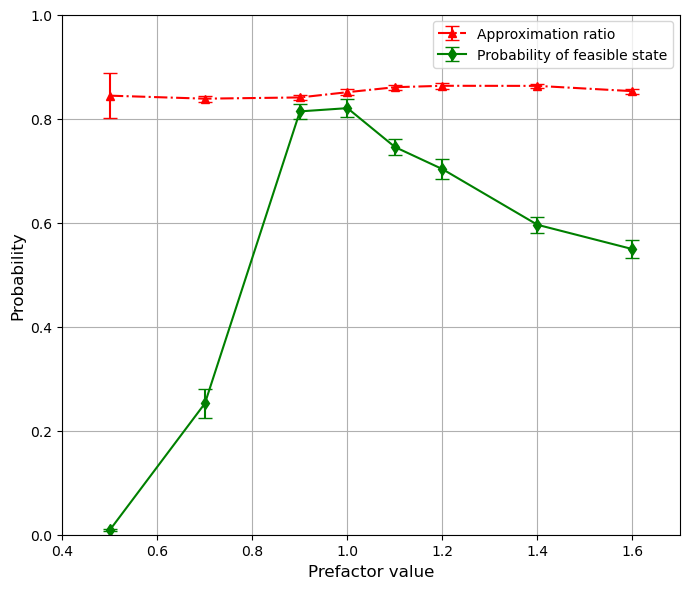}
\label{fig:prefactor_dependence 2}
\end{subfigure}\\[1ex]
\caption*{\small{The results reported are for Problem variant G (a 3x2 grid instance with 2 ambulances) with 10 runs each of 400 reads using the D-Wave Advantage 4.1. The $\lambda$ ratio value was 1.0 and the anneal time was 200$\mu$s. the prefactor is varied from 0.5 -1.6. Error bars at the 2 standard deviation level are shown. A peak in the probability of obtaining the lowest cost function state is seen when the prefactor value is in the range 1.0 - 1.2 while the probability of a feasible state is a maximum for prefactor value is 0.9-1.0.}}
\end{figure}

The value of the chain strength is captured by the prefactor value, which is related to the chain strength via the expression: chain strength = prefactor x rms(edge weight)  x $\sqrt{\text{average number of edges per qubit}}$. In Figure \ref{fig:prefactor_dependence}, the effect of varying the prefactor in the range 0.5 - 1.6 is shown.  It can be seen that a very low value for the prefactor is associated with a greatly reduced probability of a feasible state and,  since the approximation ratio is essentially  constant, a low probability of finding the lowest cost function state. The optimum values for the pre-factor are in the range 1.0-1.2 for this problem instance. Higher values for the prefactor generate a slight reduction in the probability of obtaining a feasible state and the ground state.

\begin{figure}[H]
\caption{\textbf{Quantum annealing performance metrics vs anneal time}}
\label{fig:anneal_time_dependence}
\begin{subfigure}{.5\linewidth}
\centering
\caption{Probability of lowest cost function state}
\includegraphics[width=.9\linewidth]{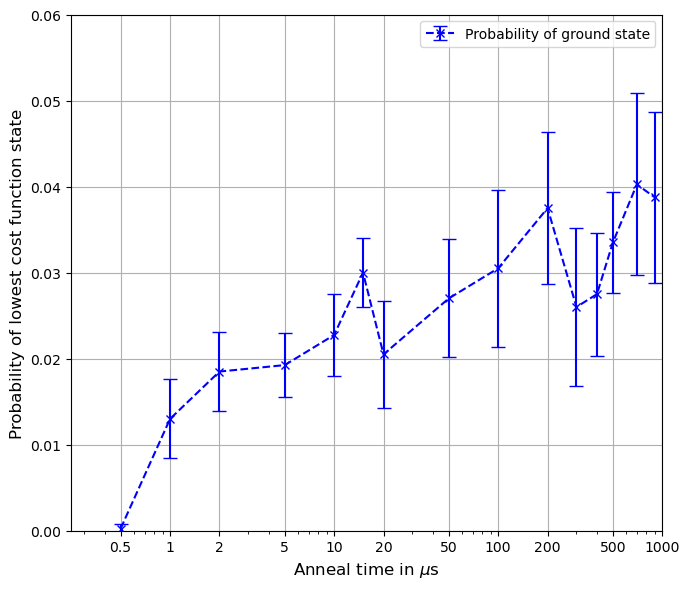}
\label{fig:anneal_time_dependence 1}
\end{subfigure}%
\begin{subfigure}{.5\linewidth}
\centering
\caption{Approximation ratio and probability of feasible state}
\includegraphics[width=.9\linewidth]{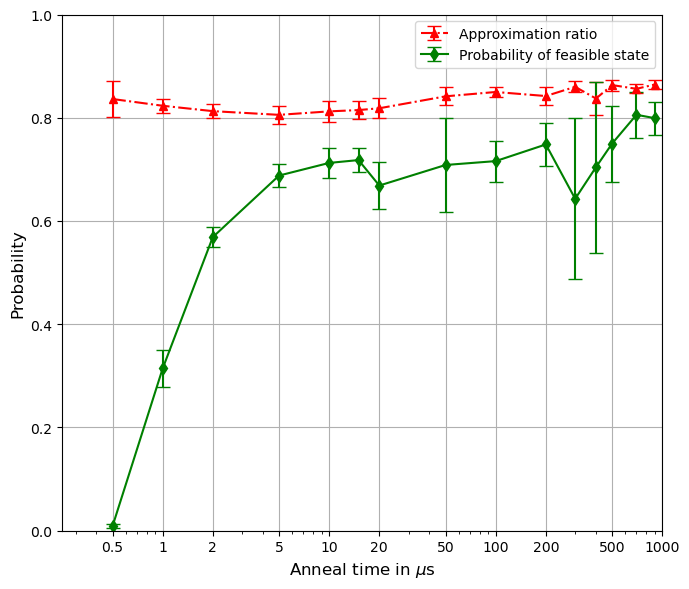}
\label{fig:anneal_time_dependence 2}
\end{subfigure}\\[1ex]
\caption*{\small{The results reported are for Problem variant G (a 3x2 grid instance with 2 ambulances) with 10 runs each of 400 reads using the D-Wave Advantage 4.1. The $\lambda$ ratio value was 1.0 and the pre-factor value was 1.0. Anneal time varies from 0.5 to 900 $\mu$s. Error bars at the 2 standard deviation level are shown. There is a slow increase in the probability of obtaining the lowest cost function state as anneal time increases in line with what might be expected from the implications of the adiabatic theorem.}}
\end{figure}

Figure \ref{fig:anneal_time_dependence} shows the impact of varying  the annealing time using a simple linear forward annealing schedule on the probability of finding the lowest cost function state.  An annealing time range of 0.5 - 900 $\mu$s was used. We see that, although not a smooth curve,  a longer annealing time improves the probability of finding the lowest cost function state by increasing the probability of a obtaining a feasible state as well as slightly improving the approximation ratio.  This is as should be expected based on the adiabatic theory supporting the annealing approach.  However it is not necessarily the best strategy to simply use the longest anneal time as really what we want to do is solve the problem as quickly as possible.  The normal measure used for this purpose is the Time to solution (``TTS''), which is usually expressed as the expectation time to achieve a 99\% probability of obtaining the optimal solution.  If the probability of finding a solution in a single read is $P_{sol}$ and the time to complete a single algorithm cycle is $T_{cycle}$, then the time to achieve a solution with 99\% confidence, $TTS_{expected}$, is given by:
\begin{equation}
    TTS_{expected} = T_{cycle} * \frac{\log(1-0.99)}{\log(1-P_{sol})}
\end{equation}

\begin{figure}[ht]
 \begin{centering}

  \caption{\textbf{Time to solution vs anneal time}}
  \label{fig:TTS}
 \includegraphics[width=0.9\textwidth, center]{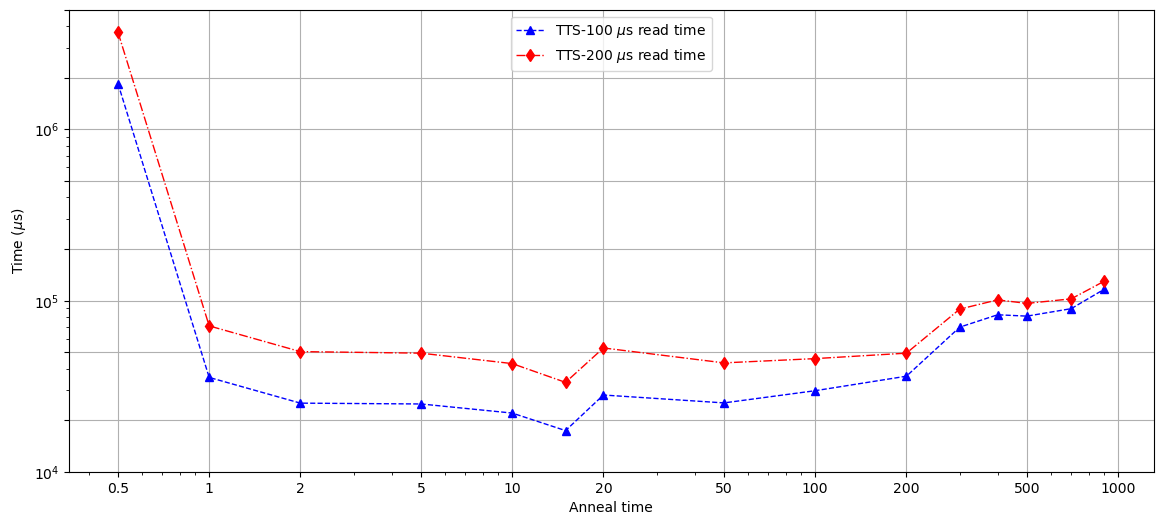}
 \end{centering}
 \caption*{\small{This figure shows the expected Time to solution (``TTS'') at the 99\% confidence level for different anneal times with the anneal cycle overhead time fixed at an assumed 100 $\mu$s and 200 $\mu$s.  The results reported are for Problem variant G (a 3x2 grid instance with 2 ambulances) with 4000 reads using the D-Wave Advantage 4.1. The $\lambda$ ratio value was 1.0 and the prefactor value is also 1.0. We find the lowest TTS is for an anneal time of 15 $\mu$s with relatively similar TTS for anneal times of 2-200 $\mu$s.  The longer anneal times ($>200 \mu$s although generating a higher probability of finding the lowest cost state are inefficient in terms of TTS.} }

\end{figure}

In the case of the D-Wave Advantage annealer, the anneal time itself is only one component of the time to complete a single anneal cycle.  In addition significant time is taken in the readout process and also in resetting the system for the next anneal-read cycle. The total time taken is to some extent dependent on the number of qubits to be read, so is not fixed. Figure  \ref{fig:TTS} plots the TTS as a function of anneal time for the 3x2 grid problem for two assumed levels of anneal cycle ``overhead'', being 100 $\mu$s and 200 $\mu$s. It demonstrates that the more efficient approach in terms of time taken is to use much shorter anneal times with anneal times in the range 2-50 $\mu$s being preferred when the process overheads are $\sim$100$\mu$s and 2-200 $\mu$s being preferred when the process overheads are $\sim$200$\mu$s.  This result reflects the fact that the total anneal cycle time is dominated by anneal process overheads for shorter anneal times. We can conclude that for this problem instance the longest anneal times ($>200\mu s$) are not efficient in terms of expected time to solution.

\vspace{2cm}

\subsubsection{Results for different problem sizes} \label{Anneal results multsize}

This section presents results for a range of increasing sizes of the target grid, i.e. problem variants B and G-J.  The operating parameter settings for each variant are shown in Table \ref{Fixed parameters for grid runs}.  They are ``optimal'' settings based on a single anneal rather than a Time to solution consideration, so have the relatively long anneal times of 200 $\mu$s.

\begin{figure}[ht]
 \begin{centering}
  \caption{\textbf{Quantum annealing performance metrics vs grid size}}
  \label{fig:problem_size_dependence}
 \includegraphics[width=0.9\textwidth, center]{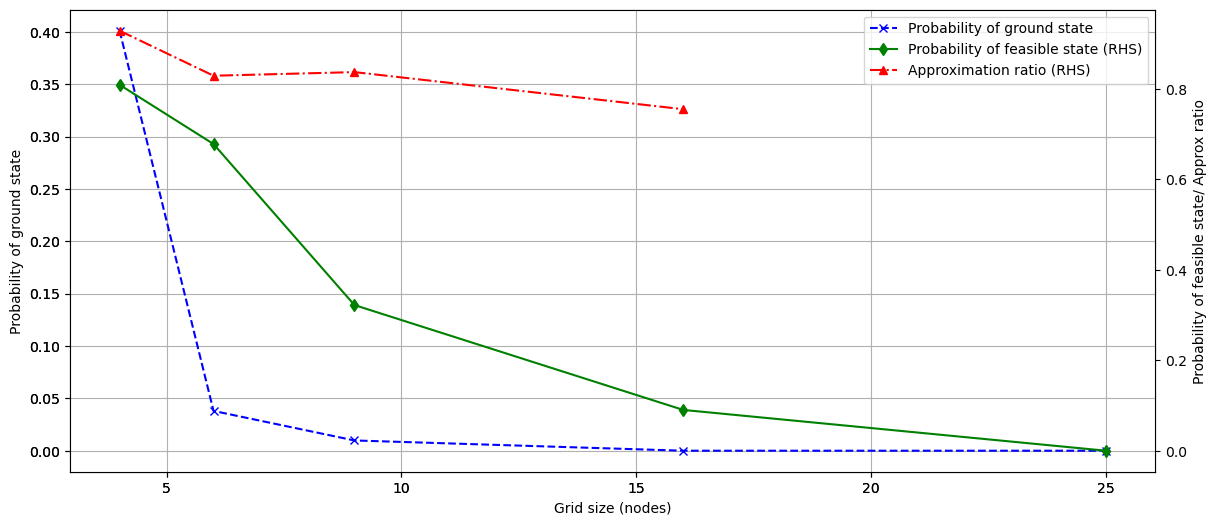}
 \end{centering}
 \caption*{\small{This figure plots performance metrics as a function of the problem size expressed in terms of the number of nodes in the problem grid.  The $\lambda$ ratio and pre-factor values are as specified in Table \ref{Fixed parameters for grid runs}. The anneal time in all cases is $200 \mu$s. The reported results are based on the mean of 5 runs each for 1000 shots.  The probability of finding a feasible state and the lowest cost function state (ground state) falls rapidly as problem size increases.}}
\end{figure}

In Figure \ref{fig:problem_size_dependence}, the probability of finding the optimum solution for the different problem sizes is plotted. We see that the rate of success in finding the optimum solution declines rapidly as the problem size increases with no solutions being obtained for problems involving 16 grid locations and above.  This is mainly attributable to the decline in the probability of finding feasible states; indeed for the 5x5 grid no feasible solutions were obtained. Although at first sight, this might be considered somewhat disappointing as an outcome, if we make use of the reverse annealing technique, we are able to significantly improve the results obtained. When a reverse anneal is undertaken the ansatz used is a single state which can be found either by taking the best ``solution'' from a series of forward anneals or a solution from using a classical algorithm.  The quantum annealer uses as its starting set up the embedded problem Hamiltonian and then reverses towards the normal equal superposition Hamiltonian.  At some chosen point the reverse anneal is stopped, the annealer is paused for a short period and then the normal forward anneal is completed.  By seeding the reverse anneal  with the lowest cost function state obtained from the forward anneal runs we were able to obtain the actual lowest cost function state for the 4x4 grid and to obtain close to the ground state solution for both the 5x5 grid and a 6x6 grid.  These latter improvements were obtained without an exhaustive search of the possible reverse annealing operating parameters and with further work the results obtained could almost certainly be improved \cite{Venturelli19a, Perdomo-Ortiz15a, Marshall19a, DWave17a}.

\begin{table}[H]
\begin{center}
\caption{\textbf{Fixed parameters for problems according to grid size }}
\vspace{0.1cm}
\begin{tabular}{|>{\centering\arraybackslash}m{3cm}| >{\centering\arraybackslash}m{2cm}| >{\centering\arraybackslash}m{2cm} |
>{\centering\arraybackslash}m{2cm} |>{\centering\arraybackslash}m{4cm} |}
 \hline
\textbf{Problem size} & \textbf{$\lambda$ ratio} & \textbf{Chain strength prefactor} & \textbf{Anneal time ($\mu$s)} & \textbf{Embedding} \\[2ex] 
 \hline
4 x 1 (4 nodes) & 0.667   & 1.41  & 200 & D-Wave random\\
\hline
3 x 2 (6 nodes) & 1   & 1.1  & 200 & Manually chosen\\
\hline
3 x 3 (9 nodes) & 1   & 1.0  & 200 & D-Wave best*\\
\hline
4 x 4 (16 nodes) & 1   & 1.0  & 200 & D-Wave best*\\
\hline
5 x 5 (25 nodes) & 1.25   & 1.41  & 200 & D-Wave best*\\
\hline
\end{tabular}
\label{Fixed parameters for grid runs}
\text{\small{* The best embedding was used from a series of trial embeddings in each case}}
\caption*{\small{The table shows the D-Wave fixed parameters used for each problem size.  These values were selected following a series of parameter tests for each problem although not all possible combinations were tested. Consequently there may be other parameter values which would give superior performance for each problem size.} }
\end{center}
\end{table}

\subsection{Quantum Annealing - challenges and developments}

Given the heuristic nature of quantum annealing (notwithstanding the theoretical backing for the basic methodological approach), it has not been subject to as much research based development as quantum gate based computing. Yarkoni et al. issued a recent review paper on the current state of quantum annealing and its application to practical problems \cite{Yarkoni2021a}. In their review, over and above the methods we have explored in this paper, they identify the potential for anneal offsets on an individual qubit basis to improve performance. The idea is that each qubit may be subject to a slow-down in the dynamics of the anneal process, which is specific to each qubit and that this can be compensated for by the precise start time of the anneal for each qubit.  The process involves a form of pre-calibration to set the offsets which is resource intensive in terms of quantum annealer time required; however the technique has been shown to increase the probability of finding the lowest cost function (or ground) state by an order of magnitude for the maximum independent set problem \cite{Yarkoni19a}.

Turning to challenges for quantum annealing, many of these relate to the capability of the current hardware; this is notwithstanding that D-Wave quantum annealers are very much more capable in terms of problem sizes that can be currently addressed by comparison with quantum gate computers.  The huge  technological achievement of the latest Advantage generation of quantum annealers must be acknowledged. However, there is still progress that is needed to enable commercially valuable problems to be clearly within the ambit of quantum annealing machines.  Three areas that are worth highlighting are the number of qubits, connectivity and coupler strength. Clearly the greater the qubit number and the higher the degree of built-in connectivity, the larger the problem size that can be encoded.  Notably, increased connectivity decreases the length of qubit chains needed to encode each logical qubit which both allows more problem variables to be represented and to decrease the likelihood of chain breaks. Currently coupler strength can only take on a quite limited range of values which means that the translation from the problem QUBO to the physical representation in the quantum annealer is approximate. A greater range and/or granularity of coupler strength would enable a more faithful realisation of the problem with likely improved performance.  In October 2021 D-Wave announced its intention to offer a further evolution of its quantum annealers - Advantage2 - which will have 7000+ qubits and connectivity of 20 (compares with 5640 qubits and connectivity of 15 in the Advantage system) \cite{DWave21c}.

\section{Discussion}
\subsection{What we have shown of interest}

In this paper we have considered a ``real-world'' optimisation problem in the form of how to locate ambulances in relation to a targeted service area represented by a grid of locations.  We initially looked at a classical heuristic algorithmic approach which performed well up to a certain problem size (10x10 grid using Tabu search) but did not find the optimal solution for larger problem sizes.   This gave a demonstration of how  optimisation problems can quickly become too large for untailored classical solvers. We then looked at the main current hybrid classical-quantum techniques for solving such a problem using both quantum gate based methods (mainly using a simulator) and quantum annealing (using a live system). We have presented these techniques  (QAOA, VQE, quantum annealing) in some detail so that those less familiar with the intricacies of quantum computing can gain a better understanding of how these methods work in practice. Our review began with some relatively basic parameter setting and extended to more advanced techniques as well as mentioning some of the more recent advances in the field.  
\newline
\newline
Specific learning from the techniques we have investigated include:

\begin{enumerate}
    \item In relation to the problem cost function Hamiltonian
        \begin{enumerate}
            \item Setting the value of the one or more penalty term multipliers optimally  is essential to maximising the probability of obtaining the lowest cost function value and state, especially for quantum annealing - grid search techniques are often adopted as a technique to achieve this.
        \end{enumerate}
    \item In relation to the Quantum gate QAOA method
    \begin{enumerate}
        \item Increasing the value of the number of steps, $p$, improves the performance although the time to solution increases due to the larger parameter space over which the optimisation occurs.
        \item The use of the XY mixer as opposed to an X mixer greatly improves the probability of obtaining the optimal solution through its restriction of the search space to the feasible or constraint satisfying space.  This is an example of the Quantum Alternating Optimisation Ansatz.
        \item There exist methods to improve QAOA performance as the number of steps increases by using a seeding approach to parameter choice for each increase in the step value. We found these methods to be much more effective than simply using increasing numbers of randomly generated initial parameter values for the rotation angles, $\beta$ and $\gamma$. The INTERP method due to Zhou et al \cite{Zhou18} worked effectively.  We also found that a simple EXTRAP (see section \ref{Efficient strategies for selecting} for details) methodology also worked nearly as well.
        \item We tried the novel approach of using 3 separate XY mixers acting on separate parts of the cost function Hamiltonian reflecting the Hamming weight constraints inherent in the chosen encoding of a 4 location 2 ambulance problem. We found that the multiple XY approach was highly effective at reducing the search space and that it was more time-efficient to limit the number of independent angle choices to \{$\beta$ =1, $\gamma = 1$\} and \{$\beta$ =2, $\gamma = 1$\}. Further we found useful additional gains from the increasing-$p$ step strategies already investigated
    \end{enumerate}
    \item In relation to quantum gate VQE method
    \begin{enumerate}
        \item Although the method does not benefit from a theoretical performance guarantee we found it worked well on different instances of the ambulance problem finding the optimal solution with a good probability
        \item As the number of layers in the ansatz increased there was no clear pattern in the performance of the algorithm
        \item Sampling methods (all qubit and causal cones) generated the solution with a lower probability than a full wavefunction simulation as is inevitable but the shortfall though material still enables a good performance from the method
        \item Running a small problem instance on an IBM QPU with a mapping designed to minimise swap gate induced infidelity  and using an all qubit measurement approach worked effectively albeit with a lower success probability due to the introduction of real-world noise
    \end{enumerate}
    \item In relation to the more mature quantum annealing technology
    \begin{enumerate}
        \item Problems of  considerably greater size can be run natively on the D-Wave Advantage quantum annealer by comparison with a quantum gate computer - we were able to investigate problems up to an 8x8 grid size albeit that no feasible solutions were found during our trial runs.
        \item Setting of the various operational parameters is critical to obtaining the best performance from this method.
        \item For a given problem instance, it is necessary to optimise each of the penalty cost term multiplier, the embedding, the pre-factor value and annealing time.  It should be noted that there are trade-offs between these operational parameters so that each cannot normally be optimally set independently, so that grid search approaches are often required.
        \item Making use of the reverse annealing technique which involved effectively seeding the annealer with a good starting ansatz, we were able to solve or get close to the lowest cost function solution for problems up to a 6x6 grid with 2 ambulances
        .
    \end{enumerate}
    
\end{enumerate}

At a higher level a number of clear lessons can be drawn.

\begin{enumerate}
    \item Quantum gate computers currently are at the stage where they can only be used to complete qualitative demonstrations of how a problem can be solved.  The corollory is that classical software and conventional computing systems are currently much better able to solve optimisation problems
    \item Quantum annealing in the form of the Advantage computer is able to solve problems of some ``real-world'' size and represents a significant gain in the size of problem that can be considered by comparison with the previous generation 2000Q system
    \item There may be opportunities to combine quantum approaches with classical methods to improve the speed of solution, particularly as the capability (size, quality) of quantum hardware improves in the next several years
    \item Quantum algorithms, particularly as regards hybrid classical-quantum approaches, are dominated by largely heuristic methodologies and although there has been a large increase in the quantity of published research in recent years, there is clearly still many areas of potential improvement to explore
    \item There needs to be significant further improvements in algorithmic methods for quantum gate computing to be able to offer commercially valuable quantum advantage in the current NISQ era; there are some encouraging signs as new refinements are developed to the existing main methods (QAOA, VQE) that this may be possible. Equally however it should be noted that there are some important challenges to be overcome including the issue of barren plateaus, an area receiving intense attention. 
    \item Quantum annealing is likely to remain the approach with the greatest near-term applicability though here too further gains in hardware capability will be necessary, particularly in terms of the degree of qubit connectivity and coupling strength sensitivity  
\end{enumerate}


\subsection{What is the outlook for technical and commercial developments?}

Given the large and still increasing resource being channeled into the field of quantum computing (mainly into quantum gate technology) by both private companies and Governments globally, rapid development progress can be expected.  However it is important to note that the technologies required for hardware advancement are inherently difficult and many gains will be required before a fault tolerant computer of commercial relevance can be produced.  There is no settled perspective on when such computers will be available:  some are suggesting the later years of the 2020s while others believe it will be the 2030s.  Relative few observers think it will take longer than this.  

As regards the NISQ era, intensive work continues to develop hybrid algorithms that can take advantage of quantum speed-ups on selected parts of problems to deliver overall commercial or scientific benefits.  It is unclear if these will be successful although it is likely that within the next few years, say by 2026, that there will be quantum gate hardware of sufficient scale and fidelity that it will not be hardware capabilities, which is the primary obstacle to the delivery of quantum advantage.

There is also the ``wild card'' of quantum annealing where future larger generations of machine may reach a critical threshold of problem size that allows commercial value to be unlocked, if not in terms of ability to solve problems than are classically solvable, then in terms of speed of solution.  It is currently unknown whether such an outcome will eventuate.

\section{Future work}

Given the broad range of quantum based methods we have covered, there is very broad scope to extend the work presented here.  Most obviously, there is scope to delve more deeply into the many variations of the algorithmic methods covered, in terms of the detailed methodologies, selection of the various meta parameters and classical techniques as well as exploring the emerging derivative methods of QAOA and VQE.

QAOA has theoretical support that as the number of layers becomes large the process tends toward adiabatic cooling and hence guarantees a ground state solution. By constrast VQE is largely a heuristic. However, the use of a of imaginary time evolution \cite{Amaro21a} does guarantee an amplification of the ground state so long as it is present with a finite probability in the initial state. Further experimentation with this algorithm may give a theoretical support to non-QAOA Ansatzes.

A further natural field of evolution is exploring the potential for problem decomposition to enable larger scale problems to be addressed through a combination of hybrid classical-quantum methods and problem decomposition.

\vspace{1cm}

\section*{Acknowledgements}

The authors would like to acknowledge the support provided by IBM as part of its quantum partner programme which included providing access to several of its quantum processors. Equally we would like to acknowledge the support provided by Amazon as part of its Activate program which includes access to its Braket quantum computer services and hardware including those of Rigetti and D-Wave.  Further T.T. would like to thank South Central Ambulance
Service NHS Foundation Trust, Bicester, UK for their assistance and Chris Bradley of The PSC, a public sector consulting firm, for the suggestion of ambulance dispatch as an interesting problem to consider.

\vspace{1cm}

\bibliography{biblibrary2}  

\newpage

\appendix

\section{Appendix} 

\subsection{Conversion from QUBO to Ising form}


The problem we discussed encoding in section \ref{Encoding the problem}  is in QUBO form where we are using the coding notation `0' and `1' but when it is run using a quantum computer it needs to take on an Ising formulation where the location variables have the values +1 and -1. This can be done for any given node by using the simple transformation $z = 1 - 2s$, where $z$ is the Ising value $\{-1, 1\}$ and $s$ is the QUBO value $\{0, 1\}$. Thus a QUBO `0' maps to an Ising `1' and a QUBO `1' maps to an Ising `-1'.  Applying the same transformation to a QUBO edge, e.g. $s_0s_1$, we obtain
\begin{equation}
    s_0s_1 \longrightarrow 0.25 * z_0 * z_1 + 0.25 - 0.25*z_0 - 0.25*z_1
\end{equation}
Naturally both expressions for the edge give the value 1 only when $s_0=s_1=1$ or equivalently $z_0=z_1=-1$; in all other cases the value 0 is obtained.

\subsection{Examples of results from multiple random angle runs using X mixer and XY mixer approaches}

\begin{figure}[ht]
\caption{\textbf{Ranked plot of cost function value and ground state probability for Problem A (5 locations, 1 ambulance) using X and XY mixers for 100 step QAOA runs}}
\begin{subfigure}{.95\linewidth}
\centering
\caption{X mixer}
\includegraphics[width=.8\linewidth]{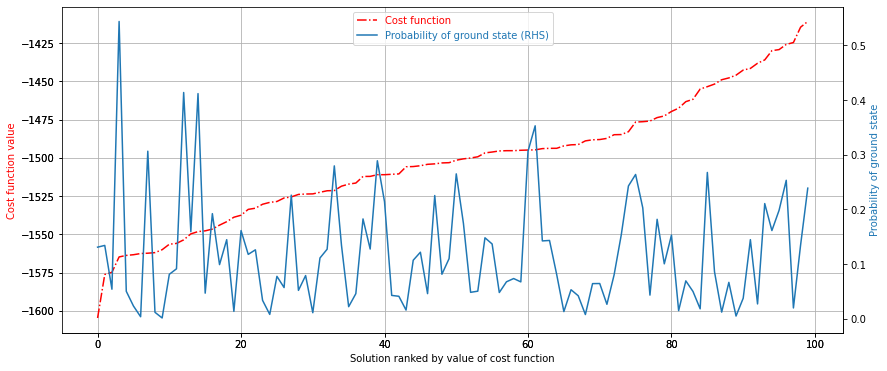}
\label{fig:sub1a}
\end{subfigure}%

\begin{subfigure}{.95\linewidth}
\centering
\caption{XY mixer}
\includegraphics[width=.8\linewidth]{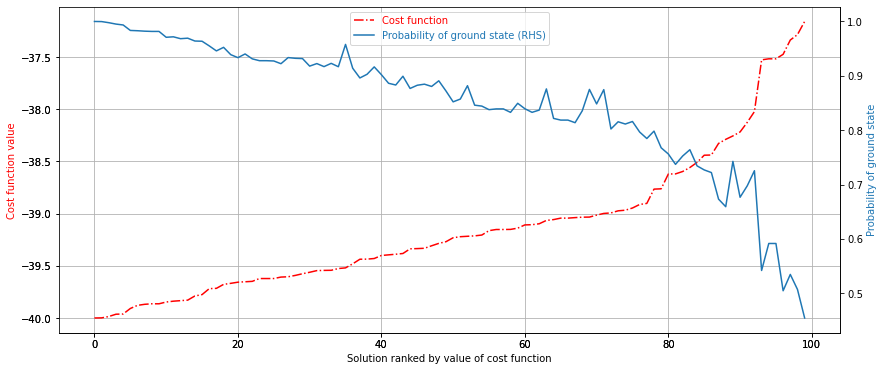}
\label{fig:sub2a}
\end{subfigure}

\label{Example plots}
\caption*{\small{}}
\end{figure}

Figure \ref{Example plots} provides 2 examples of the results obtained in terms of Cost function value and the probability of obtaining the lowest cost function state (ground state) for 100 QAOA 3 step runs where the initial angles are randomly selected from [0, $2\pi$).  The Ambulance problem is the simplest variant, problem A, with 5 locations and 1 ambulance and the classical optimiser is Nelder-Mead.   Figure \ref{fig:sub1a} uses an X mixer and initial state which is an equally weighted superposition of all possible states and figure \ref{fig:sub2a} uses an XY mixer and initial state which is a Dicke state.  In both plots the results are shown for all 100 runs ordered by cost function value.  The X mixer plot shows how the cost function is only weakly related to the probability of finding the lowest cost function state with the plot of the lowest cost function stat probability being notably ``spikey'' even for closely related cost function values.  There is however a slow increase in the probability of finding the lowest cost function state as the cost function itself declines (towards its minimum).  For the XY mixer plot, the relationship between cost function and the probability of finding the lowest cost function state is much more direct.  This closer relationship is to a significant step related to the fact that the cost function value is closer to its minimum value which for the ``best'' runs is extremely close to the minimum and accordingly the probability of the lowest cost function state approaches 1.

\subsection{QAOA Interpolation-based strategy - INTERP} \label{INTERP}


Zhou et al.\cite{Zhou18} identified a strategy for selecting the angles $\beta$ and $\gamma$ for use with QAOA as the number of steps, $p$, is iteratively increased. They called it the INTERP as it uses a form of linear interpolation to generate successive angles for $\beta$ and $\gamma$. It assumes that a good set of angles is found for the case $p=1$, for example by trying multiple sets of random chosen seed angles and optimising each to obtain a best set of angles for $p=1$.  Let us denote the $i^{th}$ angles for a QAOA with $p$ steps ($p>1$) as $[\beta_i]_p$ and $[\gamma_i]_p$ and the angles that result from an optimisation run using QAOA as $[\beta_i^*]_p$ and $[\gamma_i^*]_p$.  Then the INTERP strategy relates the seed angles for the $(p+1)$ QAOA optimisation by:

\begin{equation}
    [\beta_i]_{p+1} = \frac{i-1}{p}[\beta_{i-1}^*]_p + \frac{p-i+1}{p}[\beta_i^*]_p
\end{equation}
where $i \in \{1,2, ..., p+1\}$ and $[\beta_{0}^*]_p \equiv [\beta_{p+1}^*]_p \equiv 0$. A similar equation specifies how $\gamma$ angles are incremented as $p$ increases. The process is repeated until the chosen level of $p$ is reached or some other termination condition is reached, e.g. based on the value of the cost function.

\newpage

\subsection{SPSA optimiser gradient descent parameter choice}
\label{SPSA_met}


To use the classical SPSA and L-BFGS-B optimisers, the gradient (`$g$') is estimated by evaluating the objective function at 2 points given by an angle (parameter value) $\pm \epsilon$. $\epsilon$ must be supplied by the user and should be small compared to $\pi$ as all the parameters of the Ansatz are for rotation gates where $2\pi$ gives a full rotation. Additionally, the SPSA method selects randomly, at each iteration, which parameters of the ansatz to change, and so specifies a random direction. These parameters are then updated by an amount ${g*step\_size}$, where the  $step\_size$ is supplied by the user.
The success of the optimization varies significantly with these choices. 
 The Noisyopt module for Python includes the minimizeSPSA function which proposes asymptotically reducing $\epsilon$ and $step\_size$ after each iteration \cite{Noisyopt}. We adopted this proposal, and so supplied  initial step choices for 1) $a$,  the scaling parameter used for the gradient evaluation step size and 2) $c$, the scaling parameter for function evaluation parameter step size,  as well as the total number of iterations, $n_{iter}$, in our optimization.

For the $k^{th}$ iteration;

$${step\_size}_k  = \frac{a}{(0.01*n_{iter} +k+1)^\alpha} $$

 $${\epsilon}_k  = \frac{c}{( k+1)^\gamma} $$
 where we use the default values suggested in the Noisyopt documentation of $\alpha=0.602,\gamma=0.101$

\begin{figure}[tbp]
\begin{center}

\caption{Comparison of average EV/(Ground state energy) for two step choices used by SPSA }
    \includegraphics[width=.9\linewidth]{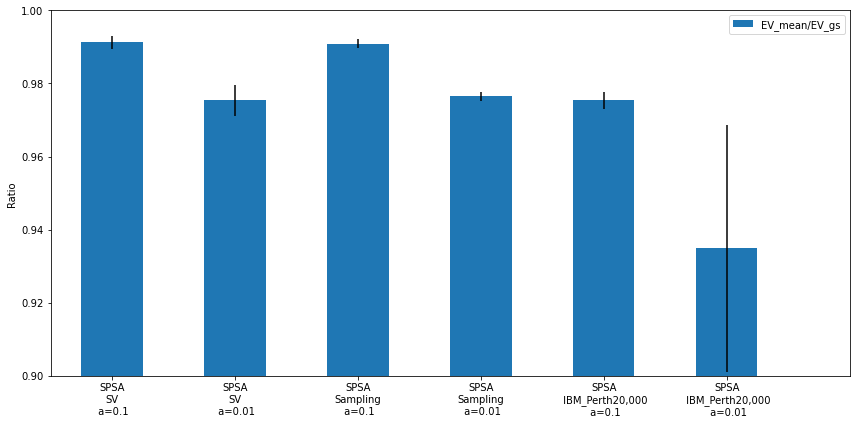}
    test
    \label{SPSA_step_choice_EV}\\
    \caption*{\small{Samples from a quantum circuit (9000 shots on the Qiskit simulator and 20,000 shots on IBM\_Perth) were used to evaluate the expected value ('EV') for a given set of circuit parameters. The SPSA optimizer used 100 iterations. For each random set of starting angles (which supply the 13 parameters for the VQE Ansatz) a final angle is selected by minimizing its EV.  The ratio EV/(Ground state energy) shown here is the average of 100  starting angles (10  starting angles on IBM\_Perth). 
     $c$, the starting perturbation size for SPSA was set equal to $a$, the starting step for the gradient evaluation. The results for $a=0.1$ and $a=0.01$ are shown. }}
\end{center}
\end{figure}

Figure \ref{SPSA_step_choice_EV} shows the impact of two choices of $a$ and $c$. For all three methods (`SV', `Sampling' and QPU),  the minimum found was statistically significantly closer to the lowest cost function state when $a = c = 0.1$, than when $a = c = 0.01$ \footnote{We also tried  $a = c = 1 \text{x} 10^{-3}$, $ a = c = 1 \text{x} 10^{-5}$, $a = c = 1 \text{x} 10^{-7}$, but these performed much less well and barely converged in 100 iterations.}.  When sampling on the QPU we chose many more shots (20,000 rather than 9,000) so as to reduce the sampling noise, but this was not enough to make up for the fidelity noise inherent in a NISQ QPU. 

A VQE heuristic method searches for a weighted set of mixed states that minimizes the average or expected energy of those states (`EV'). If the EV is \textit{lower} than the energy of the first excited state (the state whose energy, when ranked,  is one higher than the lowest cost function or ground state) then to achieve that average energy, or lower, the ground state must be represented in at least one of the states. When that condition is not met it is not guaranteed that the ground state is represented at all. For problem A (5 locations, one ambulance) the first excited state has an EV/ (ground state energy) ratio of 0.98, so any EV/ (ground state energy) lower than 0.98 cannot guarantee that the ground state probability is above zero. Similarly, when two methods both have ratios below 0.98, the one closer to 0.98 is not guaranteed to have a higher probability of measuring the lowest cost function or ground state.

\begin{figure}[ht]
\begin{center}
\caption{Figures of merit for two step choices used by SPSA}
test
\includegraphics[width=.9\linewidth]{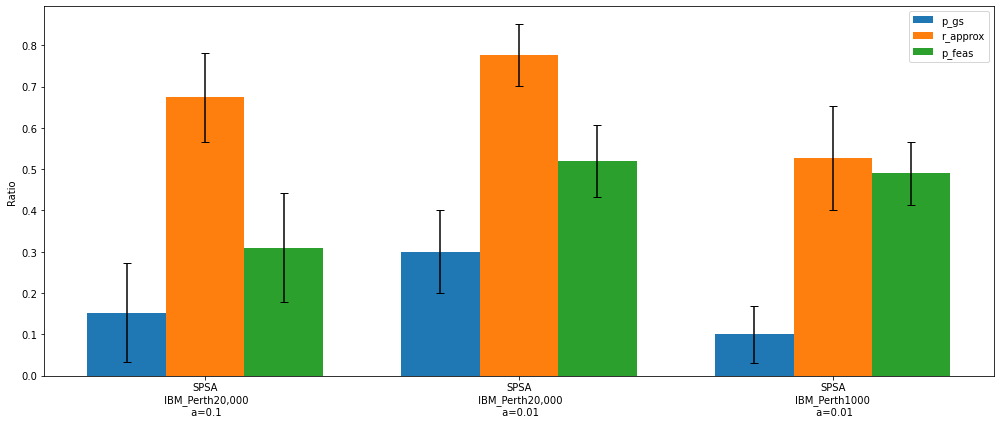}
\label{SPSA_step_choice_quality}\\
\caption*{\small{Whilst Figure \ref{SPSA_step_choice_EV} shows the EV/(Ground state energy) ratios for a VQE with SPSA optimizer and QPU IBM\_Perth. This figure shows the corresponding figures of merit defined below:  
The probability that a ground state will result from sampling an optimized Ansatz, defines p\_gs. Whilst p\_feas is the probability that a feasible state will result. r\_approx is measured in the subspace of feasible states, and is defined  in (\ref{r_appr}). }}
\end{center}
\end{figure}

As an example, selecting SPSA parameters $a=c=0.1$ improved the average EV found by the VQE run on IBM\_Perth statistically significantly to 0.975 from 0.935 for $a=c=0.01$. However, it did not improve the quality of the average solution at all: it was slightly lower than for $a=c=0.01$, albeit not statistically significantly lower (see Figure \ref{SPSA_step_choice_quality}.)

\subsection{Methods - computing hardware and software}

With the exception of section \ref{QG hardware} and \ref{QA_impl}, where quantum hardware made by IBM and D-Wave is used, the other results reported make use of conventional hardware and software.  The quantum simulator software used was a combination of Pyquil from Rigetti and Qiskit from IBM.  We undertook consistency tests from the two providers on the same problem instances and confirmed that they gave equivalent results.  In addition D-Wave Ocean software was used for the classical simulations using simulated annealing and tabu solvers undertaken in section \ref{classical_opt}. The hardware used was a mixture of Intel i7 and i5 CPU desktops and laptops.    



\end{document}